\documentclass[a4paper,10pt]{scrartcl}

\usepackage{latexsym}
\usepackage{amsmath}
\usepackage{amssymb,amsfonts}
\usepackage{graphicx}
\usepackage{cite}
\usepackage{bm,wrapfig}   
\usepackage{upgreek}

\newcommand{\beq}{\begin{equation}}
\def\eeq{\end{equation}}
\def\bea{\begin{eqnarray}}
\def\eea{\end{eqnarray}}

\title{Tunnelling Methods and Hawking's radiation: achievements and prospects}
\subtitle{Topical Review}
\author{L.~Vanzo\thanks{vanzo@science.unitn.it},
G.~Acquaviva\thanks{acquaviva@science.unitn.it}\, and
R.~Di~Criscienzo\thanks{rdicris@science.unitn.it}
\\
\\
Dipartimento di Fisica, Universit\`a di Trento,
\\
and I.N.F.N., Gruppo Collegato di Trento,
\\
via Sommarive~14, 38100~Povo, Italy}
\date{}

\begin{document}

\maketitle

\begin{abstract}
The aim of this work is to review the tunnelling method as an alternative description of the quantum radiation from black holes and cosmological horizons.  The method is first formulated and discussed for the case of stationary black holes, then a foundation is provided in terms of analytic continuation throughout complex space-time. The two principal implementations of the tunnelling approach, which are the null geodesic method and the Hamilton--Jacobi method, are shown to be equivalent in the stationary case. The Hamilton--Jacobi method is then extended to cover spherically symmetric dynamical black holes, cosmological horizons and naked singularities. Prospects and achievements are discussed in the conclusions.   
\end{abstract} 

 
\newpage
\tableofcontents

 
 \section{Introduction}
 \label{sec1}
 
Since the discovery of quantum black hole thermal radiance by 
Hawking \cite{Hawking:1974rv} (extremely good reviews and surveys are in \cite{Parentani:1992me}, \cite{Brout:1995rd}, \cite{Traschen:1999zr}, \cite{Jacobson:2003vx} and \cite{Page:2004xp}), it became pretty clear that something remarkable concerning the interface of gravity, quantum
 theory and thermodynamics was at work. In the usual picture, a radiating black hole
 loses energy and  therefore shrinks, evaporating away to a fate which is still debated. 
 Many  new ideas came out from the recognition that quantum field theory implied a
 \textit{thermal spectrum}: the most impressive probably being 't Hooft's idea of a {\it
 dimensional reduction} in quantum gravity  and the associated holographic description
 (indicating a drastic reduction in the counting of degrees of freedom in finite regions,
 which scale with area rather than volume) \cite{tHooft:1993gx,Susskind:1994vu}; and
 the {\it principle of black hole complementarity} aimed to reconcile the apparent loss of
 unitarity implied by the Hawking process with the rest of physics as seen by external
 observers \cite{Susskind:1993if}. But there were also other, more practical, issues regarding these matters, some of which
 bewildered scientists since the very beginning and that have been only partly resolved. A
 key issue is that the original derivation of Hawking radiation applied only to stationary
 black holes, while the picture above uses quasi-stationary arguments. Actually, an
 evaporating black hole is non-stationary. Nevertheless, a surprising aspect of the semi-classical result is that the radiation caused by the changing metric of the collapsing star
 approaches a steady outgoing flux at large times, implying a drastic violation of energy
 conservation. This certainly means that one cannot neglect the back-reaction problem which, however, has not been solved yet in a satisfactory way. 
A sample of other key issues have to deal with the final state of the evaporation process,
 the thermal nature and the related information loss paradox, the Bekenstein--Hawking
 entropy and the associated micro-states counting, the trans-planckian problem, and so on.  It was to address some of these questions that some
 alternative derivations and descriptions of the Hawking's emission process began to
 appear over the years 
\cite{Jacobson:1993hn,Visser:1997yu,Visser:2001kq,Kiefer:2001wn,Vaz:2002xb,Barcelo:2006uw,Unruh:2004zk,Schutzhold:2008tx,Kiefer:2007va},
 one of these being the so called \textit{tunnelling method}, the {\it leitmotiv} of this review. 

Before introducing it in full mathematical detail, we shall spend the next few paragraphs
 to spell out some general features of this method and how it compares with the general
 properties of quantum fields and states in the presence of a black hole\footnote{For those
 interested, an up to date resource for papers, books and other stuff about black hole
 physics and astrophysics is the ``Resource Letter BH-2: Black holes'' by Marolf and
 Gallo \cite{Gallo:2008bh}.}. 
\
\\

The tunnelling method has to do with the particle interpretation of the emission process,
 which would be the most natural way to explain the loss of energy suffered by the
 radiating black hole, but also poses a problem. It is that while the original derivation of
 Hawking radiation used the theory of quantum free fields propagating in a fixed
 gravitational background, for the case at hand a particle
 interpretation in the near horizon region of a black hole space-time was found to be far from unique and therefore
 inherently ambiguous. This fact is deeply rooted into the physical nature of the
 problem, which is that the general covariance of the physical laws allows many inequivalent
 choices of time, as was explained several times (see, \textit{e.g.} \cite{Birrell:1982ix, Schutzhold:2000jn}). It was
 partly to overcome such difficulties that DeWitt launched in the Seventies the ``stress-energy-tensor program'' aimed to compute the expectation values of the stress tensor and other observables to better describe quantum effects in a curved space-time
 \cite{DeWitt:1975ys}.  Thus, it was found that it is not in general possible to divide
 $\langle T_{\mu\nu}\rangle_{{ ren}}$ into  a real particle contribution and a vacuum
 polarisation part in an unambiguous  way. However, it will always be possible to choose any coordinate system, as long as it
 is regular across the horizon, and use it to define an observer dependent vacuum relative to which a
 particle definition is feasible. As we will see, with the tunnelling method we are only
 concerned with the {\it probability} that such an observer dependent notion of particle be emitted from
 the horizon of the black hole. If this probability is a coordinate scalar it
 will not depend strongly on what particle concept one employs. In fact, we can call it {\it
 a click event in a particle detector} if we like, without committing ourselves with the
 particle concept; concept which remains of great heuristic value however,  and will find its
 realisation in the choice of the particle action used to compute the tunnelling
 probability. Such an observer dependent notion of particles has been advocated also by
 Gibbons and Hawking in their treatment of de Sitter radiation \cite{Gibbons:1977mu},
 on the ground that an observer independent definition of particles is not relevant to what
 a given observer would measure with a particle detector.

The second aspect we would like to mention is the fact that, as it will be seen, the tunnelling
 method uses only the classical action of a single massless, spin-less particle and therefore
 appears to be state independent. It does what it does by relating the particle emission to
 an imaginary contribution to the classical action localised at the horizon, which only
 comes from the local geometry. This can be traced back to the work of Damour and
 Ruffini \cite{Damour:1976jd} with contributions of Padmanabhan \textit{et al.}
 \cite{padma:1999,padma:2001,padma:2002,padma:2004}, Massar and Parentani
 \cite{Massar:1999wg} and Kraus and Wilczek, Parikh and Wilczek \cite{Kraus:1994by,Kraus:1994fj,Parikh:1999mf} (see also Kraus's Ph.D. thesis \cite{Kraus:1995vr}). To what extent this is so was recently studied in \cite{Moretti:2010qd} where, for a large
 class of quantum states with Hadamard short distance singularity, it was shown that the two-point correlation function, which is directly related to the tunnelling probability, has a universal thermal appearance whose temperature is, needless to say, the Hawking temperature. 
\
\\

Now all these important findings have the drawback that they have been proved, and one
 can easily understand why, only for stationary black holes. In that case the tunnelling
 picture has not much more to say than was already known, at least for the simplest solutions, besides helping to understand black hole radiation from a different viewpoint. Where exact calculations can be done, it typically gives a less detailed picture of the radiation process since it is mainly related to a
 semi-classical emission rate. The cases where the method is more
 powerful have to do either with more intricate stationary solutions or else with truly dynamical black holes. We can list here some of its achievements and properties:

\begin{enumerate}

\item The estimation of the leading correction to the semi-classical emission rate arising
 from back-reaction to the background geometry, whose introduction by Kraus and Wilczek,
 Parikh and Wilczek \cite{Kraus:1994by,Kraus:1994fj,Parikh:1999mf} motivated the tunnelling method in a
 form that will be discussed soon; its extension to Anti--de Sitter (AdS)
 \cite{Hemming:2000as}, de Sitter (dS)
 \cite{Stotyn:2008qu,Volovik:2008ww,parikh:2002,medved:2002,desitt} and higher
 dimensional black holes; the existence of correlations among successive emission events \cite{Zhang:2009jn,Zhang:2009td,Israel:2010gk}; 
 
\item The original tunnelling method can be generalised to a Hamilton--Jacobi variant, 
 originated with the work of Padmanabhan and collaborators
 \cite{padma:1999,padma:2001,padma:2002,padma:2004} and systematically applied 
 either to stationary or dynamical black holes (see {\it e.g.} \cite{angh:2005,kernermann,mann3,Chatterjee:2007hc,Visser:2001kq,Criscienzo:2007fm,NV} for a sample of papers). For dynamical black holes this was particularly important, since even approximate quantum calculations are notoriously hard. Moreover, in this variant it manifestly preserves general covariance;

\item Supplemented with a more precise and more general mathematical definition of a
 local horizon, the Hamilton--Jacobi variant can be applied to any sort of horizon within
 this class, and in particular to cosmological and weakly isolated horizons \cite{Wu:2007ty}
 (for definitions see below).
 It can also be applied to past horizons and white holes, in which case a clear notion of
 temperature emerges in complete analogy to black holes; 

\item The tunnelling method strengthens the connection of the
 semi-classical rate with the surface gravity of the horizon even for dynamical black holes,
 provided that opportune definitions of such quantities are employed. In this respect its
 application to extremal black holes should have something to say;

\item  The tunnelling picture  can promptly give indications on what occurs in other situations. One example is the WKB approach to Unruh's radiation, reviewed in 
\cite{deGill:2010nb}.  It can also be applied to see whether naked
 singularities are going to emit radiation, or to study the decay of unstable particles that in the absence of gravity would be stable by the action of some conservation law. This is valuable given the great efforts that are normally necessary to analyse quantum effects in the presence of gravity.
\end{enumerate}

It remains to explain how the tunnelling picture works and to what extent it gives a consistent picture, at least from a logical point of view. This will cover the rest of this review whose structure, we anticipate, will essentially follow the material just indicated in listed form above.
\
\\

In Section 2, the methodological core of the topic is outlined and applied to various
 stationary solutions.  In Subsection \ref{2.1} we present the original method introduced by Parikh and Wilczek to study the black hole radiance --- namely the \textit{null geodesic method} ---
 which is subsequently applied to verify the thermal properties of a static, spherically
 symmetric Schwarzschild black hole in the scalar sector. In Subsection \ref{2.2} we comment results pertaining to statistical correlations of successive emission events, as displayed by the tunnelling method. In Subsection \ref{2.3} and Subsection \ref{2.4}, after pointing out some flaws of this approach, we extend it to the more recently conceived
 \textit{Hamilton--Jacobi tunnelling method}: this approach is examined in
 great detail in order to stress those points that --- given the vastness and heterogeneity of
 opinions in the literature --- might lead to confusion. In Subsection \ref{2.5} we present the
 extension of the Hamilton--Jacobi tunnelling method to the fermionic sector.  The
 description of the tunnelling phenomenon in non-static, but still stationary space-times is
 undertaken in rest of the section: starting from the Kerr and Kerr--Newman solutions in
 Subsection \ref{2.6}, we then conclude the section showing in Subsection \ref{2.7} more peculiar scenarios --- such as non axis-symmetric or lower and higher-dimensional solutions.

In Section \ref{3:found}, we study the mathematical foundation of the two tunnelling methods.
The Hamilton--Jacobi method, preferable in view of its manifest covariance, is examined in Subsection \ref{3.1}.  It is found to be equivalent to the null geodesic method, at least in its
 generalised form, in Subsection \ref{3.2}. The aforementioned variety of opinions also
 comprehends a number of alternative proposals to the two methods, some of which are
 summarised in Subsection \ref{3.3}.  

In Section \ref{4:dyna}, we approach the dynamical case but restricting ourselves to spherically
 symmetric non vacuum black holes. Though it may appear strange that stationary black 
holes, being a particular case of dynamical ones, are discussed first, there is a
 reason: in the stationary case there is much more control on the mathematics of
 the tunnelling picture. In particular, it is possible to use both the null geodesic method as
 well as its Hamilton--Jacobi variant to deduce the exact semi-classical thermal spectrum of
 the Hawking radiation, a fact which greatly enhances the understanding of the tunnelling
 picture in a way which is not available to dynamical situations. At the same time it
 provides the foundation for the dynamical treatment and inspires extensions to more
 general cases. The Section is further divided into several sub-sections. In Subsection \ref{4.1} we
 describe various definitions that have  been proposed over the years to characterise a
 black hole horizon and corresponding definitions of the surface gravity. Again
 few dynamically inequivalent definitions will be reviewed, putting particular emphasis on
 the one which the tunnelling method privileges. In Subsection \ref{4.2} we review material
 pertaining to spherically symmetric space-times and the form that surface gravities and
 horizons take in this highly symmetric case. In Subsection \ref{4.3} we finally come to grasp
 the tunnelling mechanism. After a careful scrutiny of the escaping paths we derive the
 basic results; the computation of the imaginary part of the action, its relation with the
 surface gravity and the ensuing quasi-thermal interpretation.

In Section \ref{5:coho}, we discuss how the tunnelling picture generalises to cosmological
 horizons, naked singularities and particle decays. We shall see that the first two cases are
 governed by a simple pole singularity while decays correspond to branch cut singularities
 of the classical action.  In Subsection \ref{5.1} we review few facts regarding Friedmann--Robertson--Walker space-times (abbr. FRW) with emphasis on the existence of horizons,
 their structure and surface gravity. In Subsection \ref{5.2} we apply the tunnelling method to
 cosmological horizons and find their emission rate. In Subsections \ref{5.3} and \ref{5.4} we
 analyse its application to particle decays and particle creation by black hole singularities.
 The former was extensively studied in de Sitter space-time, where exact results can be
 obtained, but the tunnelling method is not restricted to particular space-time choices. In
 Subsection \ref{5.5} we review the possibility that naked singularities themselves radiate
 away their (suitably defined) mass. Indications of this phenomenon are supported by the
 tunnelling method.
 
 We end with Section \ref{6:conc} where achievements, prospects and conclusions are summarised.
\newline

As a final note, since we found the literature on quantum tunnelling to be quite
 extensive\footnote{An INSPIRE search gave us more than 500 records about tunnelling from black
 holes, to date.}, the references will not exhaust all the existing literature. We struggled to
 cover a lot of papers, and especially those that appeared more meaningful to our eyes,
 being aware that this will inevitably come to reflect some personal taste. A sample of the
 vast literature is here collected for the reader's benefit, beyond those already cited. For studies on tunnelling from a general point of view see \cite{others-tunn}; tunnelling of fermions and bosons from 
 stationary black holes in asymptotically flat, de Sitter or anti-de Sitter spaces and other exotic objects in several coordinate systems were considered in \cite{others-static}; tunnelling of bosons and fermions in higher and  lower dimensions are covered in \cite{others-hilow}; for contributions to dynamical black holes see \cite{others-dyna}, for cosmology \cite{others-cosmo}. Our sincere apologies are directed to the many unwittingly omitted. 

We shall use the following conventions throughout: the metric signature is $(-,+,\cdots,+)$; the constants $c$, $G$, $k_{ B}$ and $\hbar$ set equal to unity unless explicitly stated; we use an index notation for tensor components. Greek indices are used to denote space-time
components of a tensor; mid-Latin indices as $i, j$ are typically used to denote the $0,1$ component; late-Latin indices as $m,n$ for purely spatial tensors part. If not otherwise stated, an overdot  will represent  a time derivative, and a prime a spatial derivative in the (case by case) relevant spatial coordinate.

\section{Quantum Tunnelling: Stationary Black Holes}
\label{2:stat}

According to quantum field theory, the physical vacuum -- the state with no real particles -- is in general a complex entity populated by virtual particles which constantly are created, interact among themselves and are then annihilated in this vacuum. In the absence of external fields, the vacuum is usually stable in the sense that virtual particles are not able to survive long enough to become real. However, it is now well proved that external fields are able to convert short-living particles into real ones just  supplying enough energy to the process. 

Considering particle creation by a static gravitational field endowed with a Killing vector
 $\xi^\alpha$, the energy of a particle is $\omega = - p_\alpha \xi^\alpha$, where
 $p^\alpha$ is the particle's four-momentum. For massive (massless) particles, the
 momentum $p^\alpha$ is a future-directed time-like (null) vector. Thus, the energy
 $\omega$ of a particle is always positive in the region where the Killing vector is also
 future-directed and time-like. It follows that particle creation in a space-time region with
 $\xi^\alpha$ future-directed and time-like is forbidden by energy conservation. Such
 considerations provide a heuristic argument \cite{frolov} to conclude that a static
 gravitational field can create particles only if the space-time contains a black hole. In fact,
 when a virtual particle pair is created just inside the horizon, the positive energy virtual
 particle can tunnel outside --- no classical escape route existing --- where it materialises as
 a real particle; in alternative, for a pair created just outside the horizon, the negative
 energy virtual particle, forbidden outside the horizon, can tunnel inwards \cite{parikh}.
 In either cases, the negative energy particle is absorbed by the black hole, resulting in a
 black hole mass decrease, while the positive energy particle escapes to infinity, visible to
 distant observers (Hawking radiation).

In 1974 Hawking \cite{Hawking:1974rv} proved that indeed black holes are classically
 stable objects that, due to quantum fluctuations, become unstable with respect to particle
 emission. Nowadays, several methods are known for deriving Hawking radiation and
 calculating its temperature, but none of them has been able to catch the intuitive picture
 above as the tunnelling method proposed a decade ago by Parikh and Wilczek \cite{Parikh:1999mf}:  it were not for the inclusion of back reaction, their method could be considered a semi-classical version of Hartle--Hawking's path integral derivation of black hole radiance \cite{hartle}.

\subsection{The null geodesic method}
\label{2.1}

The so called \textit{null geodesic method} --- as named by Kerner and Mann \cite{kernermann} --- derives black hole radiance from few reasonable assumptions, namely: energy conservation, detailed balance and the use of non-singular coordinates across the horizon.\\
Energy conservation requires fixing the total energy of the space-time before and after
 particle emission. Since black hole mass and volume are linked together, a mass
 reduction due to the emission of a particle translates into a size contraction; so one might
 worry how to deal with quantum fluctuations of the metric originating from such
 contraction. However, this is not a problem as far as we consider transitions between
 zero-spin geometries. In this case in fact, no graviton quantisation is involved or, said in
 other words, passing from different spherically symmetric configurations does not
 produce gravitational waves. As a consequence, the only degree of freedom remained in
 the problem is the position of the emitted particle (actually, a thin shell). Thus, to keep
 things as simple as possible, we can restrict to consider uncharged, static, spherically
 symmetric black holes emitting neutral matter,  referring to next sections treatment of the
 most general case. 

Because of Birkhoff's theorem \cite{hawkingellis}, we are dealing with Schwarzschild
 geometry which, written in the Schwarzschild frame, reads
\begin{equation}
 d s^2 = -\left(1- \frac{r_{ g}}{r}\right)d t^2 + \left(1- \frac{r_{ g}
}{r}\right)^{-1} d r^2 + r^2 (d \theta^2 + \sin^2 \theta \,d \phi^2)
\label{schwarzschild}
\end{equation}
where the \textit{gravitational radius} is $ r_{ g} = 2 M$. Computing the free fall
 acceleration of a body initially at rest in the Schwarzschild frame, we get
\begin{equation}
 a := \sqrt{a_n a^n} = \frac{M}{r^2 \sqrt{1-r_{ g}/r}}\,.
\label{acceleration}
\end{equation}
The geometry (\ref{schwarzschild}) becomes singular at the gravitational radius and the
 acceleration (\ref{acceleration}) becomes infinite in the same limit. Nonetheless, near
 $r_{ g}$ the space-time is regular ({\it i.e.} neither curvature singularities or geodesic
 incompleteness): an indication of the fact that the Schwarzschild frame ceases to be
 defined at $r=r_{ g}$ (For $r \leq r_{ g}$ systems must be non-rigid in
 the sense that the metric components $g_{\alpha\beta}$ must be functions of time, cf.
 \cite{frolov}). \\
It is well-known that the gravitational radius $r_{ g}$ represents the Killing event
 horizon of Schwarzschild black hole \cite{hawkingellis}; thus, in order to deal with
 tunnelling of a particle across it, we need to cover the region $r \sim r_{ g}$ with
 regular coordinates. A possible choice, not the only one, is represented by Painlev\'{e}--Gullstrand coordinates \cite{pg}. In order to construct this new regular-across-the-horizon frame, we consider the geodesic equations of a family of observers, freely falling
 ($d r/d\tau <0$) in the geometry (\ref{schwarzschild}), along a radial trajectory
 ($\theta, \phi$ constant):
\begin{equation}
 \frac{d t}{d \tau} = \frac{\tilde E}{1-\frac{r_{ g}}{r}} \;,\qquad
 \left(\frac{d r}{d \tau}\right)^2 + \left(1-\frac{r_{ g}}{r}\right) = \tilde
 E^2
\label{geodesics}
\end{equation}
$\tau$ representing observer's proper time and $\tilde E = E/m$ is the observer's conserved energy per unit mass. The energy parameter is related to observer's physical three-velocity $v^2 = g_{m n} d x^m d x^n /(- g_{00}) dt^2$: in fact, given the Killing vector $\xi^\mu_{(t)}$ --- generating the time-symmetry transformations of (\ref{schwarzschild}), the energy reads
\begin{equation}
 \tilde E = - u_\mu \xi^\mu_{(t)} = -u_0  = - g_{00} \,u^0 =  \frac{\sqrt{-g_{00}}}{\sqrt{1-v^2}}\,.
\label{e}
\end{equation}
We specialise to the case of observers starting at infinity with null initial velocity so that $\tilde E = 1$. For this family of observers, there exists a function $t_{ p}$ such that 
\begin{equation}
 \nabla_\mu t_{ p} =  - u_\mu \;.
\end{equation}
Taking (\ref{geodesics}) into consideration, we get the Painlev\'{e} time $t_{ p}$ in terms of Schwarzschild coordinates 
\begin{equation}
 t_{ p} = t + \int^r d r' \sqrt{\frac{r_{ g}}{r'}} \left(1-\frac{r_{ g}}{r'}\right)^{-1} = t + 2 \sqrt{r_{ g} r} + r_{ g} \ln \left(\frac{\sqrt{r} - \sqrt{r_{ g}}}{\sqrt{r} + \sqrt{r_{ g}}}\right)\,.
\label{p time}
\end{equation}
In terms of $(t_{ p}, r, \theta,\phi)$, the Schwarzschild geometry is written as
\begin{equation}
  d s^2 = -\left(1- \frac{r_{ g}}{r}\right)d t_{ p}^2 + 2 \, \sqrt{\frac{r_{ g}}{r}} \,d t_{ p}\,d r + d r^2 + r^2(d \theta^2 + \sin^2 \theta \,d \phi^2).
\label{pg schwarzschild}
\end{equation}
At fixed time, the spatial geometry described by (\ref{pg schwarzschild}) is Euclidean;
 while at any fixed radius, the boundary geometry is the same as that in
 (\ref{schwarzschild}). Most importantly, the metric is no more singular at the
 gravitational radius $r_{ g}$; the space-time is stationary, but no more static. The
 $t_{ p}$ coordinate --- being the time perceived by infalling observers who reach
 the curvature singularity $r=0$ in finite time --- remains a valid coordinate as far as it
 remains finite. 

The geodesic for a null s-wave is given by
\begin{equation}
 \dot r \equiv \frac{d r}{d t_{ p}} = \pm 1 - \sqrt{\frac{r_{ g}}{r}}\,,
\label{s wave}
\end{equation}
 with upper (lower) sign corresponding to outgoing (ingoing) geodesic. The basic idea
 behind the approach by Parikh and Wilczek \cite{Parikh:1999mf} is to study the emission of such a
 null-radial thin shell from the black hole through quantum tunnelling. Imposing energy
 conservation means that the total space-time energy is fixed and one allows the black hole
 mass to fluctuate. As showed in \cite{Kraus:1994by}, the motion of a shell of energy
 $\omega$ emitted from a black hole of initial mass $M$ is described by eq.(\ref{s wave}) in
 the space-time geometry of eq.(\ref{pg schwarzschild} with $M$ replaced by $M-\omega$. 

Differently by tunnelling in quantum mechanics, where two configurations (classical
 turning points) separated by a potential barrier are connected by a trajectory in imaginary
 time, here we have the perception that nothing similar happens. In fact, the problem with
 black hole radiation is that, as soon as a particle is displaced outside the horizon, it can
 escape along a classical trajectory. It is worth to mention how the crucial ingredient of the
 story here is played by energy conservation. As the black hole of mass $M$ and area $16
 \pi M^2$ emits a particle of energy $\omega$, its mass reduces to $M-\omega$ and its
 horizon recedes to $2(M-\omega) = r_{ g} - 2 \omega$. This amount of contraction,
 evidently, depends upon the mass of the emitted particle, so that quoting Parikh
 \textquotedblleft it is the tunnelling particle itself that secretly defines the tunnelling
 barrier'' \cite{parikh}. 

Before evaluating the black hole emission rate, we can ask to what extent is justified the
 point-particle approximation. Given that a distant observer detects a wave with frequency
 $\Omega_{{ob}}$, this has been emitted with frequency $\Omega_{{em}}
 \propto (1- r_{ g}/r)^{-1/2} \Omega_{{ob}}$. In the limit of $r\rightarrow
 r_{ g}$, where the emission process occurs, the wavelength vanishes, making the
 point-particle (WKB) approximation fully reliable. 

In the WKB approximation, the tunnelling probability rate $\Gamma_{{ em}}$ turns
 out to be equal to 
\begin{equation}
 \Gamma_{{ em}}  \sim \exp(- 2 \,\mbox{ Im} \,I)
\label{wkb}
\end{equation}
where $I$ is the action for the tunnelling process evaluated to the appropriate order. We
 shall derive later, in Subsection \ref{3.1}, the exact correspondence between $\Gamma_{{
 em}}$ and the $\exp$--function.  Eventually, one would expect to be able to read off the
 black hole radiation temperature from comparison of the the probability rate eq.(\ref{wkb}) with the  Boltzmann factor $e^{-\omega/T}$. 
The imaginary part of the action for a null s-wave outgoing positive energy particle which
 crosses the horizon outwards from $r_{ {in}}$ to $r_{ {out}}$ can be expressed
 as
\begin{equation}
 \mbox{Im}\; I = \mbox{Im} \, \int_{r_{{in}}}^{r_{{out}}} d r \,p_r =
 \mbox{Im} \, \int_{r_{{in}}}^{r_{{out}}} d r \,\int_0^{p_r} d p'_r 
\end{equation}
Changing the integration from momentum to energy thanks to Hamilton's equation $\dot
 r = d H/d p_r$ and noting that $H = M - \omega'$ with constant $M$ and
 $0\leq\omega'\leq \omega$, we have
\begin{equation}
 \mbox{Im}\; I = \mbox{Im}\ \int_{r_{{in}}}^{r_{{out}}} d r\,\int_M^{M-\omega} \frac{d H}{\dot r} = \mbox{Im} \,
 \int_{r_{{in}}}^{r_{{out}}} d r\,\int_0^{\omega} \frac{(- d \omega')
}{\dot r}\,.
\end{equation}
Using eq.(\ref{s wave}) and switching the order of integration, 
\begin{equation}
 \mbox{Im}\; I = \mbox{Im} \,\int_0^{\omega} (- d\omega') \,
 \int_{r_{{in}}}^{r_{{out}}} d r \frac{1}{1-\sqrt{\frac{2(M-\omega')}{r}}}\;.
\label{div integral}
\end{equation}
With the understanding that the particle starts at $r_{ in} = 2 M$ and materialises at
 $r_{ out} = 2(M - \omega)$, $r_{ in}> r_{ out}$, the integral over $r$ is
 divergent and needs to be appropriately regularised. A tunnelling process corresponds
 to an energy which is not in the spectrum of the Hamiltonian. We need to continue the
 energy $\omega$ to complex values, that is $\omega' \rightarrow \omega' + i \epsilon$,
 in order to ensure that positive energy solutions decay in time: 
\begin{eqnarray}
 \mbox{Im}\; I &=& \mbox{Im} \,\int_0^{\omega} (- d\omega') \, \int_{r_{{in}}}^{r_{{out}}} d r \frac{\sqrt{r}}{\sqrt{r}-\sqrt{2(M-\omega')} + i \epsilon} \nonumber \\
&=& -\mbox{Im} \, i \pi \int_0^{\omega} (- d\omega') \, 4 (M -\omega') = + 4 \pi \omega \left(M - \frac{\omega}{2}\right).
\label{action}
\end{eqnarray}
Of course, Hawking radiation can also be regarded as pair creation outside the horizon,
 with the negative energy particle tunnelling into the black hole. The calculation proceeds
 as above with slight change of signs due to the fact that anti-particles travel back in time
 (cf. \cite{Parikh:1999mf} for details). That both particle and anti-particle channels contribute to black
 hole emission is something which only affects the amplitude of the process, that is
 something which enters the proportionality factor of eq.(\ref{wkb}). In conclusion, the
 emission rate obtained is
\begin{equation}
 \Gamma \sim \exp(- 2 \,\mbox{Im} \,I) = e^{- 8\pi M \omega \left(1 - \frac{\omega
}{2 M}\right)}.
\label{non thermal spectrum}
\end{equation}
The expected Boltzmann factor $e^{-\omega/T_{ H}}$ is recovered only to order
 ${ O} (\omega)$: in this order of approximation, the black hole temperature perfectly
 coincides with standard Hawking's result, $T_{ H} = (8 \pi M)^{-1}$. The $
 O(\omega^2)$ correction arising from the physics of energy conservation, makes higher
 the effective temperature of the hole as it radiates in agreement with the well-known
 negative thermal capacity of Schwarzschild black hole. \\
Neglecting the $\omega (2 M)^{-1}$ term in (\ref{non thermal spectrum}) and invoking detailed balance\footnote{Consider an ensemble of many identical copies of the same
 quantum system. Let the energy and the number of accessible states of the system be
 fixed. The probability that a system randomly chosen out of the ensemble will be in state
 $i$ is denoted by $p_i$, with $\sum_i p_i =1$. The \textit{transition probability}
 $\lambda_{i\rightarrow j}$ denotes the conditional probability per unit time of a system
 going from state $i$ to state $j$, {\it i.e.} $p_i \lambda_{i\rightarrow j} dt$ is the probability
 of a system originally being in state $i$ and going to state $j$ within the time interval $d
 t$ \cite{balance}. From this definition, $\sum_j p_i \lambda_{i \rightarrow j} dt$
 represents the probability of transition from state $i$ to any other possible state of the
 system in time $dt$; and $\sum_j p_j \lambda_{j\rightarrow i} dt$ represents the
 probability of transition into state $i$ from any other possible state of the system in time
 $d t$. It follows that, $\dot p_i = \sum_j (p_j \lambda_{j \rightarrow i} - p_i
 \lambda_{i \rightarrow j})$ is the net rate of change of $p_i$. The \textit{detailed
 balance condition} requires that, 
\begin{equation*}
\mbox{at equilibrium,}\qquad  p_j \lambda_{j \rightarrow i} = p_i \lambda_{i \rightarrow j}\;,\qquad \forall i,j \;. \hspace{3cm} (*)
\end{equation*}
In the case at hand, the system is made of \textquotedblleft black hole $+$ radiation''. The initial state $i$ represents, for example, the black hole with $N-1$ particles (\textit{e.g.} photons). Through the physical mechanism described before, the black hole emits one more particle, so that the state $j$ becomes \textquotedblleft black hole $+$ $N$ particles''. Neglecting back-reaction, the internal degrees of freedom of the black hole do not change and by ($*$), we have that $\lambda_{i \rightarrow j} /\lambda_{j\rightarrow i} \equiv \Gamma_{{em}}/\Gamma_{{ab}} = p_N/p_{N-1}= e^{-\omega/T}$. 
This result combined with the classical constraint, $\Gamma_{{ab}} \mp \Gamma_{{em}} =\vert T_l(\omega)\vert ^2$, reproduces Planck (minus sign) and Fermi--Dirac (plus sign) distributions. Here, $T_l (\omega)$ represents the transmission coefficient of the black hole barrier which in general can depend on the energy $\omega$ and the angular momentum $l$ of the particle.}, the Planck distribution is recovered.\\
{\it A note --} It appears that by neglecting the back-reaction no member of a pair of particles created just inside the event horizon can escape it, by reason of causality. In the terminology to be used in Subsection \ref{4:dyna} this correspond to  a type-II tunnelling path, the particles created just outside forming instead a type-I path. The inclusion of the back-reaction is truly fundamental for the existence of both possibilities. Since in the dynamical case we will mainly ignore the back-reaction effect, only type-I path will be accessible to the tunnelling probability.\

\subsection{Entropy and correlations}
\label{2.2}

The exact expression for the emission rate eq.(\ref{non thermal spectrum}) can be intriguingly re-written as the change in the hole's Bekenstein--Hawking entropy \cite{bh}, 
\begin{equation}
 \Gamma \sim e^{- 8\pi M \omega \left(1 - \frac{\omega}{2 M}\right)} = e^{\Delta S_{{BH}}}\;.
\label{entropy change}
\end{equation}
This is an interesting form since it agrees with Einstein's formula for the probability of fluctuations about statistical equilibrium \cite{landau}. In this case, it refers to the probability rate with which a static Schwarzschild black hole gets out of equilibrium by emission of a scalar quantum. This formula is derived to order ${ O}(\omega)$ from the fact that  \cite{gibbons}
\begin{equation}
 \Gamma_{{ em}}\sim e^{- 2 \,\mbox{ Im} \,I} \equiv e^{-\beta \omega} =\exp \left(- \frac{\partial S}{\partial M}\, \omega\right) \simeq \exp - [ S(M) - S(M-\omega) + { O}(\omega^2)]\;, \label{entropy change 1}
\end{equation}
where the standard relation involving the inverse temperature at equilibrium, $\beta = \partial S/\partial M$, has been used. It is remarkable that Equation (\ref{non thermal spectrum}) puts forward this result to order ${ O}(\omega^2)$. Since the exp-function in Equations (\ref{entropy change}) and (\ref{entropy change 1}) can be re-written as $\exp \Delta S_{{ BH}} = \exp S_{{ final}}/\exp S_{{ intial}}$, we can re-interpret it as a factor counting the internal degrees of freedom of the black hole.
It is a long debated question how the thermal nature of Hawking radiation can be reconciled with unitarity (\textit{information loss puzzle}). From one side, unitarity is a milestone of classical and quantum physics; from the other, Hawking radiation is the most reliable result of quantum gravity derived with semi-classical techniques. This is not the place to enter the details of such a wide subject: it is enough to say that the discovery of non-thermal corrections to the black hole radiation spectrum due to the inclusion of back-reaction effects showed by Parikh and Wilczek, and the possibility of writing the emission rate in the form (\ref{entropy change}) gave hope for a possible solution of the problem. For example, let us take into consideration the emission of a quantum of energy $\omega=\omega_1 + \omega_2$ and, separately, the emission of a quantum of energy $\omega_1$ followed by the emission of a quantum of energy $\omega_2$. The function
\begin{equation}
 \chi(\omega_1, \omega_2)= \log \,\Big\vert\frac{\Gamma(\omega)}{\Gamma(\omega_1) \Gamma(\omega_2)}\Big\vert
\label{stat corr}
\end{equation}
can be considered as a measure of the statistical correlation between the two events which is expected to be zero whenever the events are independent. This is indeed the case for thermal spectrum, but also for the emission rate (\ref{non thermal spectrum}) it was argued in \cite{parikh2} to be the case. Even the inclusion in Equations (\ref{entropy change}), (\ref{entropy change 1}) and (\ref{stat corr}) of logarithmic corrections to the black hole entropy, as suggested by direct count of black hole microstates both in string theory and loop quantum gravity, seemed not to be able to show the appearance of correlations in the spectrum of black hole radiation \cite{vagenas}. \par
However, a more careful statistical treatment due to Zhang \textit{et al.}  in two significant papers \cite{Zhang:2009jn,Zhang:2009td} showed that, on the contrary, the statistical correlation eq.(\ref{stat corr}) is actually non zero and equal to $8\pi\omega_{1}\omega_{2}$, a result later confirmed by Israel and Yun \cite{Israel:2010gk}. To make clear the point, we display the black hole mass into $\Gamma(\omega)$ and, following Israel and Yun, we write the probability that a black hole of mass $M$ emits a quantum of energy $\omega$ as
\beq
\Gamma(M,\omega)=\exp[-4\pi M^{2}+4\pi(M-\omega)^{2}]
\eeq
which is the tunnelling result. Then this yields the non trivial correlation
\begin{equation}
 \chi(\omega_1, \omega_2)= \log \,\Big\vert\frac{\Gamma(M,\omega_{1}+\omega_{2})}{\Gamma(M,\omega_1) \Gamma(M,\omega_2)}\Big\vert =8\pi\omega_{1}\omega_{2}
\label{stat corr1}
\end{equation}
which is the result of Zhang \textit{et al.}. In general, one may define a measure of the correlation between events $x$ and $y$ as
\beq
\chi(x,y)=\log\frac{P(x,y)}{P(x)P(y)} =\log\frac{P(y|x)}{P(y)}
\eeq
where
\beq
P(y|x)=\frac{P(x,y)}{P(x)}
\eeq
is the conditional probability of $y$ given $x$ and $P(x,y)$ is the probability of both $x$ and $y$. Clearly, if we substitute $P(y|x)$ in place of $P(y)$ into $\chi(x,y)$, we get zero correlation for any events $x$ and $y$, because in this way one absorbs the correlation itself into the test for its existence. Thus, by the same token, if we put $\Gamma(M-\omega_{1},\omega_{2})$ into $\chi(\omega_{1},\omega_{2})$ in place of $\Gamma(M,\omega_{2})$, as it might be thought in order to take into account the mass loss accompanying the first emission, we find no correlation at all, as can be directly verified.\par  
We see that the tunnelling method leads to a very important conclusion, that the radiation does indeed contain correlations that possibly can carry off the information content of the hole.

\subsection{The Hamilton--Jacobi method}
\label{2.3}

Despite the merits of the seminal work by Parikh and Wilczek, we cannot omit to point out a couple of unpleasant features of their null geodesic method, as the fact that: (i) it strongly relies on a very specific choice of (regular-across-the-horizon) coordinates; and (ii) it turns upside down the relationship between Hawking radiation and back-reaction. As regard the former point, it should be clear how irrelevant is, in the spirit of general relativity,  the choice of coordinates: being physical observables invariant with respect to the group of diffeomorphisms (the hole temperature is such an observable), there is no reason why Painlev\'{e}--Gullstrand coordinates should be favourable with respect to other (equally well-behaved) coordinates. With respect to the latter, we notice that, in the null geodesic description, apparently there cannot be Hawking radiation without back-reaction: watching carefully, however, it is the discovery of Hawking radiation that justifies back-reaction and makes commendable the treatment
  of Hawking radiation's self-gravity. The so-called \textit{Hamilton--Jacobi method} can cope with both issues. The intent is to give a particle description of Hawking radiation, ignoring its self-gravitation, under the assumption that the emitted (scalar) particle's action does satisfy the Lorentzian Hamilton--Jacobi equation. Later we will show that the null geodesic method can do the same job using instead the reduced action. \\
As it will become clear later, this method applies to any well-behaved coordinate system across the horizon; it generalises beyond the assumption of spherical symmetry; it makes possible to include the study of tunnelling by fermionic particles. Some sceptics about Hawking radiation contest the fact that as soon as the black hole starts radiating, the same assumptions on space-time stationarity drops down, invalidating the whole derivation. Given that the departure from perfect stationarity is ridiculously small, nonetheless, as we shall show in Subsection \ref{4:dyna}, the Hamilton--Jacobi method can prove black hole evaporation even for slowly varying, time dependent, space-times.\\
Finally, we wish to notice that since the methods of tunnelling are intimately related to the physics of (some type of) horizons, we may apply them even to space-times with multiple horizons. In standard computations of Hawking radiation, this is typically a hard task, if not impossible. In fact, it is well known that, as an example, Reissner--Nordstr\"{o}m--de Sitter space does not possess regular Euclidean section for general values of mass, electric charge and cosmological constant \cite{bousso}. This means that for arbitrary values of the parameters, it is not possible to compute the Hawking temperature of event or cosmological horizons by Euclidean continuation, simply because it could be that no Euclidean section of the given Lorentzian space-time exists.
\begin{figure}[h!!!]
\begin{center}
\includegraphics[width=0.75\textwidth]{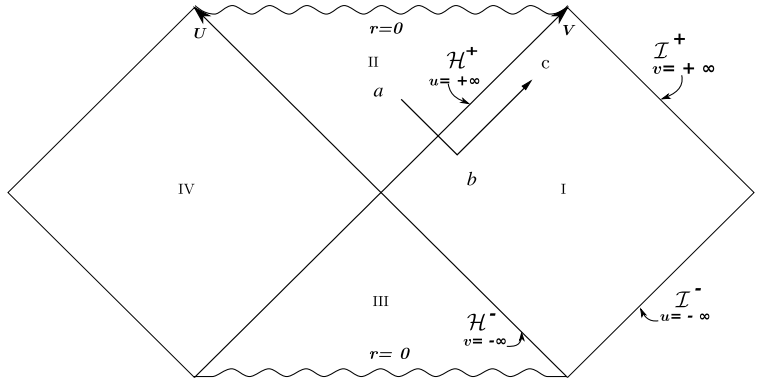}
\caption{\label{ks} Carter--Penrose diagram of the eternal Schwarzschild black hole. $\mathcal{H}^\pm$ denote future and past horizon; $\mathcal I^\pm$ future and past null infinity; $r=0$ the curvature singularity; $v,u$ represent advanced and retarded null coordinates, respectively. $\protect\overrightarrow{a b c}$ is a null piecewise continuous path from inside to outside the hole, with $\protect\overrightarrow{ab}$ running backward in time (classically forbidden trajectory).}
\end{center}
\end{figure}

In its minimal formulation, the Hamilton--Jacobi method works according to the following strategy \cite{noi}: (a) assume that the tunnelling particle's action satisfies the relativistic Hamilton--Jacobi equation,
\begin{equation}
 g^{\mu\nu}\, \partial_\mu I \,\partial_\nu I + m^2 =0
\label{hj}
\end{equation}
where $m^2$ represents the invariant mass; (b) reconstruct the whole action $I$, starting from the symmetries of the problem and the partial derivatives $\partial_\mu I$, by means of 
\begin{equation}
\label{I} 
I = \int dx^{\mu}\,\partial_\mu I\,, 
\end{equation}
where the integration is carried along an oriented, null, curve (like for example $\overrightarrow{a b c}$ in fig.\ref{ks}), to which we refer as the {\it tunnelling path}, with at least one point on the horizon; (c) split the integration 
along the null path in two pieces, one along the segment crossing the horizon ($\overrightarrow{a b}$ in fig.\ref{ks}), the remaining contribution living in the  outer domain of the space-time ($\overrightarrow{b c}$ in fig.\ref{ks}); (d) perform a near-horizon approximation in the first integral and regularise its  divergence  according to Feynman's $i\epsilon$--prescription. \\ 
Once the procedure is done, the imaginary part of the classical action, relevant for the tunnelling rate (\ref{wkb}), reads 
\begin{equation}
\mbox{Im }\, I = \frac{\pi\omega}{\kappa}\,, 
\label{ima} 
\end{equation}
where $\kappa$ and $\omega$ are the surface gravity and particle's energy, respectively.

As we have already stressed, Hawking radiation is a semi-classical result of quantum field theory in curved space-time. 
In the spirit of tunnelling, we allow particles to travel along classically forbidden
 trajectories by starting just behind the horizon onward to infinity. Thus, particles must
 travel back in time, since the horizon is locally to the future of the static external region.
 For this reason, we must implement coordinates well defined through regions $I$ and
 $II$, a requirement which automatically discards Schwarzschild-like or isotropic-like
 coordinates (it is easy to show that, in Schwarzschild space-time, isotropic
 coordinates, ${(t,\rho,\theta,\phi)}$, defined through $r(\rho):= (1+M/2\rho)^2 \rho$,
 cover regions $I$ and $IV$ of fig.\ref{ks}). The classical one-particle action becomes
 complex as an indication of the classical impossibility of the motion and gives, in this
 way, an imaginary part to the emission rate of the hole. In this sense, we can say that, of
 the actual path travelled by the tunnelling particle (\textit{e.g.} $\overrightarrow{a b c}$), only
 the infinitesimal region across the horizon plays a crucial role in the whole calculation.
 Even if null classically forbidden curves like $\overrightarrow{a b}$ do not have anything 
special with respect to other forbidden paths, their choice is preferable to computational
 purposes. Also, it will become clearer later the reason why we specifically regularise
 divergent integrals by Feynman's prescription rather than others. In consideration of our
 choices, the mass term in Equation (\ref{hj}) is irrelevant to the physics of the horizon
 (Hawking radiation) but, as we shall show in Section \ref{5:coho}, under quite general
 conditions particles masses play an important role in relation to bulk particle creation
 phenomena. 

To show in practice how the Hamilton--Jacobi method works, we are going to derive the emission rate (\ref{wkb}) and prove the identity (\ref{ima}) in the case of a scalar, uncharged, particle tunnelling from a Schwarzschild black hole, in complete analogy with what we did in the previous Section. This will give us the opportunity to unveil other debated points of the method. Later on we shall free the derivation from any dependence by special coordinate systems.

\subsubsection{Painlev\'e--Gullstrand coordinates}
In this case the space-time geometry is described by eq.(\ref{pg schwarzschild}). Because of
 eq.(\ref{p time}, expressing the relation between Painlev\'{e} and Schwarzschild times, the
 particle energy associated to a Killing observer is expressed by $\omega = -
 \partial_{t_{ p}} I$ (according to (\ref{e}), the particle energy is $\omega = - p
 \cdot \xi$ where $p_{\mu}= -\partial_{\mu}I$ is the four-momentum). Since we are
 dealing with a spherically symmetric space-time, we may neglect without fault the angular coordinates. We consider a null direction $(\Delta t_p,\Delta r)$, for which 
\begin{equation}
- \left(1-\frac{r_{ g}}{r}\right) \Delta t_{ p}^{\;2} + 2 \sqrt{\frac{r_{ g}
}{r}}\, \Delta t_{ p}\cdot \Delta r + \Delta r^2 =0 \;.
\end{equation}
In a neighbourhood of the event horizon (\textit{e.g.} $\overrightarrow{a b}$ in fig.\ref{ks}), the
 relation between $\Delta t_{ p}$ and $\Delta r$ is uniquely fixed to be $\Delta t_{
 p} = - \Delta r/2$ since, in this limit, the alternative solution $\Delta t_{ p} = -
 (1-\sqrt{r_{ g}/r})^{-1}\,\Delta r$ diverges in contrast to the physical meaning of
 the coordinate $t_{ p}$. It simply means that the segment $\overrightarrow{a b}$ is part of null direction across the horizon. The reduced Hamilton--Jacobi equation
\begin{equation}
 - (\partial_{t_{ p}} I)^2 + 2 \sqrt{\frac{r_{ g}}{r}} \, \partial_{t_{ p}} I\,\partial_r I  + \left(1-\frac{r_{ g}}{r}\right) (\partial_r I)^2 =0
\end{equation}
can be written as 
\begin{equation}
 - \omega^2 - 2 \,\omega \,\sqrt{\frac{r_{ g}}{r}} \,\partial_r I  +
 \left(1-\frac{r_{ g}}{r}\right) (\partial_r I)^2 =0\;.\label{hj3}
\end{equation}
The imaginary part of the action is
\begin{eqnarray}
  \mbox{Im}\,I &=&  \mbox{Im}\,\int_{a \rightarrow b\rightarrow c}\left(\partial_r I \,d r + \partial_{t_{ p}} I\,d t_{ p}\right) \label{st1} \\
&=&\mbox{Im}\,\int_{a\rightarrow b}\, d r\, \left(\partial_r I  + \frac{1}{2} \omega \right) \label{st2}\\
&=& \mbox{Im}\,\int\!\!\!\!\!\!\!\searrow \, d r\,\partial_r I \label{st3} \;.
\end{eqnarray}
Passing from (\ref{st1}) to (\ref{st2}), we took into consideration that: (i) only the
 classically forbidden part $\overrightarrow{a b}$ of the whole null path
 $\overrightarrow{a b c}$ contributes to complexifying the action of the tunnelling
 particle; (ii)  as far as we are concerned with the infinitesimal region of intersection
 between the horizon and the path $\overrightarrow{a b}$, $d t_{ p}$ and $dr$ are
 related to each other in the way specified above; (iii) the particle's energy goes as
 $-\partial_{t_{ p}} I$. Being $\omega$ a real positive constant, the second term in
 the integrand of (\ref{st2}) does not enter the imaginary part of the action so it drops
 down in passing to (\ref{st3}). From now on, integration along the classically forbidden
 path (\textit{e.g.} $\overrightarrow{ab}$ or, by extension, $\overrightarrow{abc}$) will be denoted simply by $\int\!\!\!\!\!\!\!\searrow$.\\
It is remarkable that also the integration over the \textquotedblleft time'' coordinate
 enters, in general, the reconstruction process of the particle's action. In this specific case,
 we see that this \textquotedblleft temporal'' contribution --- being manifestly real --- does
 not affect the result. The next subsection will show a coordinate system where instead
 the \textquotedblleft temporal'' contribution vanishes exactly. It should be clear that
 names such as \textquotedblleft temporal'' or \textquotedblleft spatial'' contributions are
 coordinate dependent terms in no way fundamentally related to the physics of the
 process, which must be covariant. The general covariance of the tunnelling method will
 be continuously emphasised as one of the main themes of this review.\par
Inserting the non-manifestly real solution of eq.(\ref{hj3} into (\ref{st3}), we get
\begin{equation}
 \mbox{Im}\, I = 2\; \mbox{Im}\,\int\!\!\!\!\!\!\!\searrow \, d r\,\frac{\omega
}{1-\frac{r_{ g}}{r}} = 2 \;\mbox{Im}\int\!\!\!\!\!\!\!\searrow\, d r\, \frac{\omega\, r}{r - r_{ g}}
 \,,\label{exp}
\end{equation}
and regularising the divergent integral according to Feynman's $i\epsilon$--prescription,
 the imaginary part of the action becomes 
\begin{equation}
 \mbox{Im}\,I = 2 \;\mbox{Im}\int\!\!\!\!\!\!\!\searrow\, d r\, \frac{\omega \, r}{r -
 r_{ g} - i \epsilon} =4 M \pi\,\omega  \,.\label{ima1}
\end{equation}
Strictly speaking $\omega$ should be computed on the horizon, because this is how the method of residues works; however this is the same thing as the energy measured at infinity since on the outgoing trajectory $\omega$ is conserved. That said, after the known identification $\kappa = 1/4M$, the identity (\ref{ima}) is fully
 recovered. Weather we had used the opposite prescription, $r\to r+i\epsilon$, a corresponding change of sign of the imaginary part would have resulted. Its meaning will be explained in Section \ref{3:found}.    

\subsubsection{Eddington--Finkelstein coordinates}

Let us introduce another reference frame without singularities on $r_{ g}$ originally
 constructed by Eddington (1924) and Finkelstein (1958), \cite{eddington}. This frame is
 fixed to radially moving photons. Since no observer can move together with photons, this
 new frame is not, strictly speaking, a reference frame. Nevertheless, this system of test
 photons proves to be very convenient, \cite{frolov}. According to (\ref{schwarzschild}),
 the equation of motion of a radial incoming photon is $dr/dt = - (1-r_{ g}/r)$. From
 the viewpoint of a distant observer, the photon, starting from $r_1$ at time $t_1$,
 arrives in $r$ ($r_{ g}< r <r_1$) at time   
\begin{equation}
 t = r_1 -r - r_{ g}\,\ln\left(\frac{r - r_{ g}}{r_1-r_{ g}}\right) + t_1\:.
\label{ef1}
\end{equation}
This expression can be opportunely re-written as $v=v_{1}$, where
\begin{equation}
 v:= t+ r_*\:, \qquad r_* := r + r_{ g}\ln\left\vert\frac{r}{r_{ g}} -1 \right\vert
\label{ef2}
\end{equation}
$r_*$ is the so-called \textit{tortoise coordinate} with $v_{1}$ a constant characterising the initial data of the photon at  $(t_{1},r_{1})$. Because of the logarithm in (\ref{ef1}), $r_*$ is defined for any $r>0$. Chosen a set of photons at fixed $t$, we may label each photon of the set through a number $v$, which will identify uniquely that photon during its whole motion: $v$ rises to the role of a new null coordinate, usually called \textit{advanced time}. After differentiation of (\ref{ef2}) and substitution in (\ref{schwarzschild}), the line element takes the so called Eddington--Finkelstein  form,
\begin{equation}
 d s^2 = - \left(1-\frac{r_{ g}}{r}\right) d v^2 + 2\, d v \,d r + r^2 (d \theta^2 + \sin^2 \theta \,d \phi^2)\:.
\label{ef metric}
\end{equation}
Because of (\ref{ef2}), the Killing vector field $\xi^\mu =\partial_t$ transforms to ${\xi'}^\mu = \partial_v$ and particle's energy as measured by such (natural) observer is simply $\omega = - \partial_v I$. Favoured by the spherical symmetry of the problem, we may consider only radial motions. Expanding the trajectory along a null direction in a neighbourhood of the horizon (\textit{e.g.} $\overrightarrow{a b}$ in fig.\ref{ks}), gives
\begin{equation}
0 = 2 \,\Delta v\cdot \Delta r\,. 
\label{ne1} 
\end{equation}
However, along the null path $\overrightarrow{a b}$, the $v$ coordinate is constant, so that 
\begin{equation}
 \Delta v\vert_{\searrow } = 0 \label{pizzi}
\end{equation}
is the right solution. The Hamilton--Jacobi equation, 
\begin{equation}
 2 \,\partial_ v I \, \partial_r I + \left(1-\frac{r_{ g}}{r}\right) (\partial_r I)^2 = 0
\label{hj1}
\end{equation}
can be re-written as
\begin{equation}
 \left(1-\frac{r_{ g}}{r}\right) \partial_r I = 2 \omega\;.
\label{hj2}
\end{equation}
According to the reconstruction assumption,
\begin{equation}
 \mbox{Im}\,I = \mbox{Im}\int_{a\rightarrow  b \rightarrow c}\left(\partial_r I \,d r + \partial_v I\,d v\right) \,,
\end{equation}
and thanks to (\ref{pizzi}) and (\ref{hj2}),  
\begin{equation}
 \mbox{Im}\,I = 2 \;\mbox{Im}\,\int\!\!\!\!\!\!\!\searrow d r \,\frac{\omega}{\left(1-\frac{r_{ g}}{r}\right)}\, + \int_{b \rightarrow c} (\dots)\:.
\end{equation}
Neglecting the real contribution coming form the classically allowed path $\overrightarrow{b c}$ and following the same procedure as in (\ref{exp}), we end with the same result as in Equations (\ref{ima}) and (\ref{ima1}). \smallskip

Among the coordinate systems of a certain importance covering the space-time region across the gravitational radius there is the so-called Lema\^{\i}tre frame.  In this frame the Schwarzschild space-time appears as truly dynamical. We shall see how the Hamilton--Jacobi method deals with it in Section \ref{tu:fm}.

We end this section pointing out the role played by the local observer. 
As discussed in \cite{stotyn}, the particle's energy $\omega=-\partial_tI$ given in the two previous examples is obviously the conserved energy as measured by an observer living at infinity.  It follows that the Hawking temperature of a Schwarzschild black hole, $T = 1/ 8\pi M$, is actually the one measured by a Killing observer at infinity. The particle's energy as detected by a Killing observer at position $\bf x_{{ob}}$ is given instead by $\omega_{{ob}} = \omega/\sqrt{- g_{00}(\bf x_{{ob}})}$. Comparing the tunnelling probability derived from (\ref{ima}) with the thermal distribution as measured by the observer $\beta_{ {ob}}\; \omega_{ {ob}} = 2\pi \omega/\kappa$, we obtain 
\begin{equation}
 T_{{ob}} \sqrt{- g_{00}(\bf x_{{ob}})} = T_{\infty} = \mbox{constant}\;,
\end{equation}
which expresses the expected result given by Tolman years ago\cite{tolman}.

\subsection{More general static solutions}
\label{2.4}

Up to now we have only considered the Schwarzschild solution in  order to keep the discussion as simple as possible and to illustrate the principles involved without unnecessary complications. However more general static solutions are of interest for a variety of reasons. In order to include a broader class, the metric can be written in a diagonal gauge as 
\beq
d s^2=-V(r)d t^2+\frac{d r^2}{W(r)} + C(r)^2 (d \theta^2 + \sin^2\theta\,d \phi^2)\, ,
\label{ab}
\eeq 
which describes what Visser \cite{visser} termed \textit{dirty black holes}. 
Black hole solutions are defined by  functions $V(r)$ and $W(r)$  having simple and positive zeroes. This is only a necessary condition to have a black hole; we must also require that the domain of outer communication be ``outside of the black hole'' {\it i.e.}, it should correspond to values of the radial coordinate larger than the horizon and extending up to spatial infinity. Interesting black holes described by such metrics can be obtained in the Einstein--Maxwell--dilaton coupled system. An example illustrating the feature is the following two-parameter family of solutions 
\beq
  d s^2=-\left(1-\frac{r_+}{r}\right)d t^2+\left(1-\frac{r_+}{r}\right)^{-1} \left(1-\frac{r_-}{r}\right)^{-1}d r^2+r^2 (d \theta^2 + \sin^2\theta\,d \phi^2) \;.
\eeq
The dilaton is $\exp(2\varphi)=(1-r_-/r)^{-1/2}$; the hole has magnetic charge $q_{ m}=3r_+r_-/16$, horizon radius $r_+=2 M$ (which defines $M$) and can be extended to a non singular, geodesically complete solution with horizons and asymptotically flat infinities. \par
As long as $V\neq W$ in eq.(\ref{ab}, it is possible to set $C(r)=r$ by an opportune radius redefinition. But for $V = W$, this is not always the case as can be seen from the following example  
\beq
  d s^2=-\frac{1-2M/r}{\sqrt{\Delta}} d t^2 + \frac{\sqrt{\Delta}}{1-2M/r} d r^2+ \sqrt{\Delta}\,r^2 (d \theta^2 + \sin^2\theta\,d \phi^2) \;.
\eeq
Here $C(r):=\Delta^{1/2} r$ with $\Delta=1+2 Mr^{-1}\sinh^2\gamma$ and $\gamma$ a real constant.  The metric is asymptotically flat with an event horizon at $r=2 M$ and an electric field due to a charge $Q=M\sinh2\gamma/2$.  This solution can be obtained from dimensional reduction of a boosted Schwarzschild solution in Kaluza--Klein theory.

The metric eq.(\ref{ab}) admits also a Painlev\'e--Gullstrand representation associated with a redefinition of the time coordinate, 
\beq
t_{ p}=t \pm \int dr\sqrt{\frac{1-W(r)}{V(r)W(r)}}\,.
\label{pain1}
\eeq 
A simple computation leads to 
\beq
  d s^{2} = -V(r)d t_{ p}^2 \pm 2\sqrt{\frac{1- W(r)}{W(r)}\cdot V(r)}\;d r \,d t_{ p}+ d r^2+ C(r)^2 \, (d \theta^2 + \sin^2\theta\,d \phi^2).
\label{BHP}
\eeq
Turning our attention back to eq.(\ref{ab}) with $C(r)\equiv r$, by computing the Einstein's tensor one sees that the stress-energy tensor must have the following form   
\beq 
T^\mu_{\;\nu} = \mbox{diag } (-\varrho(r),P(r),S(r),S(r)). 
\label{Ttt} 
\eeq 
Thus, in general, the metric eq.(\ref{ab}) has continuously distributed sources and is not asymptotically flat. If the space-time we have in mind is such that 
\begin{eqnarray}
& & V(r) = \frac{(r-r_1) \cdots (r-r_s)}{r^s} \;, \nonumber \\
& & W(r) = \frac{(r-r_1) \cdots (r-r_q)}{r^q} \;, \qquad s \leq q \;, 
\end{eqnarray}
where $r_1 \leq r_2 \leq \cdots \leq r_s \leq r_q$, then
\beq
V(r)\cdot W(r) > 0 \left\{\begin{array}{rc}
\forall r >0\;, &\qquad s=q \;,\\
\forall r > r_q\;, & \qquad s < q \;,
\end{array}\right.
\eeq
which means that outside the outermost event horizon $r=r_q$, $V(r)$
and $W(r)$ are both positive functions. In this case from Einstein's equations it is possible to prove that \cite{bob09} 
\beq
 V(r) = W(r) \quad \Longleftrightarrow \quad \varrho + P = 0 \quad
 \mbox{through space-time} \label{f=h} 
\eeq
which is satisfied only in vacuum space-times with cosmological constant.  This condition is completely equivalent to the fact that \cite{Jacob:2007} \textit{in a static space-time}, 
\beq g_{tt} g_{rr} = -1 \qquad 
\Longleftrightarrow \qquad T_{\mu\nu} n^\mu n^\nu =0\; , 
\label{Jacob} 
\eeq 
{\it for all radial null vectors} $n^{\mu}$. Another important fact regarding such dirty black holes is that as a consequence of the dominant energy condition and the Einstein's equations, if $V(r)$ has a simple zero at some $r=r_H$ then $W(r)$ also has a simple zero at the same point (for a proof of this and other properties see \cite{bob09}).\par
The metric eq.(\ref{ab}) is of interest also because it admits two inequivalent definitions of conserved energy: one is the Killing energy, $\omega=-p_{\mu}\xi^{\mu}=-\partial_tI$; the other is the Kodama energy which uses the vector field $K=\sqrt{W/V}\,\partial_t$. This vector field comes about because, in spherical symmetry (as spelled out extensively in Section \ref{4:dyna}), it has the amazing property that both $K^{\mu}$ as well as $J_{\nu}=-K^{\mu}T_{\mu\nu}$ are conserved: $\nabla_{\mu}K^{\mu}=0$, $\nabla_{\mu}J^{\mu}=0$. As a consequence it is possible to define two different notions of temperature, depending on which energy one is using. \par
As we shall prove in detail in Section \ref{3:found}, even if the metric eq.(\ref{ab}) is singular on the horizon, the tunnelling method works. Using for example the Hamilton--Jacobi version, the massless Hamilton--Jacobi equation for radial motion reads
\beq
-\frac{1}{V(r)}(\partial_t I)^2+W(r)(\partial_r I)^2=0\,.
\label{m2}
\eeq
The classical action is given by (plus/minus sign corresponding to outgoing/ingoing  particles, respectively)
\beq
I_{\pm}=-\omega t\pm\int \omega \frac{dr}{\sqrt{V(r)W(r)}}\,,
\label{m3}
\eeq
where $\omega = - \partial_t I$ represents the Killing energy. Assuming the near horizon expansion (as noted above, the occurrence of the same zero in $V(r)$ and $W(r)$ is a theorem, given Einstein's equations and some energy condition)
\beq
V(r)=V'(r_{ H})(r-r_{ H})+ \dots \;,\qquad W(r)=W'(r_{ H})(r-r_{ H})+\dots \;,
\eeq
and Feynman's prescription, we readily  obtain
\beq
\mbox{ Im}\!\int\!dI_+-\mbox{ Im}\!\int\! dI_-=\frac{\pi\omega}{\kappa}
\eeq
where
\beq\label{kvw}
\kappa=\frac{1}{2}\sqrt{V'(r_{ H})W'(r_{ H})}
\eeq
is the Killing surface gravity. One may use the Kodama energy, $\omega_{ K}=\sqrt{W/V}\,\omega$, which on the horizon takes the form $\omega_{ K}=\sqrt{W^{'}(r_{{ H}})/V^{'}(r_{{ H}})}\,\omega$ via de l'H\^opital rule; in this  case one obtains the so called Hayward's surface gravity
\beq
\kappa_{ H}=\frac{1}{2}\,W^{'}(r_{ H})
\eeq
and the corresponding temperature $T_{ H}=\kappa_{ H}/2\pi$. As we shall see in Section \ref{4:dyna}, this is the one that can be generalised naturally to dynamical situations, where there are no more Killing vectors in general. Notice however that $\omega/\kappa=\omega_{ K}/\kappa_{ H}$, so that the tunnelling probability is invariant.

\subsection{Tunnelling of fermions}
\label{2.5}

At this point of the discussion, one might ask what particles are to be found in the Hawking radiation spectrum of a black hole. Since a black hole has a well defined temperature, in principle it should radiate all the standard model particles like a black body at that temperature (ignoring grey body
factors). The emission spectrum, therefore, is expected to contain particles of all spins, in particular fermions as well. The basic reference here is the seminal paper by Kerner and Mann \cite{mannf}, to which others followed for both stationary and dynamical black holes\cite{kerner2,yale,Yale:2010tn,reh,DiCriscienzo:2008dm,Li:2008zz}.  The fact that massless fermions are emitted with the same temperature as massless bosons is not a trivial result, given the fact that fermionic and bosonic vacua are in general distinct. It is important to mention that these expectations have been recently extended to include spin-$1$ bosons; and that the Hawking temperature does not receive higher order corrections in $\hbar$ beyond the semi-classical one \cite{Yale:2010tn,Chatterjee:2009mw,Yale:2011zz} as originally proposed in \cite{Banerjee:2008cf} (see also \cite{Banerjee:2009wb,Majhi:2008gi,Majhi:2009uk,Banerjee:2010xi} for further extensions). \par
Now, what do we take as the action of fermionic particles? Undoubtedly the most convenient choice would be to take the phase of the spinor wave function which, as is well known, satisfies the Hamilton--Jacobi equation as for spin-less particles. However, one may also consider the action from another point of view. Virtually, all known variational formulations of the motion of spinning particles have an action of the form \cite{Barducci:1976qu,Berezin:1976eg,Cognola:1981qw}

\begin{equation*}
 I_{ f} = I_0 + ({ spin\;\; corrections})
\end{equation*}
where $I_0$ is the kinetic term equal to the classical action of scalar particles as considered in previous sections, and the correction terms contain the coupling of the spin degrees of freedom with the spin connection of the manifold. These can either be derived from covariance considerations or more directly by the semi-classical treatment of the Dirac equation itself. Some of these terms give additive corrections to the particle four-momentum but in no case the remaining terms contain horizon singularities, as they are only responsible for spin precession effects. In the following course, therefore, we shall ignore them. Next, we neglect any change of angular momentum of the black hole due to the spin of the emitted particle. For zero-angular momentum black holes with mass much larger than the Planck mass ($m_{ Pl} \sim 10^{-5} \,{ gr}$) this is a good approximation. Statistically, as many particles with spin in one direction will be emitted as particles with spin 
 in the opposite direction, producing no net change in the black hole angular momentum (although second-order statistical fluctuations will be present in general). 

As an example, we consider the tunnelling of fermionic particles across the event horizon of a spherically symmetric, uncharged, static black hole described by the metric (\ref{ab}) with $C(r) = r$; and compute the emission rate in a singular Schwarzschild-like coordinate frame. We refer to the original paper \cite{mannf} for analogous discussions in generalised Painlev\'{e}--Gullstrand and Kruskal--Szekeres coordinates. 

Temporarily re-introducing $\hbar$, the Dirac equation in curved space-time reads,
\begin{equation}
\left(\gamma^\mu D_\mu + \frac{m}{\hbar}\right) \Uppsi(t,r,\theta,\phi) = 0, \label{Dirac} 
\end{equation}
where
\begin{eqnarray}
D_\mu &=& \partial_\mu + \frac{i}{2} {\Gamma^\alpha_{\;\mu}}^{\;\beta} \,\Sigma_{\alpha\beta} , \label{diffop} \\ 
\Sigma_{\alpha\beta} &=& \frac{i}{4} [\gamma_\alpha,  \gamma_\beta]_-.\label{Comm} 
\end{eqnarray}
The $\gamma^\mu -$ matrices satisfy the Clifford algebra,
\begin{equation}
[\gamma_\alpha,\gamma_\beta]_+ = 2 g_{\alpha\beta} \mathbb{I},
\end{equation}
where $\mathbb{I}$ is the $(4\times 4)$-identity matrix. For this case, we pick the $\gamma$ matrices 
\begin{eqnarray*}
 \gamma ^{t} &=&\frac{i}{\sqrt{V(r)}}\left( \begin{array}{cc}
{\bf 1}& {\bf 0} \\ 
{\bf 0} & -{\bf 1}%
\end{array}%
\right) \;\;
\gamma ^{r} =\sqrt{W(r)}\left( 
\begin{array}{cc}
{\bf 0} &  \sigma ^{3} \\ 
 \sigma ^{3} & {\bf 0}%
\end{array}%
\right) \\
\gamma ^{\theta } &=&\frac{1}{r}\left( 
\begin{array}{cc}
{\bf 0} &  \sigma ^{1} \\ 
 \sigma ^{1} & {\bf 0}%
\end{array}%
\right) \;\;
\gamma ^{\phi } =\frac{1}{r\sin \theta }\left( 
\begin{array}{cc}
{\bf 0} &  \sigma ^{2} \\ 
 \sigma ^{2} & {\bf 0}%
\end{array}%
\right)
\end{eqnarray*}%
where $\sigma$'s are the Pauli matrices satisfying usual relations, 
\begin{equation}
 \sigma_i  \sigma_j = {\bf 1} \delta_{ij} + i \varepsilon_{ijk} \sigma_k ,\qquad i,j,k =1,2,3\;. 
\label{pauli}
\end{equation}
The matrix for $\gamma^{5}$ is instead
\begin{equation*}
\gamma ^{5}=i\gamma ^{t}\gamma ^{r}\gamma ^{\theta }\gamma ^{\phi }=i\sqrt{\frac{W(r)}{V(r)}}\frac{1}{r^{2}\sin \theta }\left( 
\begin{array}{cc}
{\bf 0} & -{\bf 1} \\ 
{\bf 1} & {\bf 0}%
\end{array}%
\right)\:.
\end{equation*}
For the spin-up (positive $r$-direction) Dirac field, we employ the {\it ansatz}
\begin{equation}
\Uppsi _{\uparrow }(t,r,\theta ,\phi ) = \left( \begin{array}{c}
H(t,r,\theta ,\phi ) \\ 
0 \\ 
Y(t,r,\theta ,\phi ) \\ 
0%
\end{array}%
\right) \exp \left[ \frac{i}{\hbar }I_{\uparrow }(t,r,\theta ,\phi )\right]\;.
\label{spinupbh} 
\end{equation} 
We will only solve the spin-up case explicitly since the spin-down (negative $r$-direction) case is fully analogous. Employing the {\it ansatz} (\ref{spinupbh}) into the Dirac equation (\ref{Dirac}) results in 
\begin{eqnarray}
-\left( \frac{i \,H}{\sqrt{V(r)}}\,\partial _{t} I_{\uparrow } + Y \sqrt{W(r)} \,\partial _{r} I_{\uparrow }\right) + H m &=&0 \\
-\frac{Y}{r}\left( \partial _{\theta }I_{\uparrow }+\frac{i}{\sin \theta } \, \partial _{\phi }I_{\uparrow}\right) &=& 0 \\
\left( \frac{i \,Y}{\sqrt{V(r)}}\,\partial _{t}I_{\uparrow } - H \sqrt{W(r)}\,\partial_{r}I_{\uparrow}\right) + Y m &=& 0 \\
-\frac{H}{r}\left(\partial _{\theta }I_{\uparrow } + \frac{i}{\sin \theta }\,\partial _{\phi }I_{\uparrow }\right) &=& 0
\end{eqnarray}%
to leading order in $\hbar $. As we assume that the action takes the form
\begin{equation}
I_{\uparrow }=- \omega\, t + \mathcal F(r) + J(\theta ,\phi )  \label{actbh}
\end{equation}%
these yield the set of equations
\begin{eqnarray}
\left( \frac{i\, \omega\, H}{\sqrt{V(r)}} - Y \sqrt{W(r)}\, \mathcal F^{\prime }(r)\right) +m\,H &=&0
\label{bhspin5} \\
-\frac{H}{r}\left( J_{\theta }+\frac{i}{\sin \theta }J_{\phi }\right) &=&0
\label{bhspin6} \\
-\left( \frac{i\,\omega\, Y}{\sqrt{V(r)}} + H\sqrt{W(r)}\, \mathcal F^{\prime }(r)\right) +Y\,m &=&0
\label{bhspin7} \\
-\frac{H}{r}\left( J_{\theta } + \frac{i}{\sin \theta }J_{\phi }\right) &=& 0
\label{bhspin8}
\end{eqnarray}
Regardless of $H$ and $Y$, Equations (\ref{bhspin6}) and (\ref{bhspin8}) result in $ J_{\theta }+ i(\sin \theta)^{-1} J_{\phi } =0$, implying that $J(\theta ,\phi )$ must be a complex function. The same solution for $J$ is obtained for both the outgoing and incoming cases. Consequently the contribution from $J$ cancels out upon dividing the outgoing probability by the incoming probability as in Equation (\ref{hh}); and we can ignore $J$ from this point. Equations (\ref{bhspin5}) and (\ref{bhspin7}) (for $m=0$) have two possible
solutions: 
\begin{eqnarray*}
H &=& -i\, Y \;, \qquad \mathcal F^{\prime }(r) \equiv \mathcal F_{ out}'= \frac{\omega}{\sqrt{V(r)W(r)}} \\
H &=& i\, Y \;,\qquad \mathcal F^{\prime }(r)\equiv \mathcal F_{ in}^{\prime }(r)= - \frac{\omega}{\sqrt{V(r)W(r)}}
\end{eqnarray*}
where $\mathcal F_{ out, in}$ corresponds to outward, inward solutions. The overall tunnelling probability is $\Gamma \sim \exp -2 \mbox{ Im}\; \left( \mathcal F_{ out} - \mathcal F_{ in} \right)$, with 
\begin{equation}
\mathcal F_{ out}(r)= - \mathcal F_{ in} (r) = \int\!\!\!\!\!\!\!\searrow d r \,\frac{\omega}{\sqrt{V(r)W(r)}}\:.
\end{equation}%
Let us remind readers that under the dominant energy condition and Einstein equations, the functions $V$, $W$ have the same zeroes. Therefore near $r_H$ we have, to first order,
\[
V(r)W(r)=V^{'}(r_{ H})W^{'}(r_{ H})(r-r_{ H})^2+\cdots
\]
and we see that there is a simple pole with a determined coefficient. Using the Feynman's prescription we obtain
\beq\label{ant}
\mbox{ Im}\;\left( \mathcal F_{ out} - \mathcal F_{ in} \right) =\mbox{ Im}\int\!\!\!\!\!\!\!\searrow d r \,\frac{2\omega}{\sqrt{V(r)W(r)}}=\frac{\pi\omega}{\kappa}
\eeq
where the surface gravity eq.(\ref{kvw}) (see also eq.(\ref{kappa anvz}) is recovered, namely
\beq\label{dirtyk}
\kappa= \frac{1}{2}\sqrt{V^{'}(r_{ H})W^{'}(r_{ H})}\;.
\eeq
eq.(\ref{ant} is a special case of a general identity we shall prove in Section \ref{3:found}. Alternatively, regularising as fully explained in Equation (\ref{anvz1}) we end with the same outgoing action of (\ref{angheb}) and the same surface gravity.  
Solving equations (\ref{bhspin5}) and (\ref{bhspin7}) for $H$ and $Y$ in the case that $m\neq 0$ leads to the result:
\begin{equation*}
\left( \frac{H}{Y}\right) ^{2}=\frac{-i\, \omega +\sqrt{V(r)}\, m}{i \,\omega + \sqrt{V(r)}\,m}
\end{equation*}%
and approaching the horizon, we see that 
$\lim_{r\rightarrow r_{ H}}\left(H/Y\right) ^{2} = -1$. Following a procedure similar
 to what was done above, we obtain the same result for the Hawking temperature as in the massless case. The spin-down calculation is very similar to the spin-up case discussed above. Other than
 some changes of signs, the equations are of the same form as the spin-up case. For both
 the massive and massless spin-down cases the Hawking temperature $\kappa/2\pi$, with
 $\kappa$ provided by (\ref{dirtyk}), is obtained, implying that both spin-up and
 spin-down particles are emitted at the same rate. This is consistent with the initial
 assumption that there are as many spin-up as spin-down fermions emitted.\medskip

With this calculation, we basically end the whole story about tunnelling in static,
 spherically symmetric black hole space-times. We have seen how the tunnelling picture
 arose and developed in a variety of different proposals; in Section \ref{3:found} we shall confront with more technical aspects as the equivalence between so called null geodesic and Hamilton--Jacobi methods and their mathematical foundations. For now, however, it seems us to be more important to move on and see how
 the tunnelling method works in conditions where, for example, time
 starts to play an active role.

\subsection{Axis-symmetric stationary black holes}
\label{2.6}

The generalisation to a less symmetric scenario than the spherical one has not only obvious motivations, being the static spherically symmetric situation not so realistic, but also of fundamental relevance. Accretion processes occurring naturally in astronomical  stages are able to spin up a black hole \cite{bard,tamb} as well as to determine the growth in time of the black hole itself. In this subsection we are not concerned with the latter effect (being substantially a non-stationary process, it will be treated in Section \ref{4:dyna}), but rather we would like to point out some efforts that have been made to extend the tunnelling approach to stationary black hole space-times.

The most immediate generalisation of the Schwarzschild spherical symmetry is the axis-symmetric one, {\it i.e.} the Kerr solution for a rotating body with mass $M$ and specific angular momentum $a=J M^{-1}$.  The subsequent inclusion of electric charge brings us to the Kerr--Newman solution, which still observes the axial symmetry and can be seen as a rotating Reissner--Nordstr\"{o}m black hole.  We will apply the tunnelling method in order to retrieve the emission probability for scalar and fermionic particles in these cases, highlighting some features that Kerr and Kerr--Newman solutions show.

\subsubsection{Kerr black hole}  The Boyer--Lindquist form for the metric of a stationary rotating black hole --- corresponding to the choice of a time-independent reference frame which is asymptotically a Lorentz frame at infinity --- is given by
\begin{eqnarray}
\label{boy}
 d s^2 = &-\left( 1-\frac{2 M r}{\Sigma} \right) dt^2- \frac{4 M r a \sin^2\theta}{\Sigma}d t\, d\bar{\phi}+ \frac{\Sigma}{\Delta}d r^2 +  \Sigma \,d\theta^2 +\nonumber
\\\nonumber
\\
 &+ \frac{(r^2+a^2)^2 - a^2 \Delta\,\sin^2\theta}{\Sigma}\,\sin^2\theta\,d\bar{\phi}^2
\end{eqnarray}
where we have defined,
\begin{equation}
 \Sigma := r^2+a^2 \cos^2\theta \,,\qquad \Delta :=r^2+a^2-2 M r\;.
\label{def}
\end{equation}
The roots of the equation $\Delta(r)=0$, $r_{\pm} = M \pm \sqrt{M^2-a^2}$, identify the event (outer, plus sign) horizon and the Cauchy (inner, minus sign) horizon. \\
In order to study the tunnelling process of massless particles from the rotating black hole, however, we consider a metric where the world lines are those of photons with energy $\omega$ moving at infinity with constant $\theta$ and whose projection of the angular momentum on the rotation axis of the black hole is $L_z = a \omega \sin^2\theta$.  In these \textit{Kerr ingoing coordinates} \cite{kerr}, obtained from the transformation
\begin{equation}
 d v = d t + \frac{(r^2+a^2)}{\Delta} d r\ \ \ ,\ \ \ d\phi = d\bar{\phi} + \frac{a}{\Delta} d r
\end{equation}
the line element (\ref{boy}) reads
\begin{eqnarray}
\label{kerr}
 d s^2 =& -\left( 1-\frac{2 M r}{\Sigma} \right) d v^2 + 2 d v \,d r + \Sigma \,d\theta^2 - \frac{4 a M r \sin^2\theta}{\Sigma}\, d\phi\, d v  \nonumber
\\\nonumber
\\
 &- 2 a \sin^2\theta \,d\phi\,d r + \frac{(r^2+a^2)^2 - a^2 \Delta\,\sin^2\theta}{\Sigma}\sin^2\theta\,d\phi^2  \;.
\end{eqnarray}

We recall that in a rotating black hole the \textit{static limit surface} $r_{ st}$, that is the external boundary of the ergosphere given by the equation $g_{tt}=0$, does not coincide with the event horizon. Some remarks regarding the validity of the semi-classical approach have been raised \cite{jiang} because the geometrical optical approximation is reliable in proximity of $r_{ st}$, not the event horizon, where the tunnelling is supposed to occur. In some works \cite{angheben,jiang} the problem is circumvented thanks to a co-rotating change of coordinates $\phi \rightarrow \phi - \Omega^{ H}\, t$, where $\Omega^{ H}= a/(r_+^2+a^2)$ is the angular velocity of the horizon.  In the following we will show that this is not necessary and the right result is recovered automatically.

As fully explained in the previous sections, in the semi-classical approximation, the relevant quantity that plays a role in the
 calculation of the transition probability through the horizon is the classical action $I$
 satisfying the relativistic Hamilton--Jacobi equation. Being the metric eq.(\ref{kerr})
 independent of $v$ and $\phi$, a suitable {\it ansatz} for the form of the action is given by $I
 = -\omega\, v + J\, \phi + \mathcal F(\theta, r)$.  We are going to see that the
 imaginary contribution is due to the $r$-dependent part of the action, which will produce
 a pole in correspondence of the horizon. Expanding the Hamilton--Jacobi equation with
 the {\it ansatz} for $I$ above, we obtain
\begin{eqnarray}
 &a^2 (m \csc\theta - \omega  \sin\theta)^2+2\left[a^2 m-\left(a^2+r^2\right) \omega \right]\mathcal F_r +\nonumber
\\ 
 &+ \left[a^2+r (-2 m+r)\right]\mathcal F_r^2 + \mathcal F_\theta^2 =0
\end{eqnarray}
where subscripts $r, \theta$ are for $\partial_{r,\theta}$. Solving for $\mathcal
 F_r(\theta,r)$, we immediately note that the function to be integrated is also $\theta$-dependent:
\begin{equation}
   \mathcal F_r(\theta,r)=-\frac{X(r)}{ \Delta(r)}\pm\frac{\sqrt{ X(r)^2- \Delta(r) \left[a^2 (M \csc\theta-\omega  \sin\theta)^2+ \mathcal F_\theta(\theta ,r)^2\right]}}{\Delta(r)}
\end{equation}
where $X(r)\equiv a^2 M-\left(a^2+r^2\right) \omega$.  A way to deal with this
 dependence is to simply fix a constant value $\theta=\theta_0$ and show that eventually
 the result has no effective dependence on the choice of $\theta_0$ \cite{kernermann}. 
 Actually there is no need to fix $\theta$ because in order to apply the method we
 consider the regime in which the tunnelling occurs near the horizon, $\Delta(r_+)=0$: in
 this situation the term that brings the $\theta$-dependence drops out and the function
 depends only on $r$. The root of this result is the complete separability of the Hamilton--Jacobi equation in Kerr space-time. Recasting all the expressions in terms of $r_+$ and $r_-$, so that
 $X(r)=(r_++r_-)a^2/2 -\omega(r_+^2+a^2)$ and $\Delta(r)=(r-r_+)(r-r_-)$, we get
\begin{equation}
 \mbox{ Im}\,\mathcal F(\theta,r)=  -\mbox{ Im}\,\int\!\!\!\!\!\!\!\searrow \,\frac{X(r) + \sqrt{
 X^2(r) - (r-r_+)(r-r_-)\ (\dots)}}{(r-r_-)} \frac{d r}{(r-r_+)} 
\end{equation}
where $(\dots)$ contains the whole $\theta$-dependence. Regularising the integral by
 Feynman's prescription, 
\begin{equation}
  \mbox{ Im}\,I = - 2 \pi\, \frac{X(r_+)}{r_+-r_-} =\ \pi \left[ 2\omega\frac{(r_+^2+a^2)}{(r_+-r_-)} - \frac{a^2 (r_++r_-)}{(r_+-r_-)} \right]\;.
\end{equation}
Reintroducing $J = a (r_+ + r_-)/2$ and the angular velocity of the horizon $\Omega^{ H}$, we can rearrange the terms and obtain the tunnelling probability rate:
\begin{equation}
 \exp\left[ -2\, \mbox{ Im}\, I \right] = \exp\left[ -\beta_{ Kerr} (\omega - \Omega^{ H} J) \right]
\end{equation}
where the inverse temperature is $\beta_{ Kerr} = 4\pi (r_+^2 + a^2)/(r_+ - r_-)$.  In the limit $a \rightarrow 0$ we recover the result for the Schwarzschild black hole.

\subsubsection{Kerr--Newman black hole}  The tunnelling of scalar and fermionic, electrically charged, particles in the family of Kerr--Newman space-times does not present new features with respect to previous discussions, the black hole charge $Q$ and specific angular momentum $a$ parameters being treated on the same footing. In the following, we are going to consider first the tunnelling process for a scalar particle of charge $q$ and energy $\omega$, then for a fermion. In both the examples, we implement singular Boyer--Lindquist coordinates $(t,r,\theta,\phi)$
\begin{eqnarray}
\label{boy2}
  d s^2 =& -\left( \frac{\Delta -a^2\sin^2\theta}{\Sigma} \right)\,d t^2 -\frac{2 (r^2+a^2-\Delta)\ a \sin^2\theta}{\Sigma}\,d t\, d\phi + \frac{\Sigma}{\Delta}\,d r^2 + \Sigma\,d\theta^2\nonumber
\\\nonumber
\\
&\hspace{2cm}  + \frac{(r^2+a^2)^2 - a^2 \Delta\,\sin^2\theta}{\Sigma}\,\sin^2\theta\,d\phi^2
\end{eqnarray}
where both incoming and outgoing trajectories are expected to contribute to the emission rate, according to our master Equation (\ref{vie2}). The symbols in (\ref{boy2}) are only slightly changed with respect to above: $\Delta := r^2 + a^2+ Q^2-2 M r$.  The static limit surface is now located at $r_{ st} = M+\sqrt{M^2-Q^2-a^2\cos^2\theta}$, while the event horizon, still the greater root of $\Delta(r)=0$, is $r_+=M+\sqrt{M^2-Q^2-a^2}$.

\paragraph{Tunnelling of scalar particles} As in the previous case, the action can be separated as $I=-\omega t  + J \phi + \mathcal F(\theta,r)$ and further separation of the last terms occurs near the horizon. We explicit the metric functions inside Hamilton--Jacobi equation
\begin{equation}
 g^{\mu \nu}\Big( \partial_{\mu} I - q A_{\mu} \Big)\Big( \partial_{\nu} I - q A_{\nu} \Big)=0
\end{equation}
where the vector potential one-form is given by $A =-Q\ r\ \Sigma^{-1} \Big( d t - a \sin^2\theta d\phi  \Big)$, to obtain
\begin{eqnarray}
   0=&-\frac{\left[(r^2+a^2)^2\ \partial_t I +a\ \partial_{\phi}I- q Q r\right]^2}{2 \Sigma \Delta} + \frac{\left( \partial_{\phi}I+a\sin^2\theta\ \partial_t I \right)^2}{2\Sigma\sin^2\theta}\ +\nonumber
\\\nonumber
\\
&+\frac{\Delta}{2 \Sigma}\left( \partial_rI \right)^2+\frac{\left( \partial_{\theta}I \right)^2}{2\Sigma} \:
\end{eqnarray}
Making use of the {\it ansatz} for $I$ and after some manipulations, we solve for $\mathcal F_r(\theta,r)$,
\begin{equation}\label{F}
 \partial_r \mathcal F_{ in, out}(\theta,r)= \pm \frac{\sqrt{\big[\left(a^2+r^2\right) \omega -a^2 M- q Q r\big]^2-\Delta(r) (\dots)}}{\Delta(r)}
\end{equation}
where the plus (minus) sign corresponds to outgoing (incoming) trajectories. Again, the $(\dots)$ under the square root contains all the $\theta$-dependence and near the horizon it will be negligible.  Equation eq.(\ref{F}) has to be integrated along the path crossing the horizon, so let us write $\Delta(r)=(r-r_+)(r-r_-)$ in order to show explicitly the pole in the integrand and choose the plus sign to select a path that comes out of the black hole.  The imaginary part due to the pole, rearranging a little bit the terms, is given by
\begin{equation}
 \frac{ \pi}{2} \left( \frac{r_+^2+a^2}{r_+-M} \right) \left( \omega - \frac{a}{r_+^2+a^2} M a - \frac{q Q r_+}{r_+^2+a^2}\right)
\end{equation}          
where we easily identify the angular velocity $\Omega^{ H}$, the angular momentum parameter $J$ and the term arising from the presence of a charge parameter: this last term is expressible in terms of the electric potential $\Phi = Q r_+/(r_+^2+a^2)$ of the black hole.  In order to retrieve the final expression, we take into account also the contribution coming from the ingoing trajectory, which is given by the very same procedure as before applied to the function Equation eq.(\ref{F}) with the minus sign.  The imaginary contribution is found to be equal and opposite in sign so the tunnelling probability will be given by
\begin{equation}
\label{ratio}
 \frac{\Gamma_{ em}}{\Gamma_{ ab}} =  \exp\left[-2 \,\mbox{ Im}\,\left(\mathcal F_{ out} - \mathcal F_{ in}\right) \right] = e^{-4\,\mbox{ Im}\,\mathcal F_{ out}}\;.
\end{equation}
A more rigorous derivation of this formula will be given in the next section. The final expression for the tunnelling probability can be written as
\begin{equation}
 \Gamma_{ em} \propto \exp\Big[ -\beta_{ KN} \left( \omega - \Omega^{ H} J - q \Phi \right) \Big] \label{cazzo}
\end{equation}
from which the temperature can be read\ $\beta_{ KN}^{-1}=T_{ KN} = (r_+ - r_-)/4\pi (r_+^2+a^2)$. It is easy to see that two subsequent limits $Q\rightarrow0$ and $a\rightarrow0$ lead us first to the Kerr and then to the Schwarzschild solutions.

\paragraph{Tunnelling of fermions} As a further extension, it is interesting to consider the tunnelling of gravitinos (spin-$3/2$ fermions predicted in supergravity theories \cite{brandt}) in Kerr--Newman black hole space-times \cite{kerner2, yale}. The result is analogous to the scalar case, suggesting again evidence for the universality of black hole radiation, and follows the same line of Section \ref{2.5}, where now only little more shrewdness in the choice of the representation for the Dirac matrices is needed. 

The equation of motion of spin-$3/2$ particles, namely the Rarita--Schwinger equation, can be written as 
\begin{eqnarray*}
 i\ \gamma^{\nu} \left( D_{\nu} + i q A_{\nu} \right) \Uppsi_{\mu} + \frac{m}{\hbar} \Uppsi_{\mu} = 0\label{r-s}\\
 \gamma^{\mu} \,\Uppsi_{\mu} = 0\label{cons}
\end{eqnarray*}
where $\Uppsi_{\mu} \equiv \Uppsi_{\mu\,b}$ is a vector valued spinor of charge $q$ and mass $m$; matrices $\gamma^{\mu}$ satisfy the Clifford algebra; and the covariant derivative is defined as in (\ref{diffop}) and (\ref{Comm}).
The Rarita--Schwinger equation presents as a Dirac equation applied to every vector index of the field $\Uppsi$ and a set of constraints  preventing ghosts propagation.  In the present context, the {\it anstatz} for the wave function is $\Uppsi_{\mu}=(a_{\mu}, b_{\mu}, c_{\mu},d_{\mu})^{ T}\, e^{i I/\hbar}$, where $ T$ is for transposition. It turns out that the Dirac equation can be solved for the action $I$ independently from the components $a_{\mu},...,d_{\mu}$; conversely, $\gamma^\mu \Uppsi_\mu =0$ yields a set of constraints for the components $a_{\mu},...,d_{\mu}$, independently by the action $I$. Thus we may conclude that to calculate the Hawking temperature of gravitinos emitted from the Kerr--Newman black hole, we only need to solve for the Dirac equation in precise the same way as in Section \ref{2.5}; and that, being the action unaffected by the constraints, fermions with every spin will be emitted at the same temperature. We refer the interested reader to the original paper \cite{kerner2} for the precise form of the Dirac matrices suitable to the geometry at hand.  

\subsection{Further extensions}
\label{2.7}

In the literature, the tunnelling method has been extensively applied to generalizations of the stationary black holes.  Among the plethora of models that have been considered, we choose to focus our attention on few interesting cases, comprehensive of non axis-symmetric solutions as well as higher and lower dimensional cases. Our main interest in considering such extensions is to point out features that would be difficult to get by quantum field theory techniques.

\paragraph{Rotating and accelerating black holes} The metric \cite{reh,grif,podo} can
 be written in the form
\begin{equation}
   d s^2 = -f(r,\theta)\ d t^2 + \frac{d r^2}{g(r,\theta)} + \Xi(r,\theta)\ d\theta^2 + K(r,\theta)d\phi^2-2H(r,\theta)\ d t d\phi
\end{equation}
where
\begin{eqnarray*}
   f(r,\theta) &:= \frac{Q - P a^2  \sin^2\theta}{\Sigma\, \Omega^2 }\;,\quad g(r,\theta) := \frac{Q\ \Omega^2}{\Sigma}\;,\quad  \Xi(r,\theta) := \frac{\Sigma}{ P\, \Omega^{2}}\nonumber\\
   K(r,\theta) &:= \frac{\sin^2\theta \left[ P (r^2+a^2)^2 - Q\ a^2\ \sin^2\theta \right]}{\Sigma \,\Omega^2}\,,\quad  H(r,\theta) := 2 \ a \sin^2\theta \,\frac{ P(r^2+a^2)-Q }{\Sigma \,\Omega^2} 
\end{eqnarray*} 
and further,
\begin{eqnarray*}
   \Omega &:= 1-\alpha\ r \cos\theta\ \;,\quad Q := \left[ r^2 - 2 M r + (a^2+e^2+g^2) \right](1-\alpha^2r^2) \nonumber\\
   \Sigma &:= r^2+a^2\cos^2\theta \;,\quad  P := 1-2\alpha M \cos\theta + \alpha^2 ( a^2+e^2+g^2 ) \cos^2\theta
\end{eqnarray*}
This metric is described by the parameters $M, a, \alpha, e$ and $g$ -- respectively the mass, the angular momentum, the acceleration, the electric and magnetic charges of the black hole. The vector potential for these solutions is given by $A = -\Sigma^{-1} [e r ( d t - a\sin^2\theta d\phi) + g\cos\theta ( a d t -(r^2+a^2) d\phi)]$.
Inspecting $\Omega = Q = 0$, we see that, apart from the inner and outer horizons located at $r_{\pm}=M\pm(M^2-a^2-e^2-g^2)^{1/2}$, this black hole has two acceleration horizons $r_{\alpha\pm}=\pm\alpha^{-1}$ and $r_{\alpha}=(\alpha \cos\theta)^{-1}$.  The calculation of the tunnelling probability from the outer horizon gives a temperature
\begin{equation}
 T_+ = \frac{1}{2\pi}\, \frac{(1-\alpha r_+)^3(1+\alpha r_+\sqrt{M^2-a^2-e^2-g^2})}{2M(M+\sqrt{M^2-a^2-e^2-g^2})-e^2-g^2}
\end{equation}
while for the acceleration horizon, given by the same expression with the substitution $r_+ \rightarrow r_\alpha$, one gets a vanishing temperature, a confirmation that \textquotedblleft the acceleration horizon is not a \textit{black hole horizon} essentially\textquotedblright\cite{reh}.

\paragraph{Kerr--G\"{o}del space-time} A second example comes from the Kerr solution embedded in five-dimensional G\"{o}del universe, that is the \textit{Kerr--G\"{o}del} solution \cite{gimon,mann3}.  Among several useful ways to write down the metric for this solution, we can start from the lapse-shift form 
\begin{equation}
   d s^2 = -N^2 d t^2 + g(r) \Big(\varsigma_3 - \frac{a(r)}{g(r)} d t\Big)^2 + \frac{d r^2}{V(r)} + \frac{1}{4}\, r^2 \Big(\varsigma_1^2+\varsigma_2^2\Big)
\end{equation}
and the vector potential $A = \sqrt{3} \,j\, r^2\, \varsigma_3/2$ where
\begin{eqnarray*}
   N^2 &:= \frac{r^2 V(r)}{4 g(r)} \;, \qquad g(r) := -j^2 r^4 + \frac{1}{4}(1-8M j^2)r^2 + \frac{M l^2}{2r^2} \nonumber \\
   V(r) &:= 1- \frac{2M-16j^2M^2-8j M l}{r^2} + \frac{2 M l^2}{r^4}\;,\qquad a(r) := j r^2 + \frac{M l}{r^2}\,,
\end{eqnarray*}
the $\varsigma_i$ are the right-invariant one-forms on $SU(2)$
\begin{eqnarray*}
  \varsigma_1 &= \sin\phi\ d\theta - \cos\phi\sin\theta\ d\psi\,, \quad  \varsigma_2 = \cos\phi\ d\theta + \sin\phi\sin\theta\ d\psi\,, \quad \varsigma_3 = d\phi + \cos\theta\ d\psi
\end{eqnarray*}
and the parameters $j$ and $l$ are respectively the G\"{o}del parameter (determining the rotation of the space-time) and the black hole rotation parameter\footnote{The five-dimensional Kerr solution has usually two rotation parameters $l_1,l_2$ \cite{mp}: in the case considered here we have $l_1=l_2=l$.}: when $l=0$ the metric reduces to the Schwarzschild black hole embedded in G\"{o}del universe; when $m=l=0$ it reduces to the five-dimensional G\"{o}del universe; for $j=0$ we recover the five-dimensional Kerr black hole. Peculiarity of this solution is the existence of closed time-like curves inherited from the G\"odel universe.  In fact, in the suitable range of values for which $g(r)<0$, we have $g_{\phi\phi}<0$ and $\partial_{\phi}$ becomes time-like: in this case the curve with constant $t,r$ and $\theta$, being $\phi$ a periodic coordinate, becomes a closed time-like curve.  The main global properties of this space-time are the (possible) existence of: (i) a {\it closed time-like curve horizon} defined by the condition $g(r)=0$ (being $N^2$ divergent there, it means that nothing can cross the closed time-like curve horizon starting from the region without closed time-like curves); (ii) black hole horizons located at the positive roots of $V(r)=0$; (iii) the ergosphere given by $f(r)=0$, that is $r_{ erg}=\sqrt{2M}$.

We consider the parametrisation $J=j\sqrt{8M}$ with $J^2<1$: in this case the closed time-like horizon will be the outermost surface beyond which closed time-like curves occur. In this region of parameter space, the event horizon of the Kerr black hole can be thought simply as embedded in a G\"{o}del space-time \footnote{It is not clear what  the parameter region $J^2>1$ represents physically. This case, when the closed time-like horizon is the innermost surface, is analysed in \cite{mann3}.}

We want to show the result for the tunnelling in this solution using the null-geodesics method, so it is more convenient to cast the metric in a Painlev\'{e} form \cite{mann3}: first of all we move to co-rotating coordinates, choosing geodesics with zero angular momentum $\phi=\Omega\ t$ and, for convenience, also $d\theta=d\psi=0$; then we pass in Painlev\'{e}-like form with the transformation
\begin{equation}
 t\ \rightarrow\ t_{ p} = t-2 \int d r \, \frac{\sqrt{g(r)(1-V(r))}}{r V(r)}
\end{equation}
The final form of the reduced line element is
\begin{equation}
  d s^2 = - \frac{r^2 V(r)}{ 4 g(r)}\, d t_{ p}^2 +r\,\frac{\sqrt{ 1-V(r)}}{g(r)}\,d r \,d t_{ p}+ d r^2
\end{equation}
The radial null geodesic equation to be used in order to calculate the imaginary part of the action is given by
\begin{equation}
\label{geod}
 \dot{r} = \frac{r}{2 \sqrt{g(r)}}\left(\pm 1 - \sqrt{1-V(r)}\right)
\end{equation}
where the plus (minus) corresponds to outgoing (incoming) trajectories.  The expression develops a divergence at the horizon, leading to a pole in the integral for the reconstruction of the action $I$, with residue $4\sqrt{g(r)} /r_{ H} V'(r_{ H})$ on the horizon.  The tunnelling probability turns out to be
\begin{equation}
 \Gamma_{ em} \propto \exp \left( - \frac{8 \pi \sqrt{g(r)}}{r_{ H} V'(r_{ H})} \;\omega \right)
\end{equation}
from which the temperature can be read off
\begin{equation}
 T_{ KG} = \frac{M\Big[ r_H^2 \left( 1-8j^2M-4Ml \right)-2l^2 \Big]}{\pi r_H^3 \sqrt{ -4j^2r_H^6+(1-8j^2M)r_H^4+2Ml^2 }}\;.
\end{equation}
For $l=0$ it reduces to the Schwarzschild--G\"{o}del case \cite{klemm}.  The application of the method to the closed time-like horizon gives a vanishing temperature: in fact, from Equation eq.(\ref{geod}) we see that $\dot{r}\rightarrow \infty$ when $g(r)\rightarrow 0$, so the integral has no poles there.  This means that there is no tunnelling through the closed time-like curve horizon, a very interesting result that would be difficult to obtain by quantum field theoretic methods.

\paragraph{Taub--NUT--AdS black hole} As a third example, we recover the temperature associated to the horizon of a \textit{Taub-NUT-AdS} black hole \cite{kernermann}, whose general form for the metric can be written as \cite{aste}
\begin{equation}
\label{tna}
 d s^2 = -F(r) \left[d t + 4n^2\ f^2_k\left(\frac{\theta}{2}\right) d\phi \right]^2 + \frac{d r^2}{F(r)} + (r^2+n^2)(d\theta^2 + f^2_k(\theta) d\phi^2)
\end{equation}
where
\begin{equation*}
  F(r) := k (r^2-n^2)(r^2+n^2)^{-1} + (r^2+n^2)\Big[ -2Mr+\mathfrak{l}^2 (r^4+6n^2r^2-3n^4) \Big]
\end{equation*}
$\Lambda = -3\ \mathfrak{l}^{-2}$ is the cosmological constant, $n$ is the so-called NUT charge and the parameter $k$ determines the explicit form of the function $f_k(\theta)$,
\begin{equation*}
 f_k(\theta) :=\left\{ 
                 \begin{array}{ll}
                  \sin\theta\,, &\qquad \mbox{if}\ k=1, \\
                  0 \,,&\qquad \mbox{if}\ k=0, \\ 
                  \sinh\theta \,,&\qquad \mbox{if}\ k=-1
                 \end{array}
                \right.
\end{equation*}
The Taub--NUT solution represents a generalisation of the Schwarzschild space-time (for details see \cite{aste,hawkingellis}) and, as in the previous example, also these solutions possess closed time-like curves for some ranges of the parameters involved.  Here, we shall explicitly refer to the case $k=-1$ and $4n^2 \leq \mathfrak{l}^2$, that is a hyperbolic subclass that does not contain closed time-like curves. In this case, in order to apply the Hamilton--Jacobi tunnelling method we expand the metric functions near the horizon, $F(r) \simeq F'(r_+)\ (r-r_+)$. The metric eq.(\ref{tna}) can now be written in the near horizon approximation as
\begin{equation}
 d s^2 = -F'(r_+) (r-r_+) d t^2 + \frac{d r^2}{F'(r_+) (r-r_+)} + G(r_+,\theta) d\phi^2\;.
\end{equation}
The Hamilton--Jacobi method can be applied straightforwardly, leading to a $\theta$-independent expression for the tunnelling probability $\Gamma_{ em} \sim \exp \left(- 4 \pi \omega / F'(r_+)\right)$ and hence a temperature $T_{ TN} =  F'(r_+)/4\pi$.  This result can be compared to the one calculated by Wick rotation \cite{aste,ste} (to compare the two results, it is sufficient to Wick-rotate the NUT charge $n^2=-N^2$).

\paragraph{BTZ black hole} As a further application of the tunnelling method in stationary black hole space-times, we consider the Ba\~{n}ados--Teitelboim--Zanelli (BTZ) solution \cite{btz,med,li}, which represents a $(2+1)$-dimensional rotating black hole with negative cosmological constant.  The line element is written as
\begin{eqnarray}
 ds^2 = &-N^2\ d t^2 + \frac{d r^2}{N^2}+ r^2 \left( N^{\phi} d t + d\phi \right)^2
\end{eqnarray}
where
\begin{equation*}
 N^2 :=  \frac{r^2}{\mathfrak{l}^2} - M + \frac{J^2}{4 r^2} \;,\qquad N^{\phi} := -\frac{J}{2r^2} ,
\end{equation*}
with the cosmological constant $\Lambda = -\mathfrak{l}^{-2}$.  The line element shows two horizons in correspondence of the zeros of $N^2$, which can be written as
\begin{equation}
 \label{zero}
 r^2_{\pm} = \frac{1}{2} M \mathfrak{l}^2\ \left[ 1\ \pm\ \sqrt{ 1-\frac{J^2}{M^2 \mathfrak{l}^2}} \right]
\end{equation}
so that $N^2 =(r^2-r_+^2)(r^2-r_-^2)/\mathfrak{l}^2 r^2$.
We maintain the condition $M\mathfrak{l}>J$, which ensures non-extremality.  The two Killing vectors of the space-time, $\partial_t$ and $\partial_{\phi}$, allow the separation of the action as $I = -\omega t + J \phi + \mathcal F(r)$. The Hamilton--Jacobi equation in this case reduces to
\begin{eqnarray}
 \frac{J^2}{r^{2}}\ -\ \frac{(\omega + J N^{\phi})^2}{N^2}+N^2 \mathcal F_r(r)^2 = 0
\end{eqnarray}
from which the function $\mathcal F'(r)$ is easily recovered
\begin{equation}
 \mathcal F'_{\pm}(r) = \pm\ \frac{\sqrt{(r \omega +J r N^{\phi}-J N)(r \omega +J r N^{\phi}+J N)}}{r N^2}
\end{equation}
The calculation of the imaginary part of the action goes on taking into account that, being the metric singular on the horizon $r_+$, we have to consider both outgoing and ingoing trajectories.  From the considerations already made, we know that it is sufficient to calculate the contribution coming from the outgoing path and then consider $- 2\, \mbox{ Im} (\mathcal F_+ - \mathcal F_-) = - 4 \mbox{ Im} \,\mathcal F_+$.  Using Equation eq.(\ref{zero}) to show explicitly the pole of the integrand, the result is found to be
\begin{equation}
 \Gamma_{ em}\sim \exp \left[ -4 \,\mbox{ Im}\,(\mathcal F_+)\right] \sim \exp \left[ -\frac{2 \pi \mathfrak{l}^2 r_+}{r_+^2 - r_-^2} \left(\omega - \frac{J^2}{2 r_+^2} \right) \right]
\end{equation}
We recognise the angular velocity of the horizon $\Omega^{ H} = J/2r_+^2$ and infer that the BTZ temperature is $T_{ BTZ} =(r_+^2-r_-^2)/2 \pi \mathfrak{l}^2 r_+$.

\paragraph{Topological black holes}  In the simplest form these are asymptotically Anti-de Sitter metrics of the form eq.(\ref{ab}) with $C(r)=r$ and
\beq
V(r)=W(r)=k-\frac{2 M}{r}+\frac{r^2}{L^2}, \quad k=0,-1
\eeq
except that the transverse manifold is either a torus (for $k=0)$ or a compact Riemann surface with higher genus (for $k=-1)$ \cite{peld96-13-2707,mann97-14-L109,LV,Brill:1997mf}; the case $k=1$ is the Schwarzschild--AdS solution with spherical symmetry and spherical horizon.   In all cases the horizon inherits the topology of the transverse manifold but, independently of this, the horizon pole responsible for the imaginary part of the action is given by the outermost zero of the function $V(r)$, or equivalently as a root of the cubic algebraic equation $rV(r)=0$.  The derivation of the tunnelling probability given for spherical symmetry does not actually depend so much on the symmetry of the transverse manifold, what really matters being the warped form $d s^{2}=d\gamma^{2}+r^{2}d\sigma^{2}$. In this way one recovers the Hawking temperature, $T_{ H}=V'(r_{ H})/4\pi$, from the tunnelling picture for AdS black holes of any topology.  It is worth to mention at this point that the local character of the tunnelling method makes the temperature and the rate independent of the effects of ``large isometries'', when they are used to construct new solutions by the quotient method (identification of points). Thus, for example, the temperature of the $RP^{3}$ geon is the same as the Schwarzschild black hole. The universality of the Hawking effect is all the more evident in the tunnelling picture.

\paragraph{Extremal black holes} It may happen that for certain values of the macroscopic black hole parameters the corresponding surface gravity vanishes; such solutions are called extremal black holes since, as a rule, these parameters are on the line separating black holes from naked singularities. Such is the case of the Reissner--Nordsrt\"om solution, which describes a black hole if and only if $M^{2}\geq Q^{2}$, or of the Kerr solution for values $M^{2}\geq J^{2}/M^{2}$, or of the more general Kerr--Newmann family. These are the extremal black holes of general relativity without cosmological term $\Lambda$. There are many other examples either in the presence of $\Lambda$, or in theories involving coupled gravitational, electromagnetic and scalar fields \cite{maeda,Garfinkle:1990qj}, but we will not cover all such solutions since for certain values of the dilaton coupling (the term $\exp(-2 \mathfrak a\varphi)F^2$ in the Lagrangian), the thermodynamic description breaks down \cite{Holzhey:1991bx,Preskill:1991tb}. As a warning, the extremal black holes considered here must be distinguished from the extremal solutions which are obtained by means of limits taken on the parameters of the metric (such as the limit $Q\to M$ in the Reissner--Nordstr\"om solution) further  accompanied by coordinate changes becoming singular in the limit (see, {\it e.g.} \cite{Caldarelli:2000wc} and references therein). Such is the case of the charged Nariai solution \cite{nariai,GP,bousso1} describing de Sitter black holes with the maximum mass for a given charge, which are finite temperature objects.\par 
As a rule the horizon of extremal black holes is at infinite spatial distance, corresponding to a double pole in the metric component $g_{rr}$, but there are exceptions. One example is the family of extremal black hole solutions in dilaton gravity with electromagnetic coupling $\exp(-2 \mathfrak a\varphi)F^{2}$, $\mathfrak a>1$, which reads \cite{Garfinkle:1990qj}
$$
d s^{2}=-\left(1-\frac{r_{0}}{r}\right)^{\frac{2}{1+\mathfrak a^{2}}}d t^{2}+\left(1-\frac{r_{0}}{r}\right)^{-\frac{2}{1+\mathfrak a^{2}}}d r^{2}+r^{2}\left(1-\frac{r_{0}}{r}\right)^{\frac{2\mathfrak a^{2}}{1+\mathfrak a^{2}}}d\Omega^{2}\;.
$$
This metric has zero surface gravity, an event horizon at finite spatial distance, but also a curvature singularity at  $r=r_{0}$.\par

The tunnelling method has been originally applied to extremal black holes in \cite{angheben,kernermann}, where several problems were pointed out. For example, 
consider an extremal Reissner--Nordstr\"om solution
\beq
d s^{2}=-\left(1-\frac{M}{r}\right)^{2} d t^2+\left(1-\frac{M}{r}\right)^{-2}d r^{2} +r^{2}d\Omega^{2}
\eeq
which can be either magnetically or electrically charged, with charge $Q=\pm M$ in both cases. Kerner and Mann \cite{kernermann} showed that there is a breakdown of the WKB method in the null geodesic approximation, invalidating the method. The fact is that the quantity $r_{ out}$, so crucial for the null geodesic method, actually does not exist since upon emitting a neutral particle the extremal hole becomes a naked singularity. They also showed that both methods yield a divergent real part to the action, which is suggestive of a full suppression of particle emission since then there is no trace of an imaginary part. Further, taking into account self-gravitating effects, they were able to show that the temperature approaches zero the closer the original black hole (that is prior to the emission of a particle) is to extremality. Angheben \textit{et al.} \cite{angheben} found the same result using the proper spatial distance to treat covariantly the horizon pole. On the other hand, using the radial coordinate one can easily find a prescription giving a finite result. From the metric one is led to study the integral
$$
\text{Im} \int\!\!\!\!\!\!\!\searrow dr \,\omega\,\frac{r^2}{(r-M)^{2}}\;.
$$
Note the double pole, still present when using regular advanced time coordinates. Making  use  of the formula
$$
\int \frac{dx}{(x \pm i0)^{2}}f(x)=P\int\frac{dx}{x^2}f(x)\pm i\pi f'(0) 
$$
one gets a non vanishing imaginary part. Using instead the principal value, the integral will be real. How should one interpreted these results? It is known that an extremal Reissner--Nordstr\"om black hole will not radiate away its mass by emission of neutral particles\footnote{It has been argued in \cite{Kraus:1994fj} that the emission vanishes even for charged particles if self-gravitation is taken into account.}, as can be seen either by taking the extremal limit of the emission rate of non extremal black holes, or by computing the spectrum starting directly with the extremal solution. Therefore the temperature vanishes and one has to use the principal value prescription in order to be consistent with the (free) quantum field theory result, or alternatively a  regularisation of the double pole giving a null result. The formal reason is easy to understand: for a simple pole the horizon is reached logarithmically fast as $~\log(r-r_{H})$, so to pass the horizon inward necessarily 
generates an imaginary part $\pm i\pi$, while for a double pole the behaviour of the integral is $(r-m)^{-1}$ which can assume negative real values. The physical reason is that revealed by Kerner \& Mann: there is no horizon's interior in a naked singularity. Of course one has to use the same prescription even for the emission of charged particles, so these too will not be emitted. It is of interest that the hole still emits charged particle in the super-radiance channel, with total luminosity \cite{Gibbons:1975kk,Vanzo:1995bh}
\beq\label{lumi}
L=-\frac{1}{2\pi}\sum_{l\geq 0}(2l+1)\int_{m}^{q\Phi_{+}}\Gamma_{l}(\omega)\omega d\omega
\eeq  
where $q$ and $m$ are the particle's charge and mass, respectively; $\Phi_{+}=Q/M$ is the electrostatic potential at the horizon; and $\Gamma_{l}(\omega)$ the absorbitivity of the hole. Note that (i) $\Phi_{+}$ is the sign of the hole charge since $M=|Q|$ in the extremal case; and (ii) the range of integration is within the super-radiance region, $\omega-q\Phi_{+}<0$. Therefore the hole will preferentially emit charges of the same sign of the hole charge, thereby driving it away from extremality. We can interpret this result as cosmic censorship at work.\par
Under further conditions $\omega M\gg l(l+1)$, $m M\gg 1$ (in Planck's units) the absorbitivity can be computed in the WKB approximation, which gives
\beq
\Gamma_l(\omega)=-\exp\left(-\frac{\pi m^{2}M^{2}}{q Q}\right), \quad q Q>0\;.
\eeq
Similar results and identical conclusions can be obtained for extremal rotating black holes. Here the sum  in eq.(\ref{lumi} is extended to cover the azimuthal quantum number $\mathfrak m$ and the upper limit of integration is to be replaced with $\mathfrak m \Omega^{H}$, so that the hole will not emit radiation in the s-wave channel.\par
We conclude that the tunnelling methods, at least in the present form, cannot describe the super-radiance regime of black hole evaporation but are otherwise consistent with their stability as predicted on other grounds. Perhaps this means that with super-radiance we are dealing with a bulk phenomenon that should be treated by analogy with the Schwinger's pair creation effect by an electric field.
 
\paragraph{Higher order corrections} As our last point we report on a case which has been put forward in \cite{Banerjee:2008cf,Majhi:2008gi} and further extended in \cite{Banerjee:2009wb,Majhi:2009uk,Banerjee:2010xi}, regarding corrections of higher order in $\hbar$ to the Hawking temperature. The argument starts from the semi-classical expansion
\beq
I=I_{0}+\sum_{n\geq 1}\hbar^{n}I_{n}
\eeq
and the assumption that the field obeying the Klein--Gordon equation in the black hole background is as usual $\varphi=\exp \left(iI/\hbar\right)$. It follows from this that all $I_{n}$ obey the same partial differential equation, giving rationale to the further assumption that $I_{n}\propto I_{0}$ for every $n$. Therefore, writing $I_{n}=c_{n}I_{0}$, the action takes the form
\beq\label{II0}
I=\left(1+\sum_{n\geq 1}c_{n}\hbar^{n}\right)I_{0}
\eeq
where $I_{0}$, to fix ideas, obeys the Hamilton--Jacobi eq.(\ref{m2} and $\{c_n\}_n$ remain unpredictable quantities. It is clear that the tunnelling probabilty $\Gamma=\exp \left(-2\mbox{ Im}I\right)$ will predict a modified temperature
\beq\label{BM}
T=\frac{T_{0}}{(1+\sum_{n\geq 1}c_{n}\hbar^{n})}
\eeq
where 
\beq
T_{0}=\frac{1}{4\pi}\sqrt{V'(r_{ H})W'(r_{ H})}=\frac{\kappa}{2\pi}
\eeq
is the uncorrected black hole temperature associated to the Killing surface gravity. Obviously this is not changed by simply passing to Kodama--Hayward surface gravity and energy. 
The procedure looked very suspicious to several authors  \cite{Yale:2010tn,Chatterjee:2009mw,Wang:2009zzw}, who did not find any higher order correction neither from scalar emission nor for bosons or fermions. The contradiction was settled by Yale \cite{Yale:2011zz}, who noted that the result eq.(\ref{BM}) assumes an incorrect  definition of energy, namely $\omega_{0}=-\partial_{t}I_{0}$, while Hamilton--Jacobi theory would require $\omega=-\partial_{t}I$ instead. Then, from eq.(\ref{II0})
\beq
\omega=\left(1+\sum_{n\geq 1}c_{n}\hbar^{n}\right)\,\omega_{0}   
\eeq
so that
\beq
T=\frac{\omega}{2\mbox{ Im}I}=\frac{\omega_{0}}{2\mbox{ Im}I_{0}}=T_{0}
\eeq
and all supposed corrections disappear. Of course, one should also allow complex solutions of the array of semi-classical equations, so even the assumption $I_{n}\propto I_{0}$ for all $n$ is really unjustified.

\section{Analytic Continuation Arguments}
\label{3:found}
\setcounter{equation}{0}

In the previous sections, we have described the Hamilton--Jacobi strategy in a list of four
 steps, from (a) to (d). It is clear, however, that, at least at first sight, not all of them stay
 on equal footing: besides some irremissible (\textit{e.g.} postulation of Hamilton--Jacobi
 equation) or very natural requirements (\textit{e.g.} trajectory splitting as
 $\overrightarrow{abc}$ into $\overrightarrow{a b} + \overrightarrow{b c}$), we find
 other less tolerable points. Why, in fact, should it be that particles traveling along
 classically forbidden trajectories from inside the black hole to outside must follow null
 paths? And which fundamental principle suggests us to regularise divergent integrals
 according to one prescription rather than others? By the end of the day, we shall show
 that --- in contrast to the common sense --- even the Hamilton--Jacobi equation is an
 accessory requirement. A certain experience in the field tells us how these points can
 result in some sense cryptic to the same experts. 
With the purpose of clarifying some of the points mentioned above, we are going to
 outline the foundations of the  Hamilton--Jacobi method in order to point out what is
 fundamental and what is only an additional assumption; and comparing, if possible,
 alternative solutions attempted in literature with present proposals.

\subsection{Foundation of Hamilton--Jacobi method}
\label{3.1}

\begin{figure}[h]
\begin{center}
\includegraphics[width=0.75\textwidth]{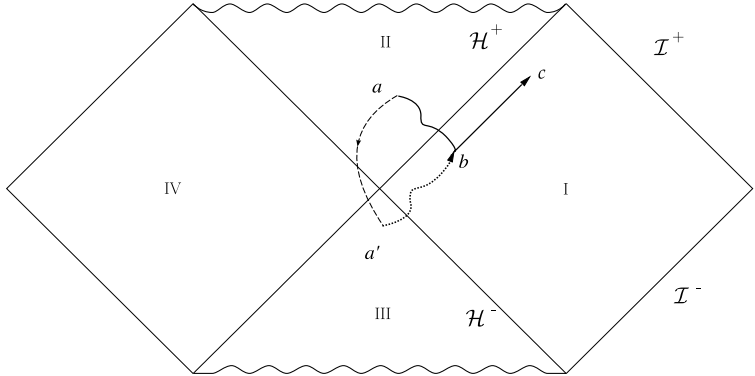}
\caption{\label{ks1} A section of the complexified $U,V$ plane of the eternal
 Schwarzschild black hole. $\theta, \phi$ coordinates are suppressed. 
$\protect\overrightarrow{a b c}$ is a path from inside to outside the hole, with
 $\protect\overrightarrow{a b}$ a classically forbidden trajectory and $\protect\overrightarrow{a'b}$
 representing its reflected trajectory.}
\end{center}
\end{figure}

Let us consider the motion of a scalar particle from region $II$ to region $I$ in the
 eternal version of Schwarzschild black hole ($\overrightarrow{a b}$ in fig.\ref{ks1}). This
 motion is classically forbidden since the particle should travel back in time to follow it.
 Notice that, in general, nothing we can say about the causal nature of the forbidden path.
 However, if the coordinates of the starting point $a$ are displaced to complex values,
 then such an allowed path exists.
Thus, the reader must think to fig.\ref{ks1} as a section of the complexified Schwarzschild
plane with coordinates $(\theta,\phi)$ constant and suppressed. By analytically
continuing the point $a$ in the complexified Schwarzschild space, the amplitude to
propagate to $c$ from a real point $a$ in region $II$ can be related to the amplitude to
propagate to $c$ from a reflected point $a'$ in region $III$ \cite{hartle}. This latter
process is just the time-reversed of absorption of a particle by the black hole. In this way,
the emission probability for a black hole is related to the probability for it to absorb. All
that we need in order to prove the case, is to take as a fundamental assumption the analyticity of the particle action in the complex $(U,V)$ plane as a function of the space-time coordinates of point $a$ \cite{Vanzo:2011nd}.  

In region $II$ ($III$), the Kruskal--Szekeres coordinates $U, V$ are both positive (negative) definite. So, let us rotate $U$ and $V$ from positive to negative values  by posing
\begin{equation}
 \tilde U = e^{i \vartheta} \,U \;, \quad \tilde V= e^{-i \vartheta}\, V\;, \qquad \vartheta\in[0, \pi]\;,
\label{analytic}
\end{equation}
a choice which clearly preserves the invariant product $U V = (1-r/r_{ g}) e^{r/r_{ g}}$. The integral of $d I$ over the tunnelling path $a\to b\to c$ will now be replaced by the integral over the path (say $\gamma$) consisting of the semi-circle $a\to a^{'}$ with $\vartheta$ ranging from $0$ to $\pi$ (over which $|U|$, $|V|$ are constants) plus the integral over the path $a^{'}\to b\to c$ which crosses the past horizon (over which $\vartheta$ is constant). In formulas
\begin{eqnarray}
 I &\equiv& \int\!\!\!\!\!\!\!\searrow d I
 =  \int_{\gamma}(\partial_{\tilde U} I\, d\tilde U + \partial_{\tilde V} I\, d\tilde V) \nonumber \\
&=& \int\!\!\!\!\!\!\!\nearrow (\partial_U I \, d U + \partial_V I \,d V) - i \int_{0}^{\pi} d\vartheta (\tilde V \,\partial_{\tilde V} - \tilde U \,\partial_{\tilde U}) I \label{bho} \;,
\end{eqnarray}
where, this time, the upward arrow denotes integration along the classically allowed trajectory $\overrightarrow{a'bc}$. Since in region $I$, $U = - e^{-\kappa u}$ and $V= e^{+ \kappa v}$, where $u,v$ are retarded and advanced time coordinates and $\kappa$ is the horizon surface gravity, $\kappa = 1/4 M$, we have that $\partial_t = \kappa(V \partial_V - U \partial_U)$. Hence we obtain
\begin{equation}
 \int\!\!\!\!\!\!\!\searrow d I = \int\!\!\!\!\!\!\!\nearrow d I - \frac{i \pi }{\kappa}\, \partial_t I =   \int\!\!\!\!\!\!\!\nearrow d I + \frac{i \pi \omega}{\kappa}\;,
\label{vie} 
\end{equation}
with $\omega= - p \cdot \xi = - \partial_t I$ the conserved Killing energy of the emitted particle. Taking into consideration only the imaginary part of the action, which is the relevant part for tunnelling purposes, we get   
\begin{equation}
-2\, \mbox{ Im}\,  \left(\int\!\!\!\!\!\!\!\searrow d I - \int\!\!\!\!\!\!\!\nearrow d I\right) = -\frac{2\pi \omega}{\kappa}\;.\label{vie2}
\end{equation}
Since the geometry is static, it is left invariant by time inversion $\hat T$:
\begin{equation}
  \,\mbox{ Im}\,\int\!\!\!\!\!\!\!\nearrow d I = \hat T \; \mbox{ Im}\,\int\!\!\!\!\!\!\!\nearrow d I = \mbox{ Im}\,\int_{b\rightarrow a } d I
\label{t inversion}
\end{equation}
where $b\rightarrow a$ is the time inverse of the path $a'\rightarrow b$ (fig.\ref{ks1}). With a justified abuse of terminology, $\int\!\!\!\!\!\!\!\nearrow$ will denote also the integration along the time-reversed path. From a physical point of view, this is the path followed by an incoming particle absorbed by the black hole, so that  exponentiating (\ref{vie2}), we can write
\begin{equation}
 \Gamma_{{em}} = \Gamma_{{abs}} \,e^{-\frac{2\pi \omega}{\kappa}} \;, \label{hh}
\end{equation}
in agreement with the result of Hartle and Hawking \cite{hartle} derived by path--integral methods. \\
The rotation of the $(U,V)$ coordinates in the complex plane, Equation (\ref{analytic}), has been chosen in the form given because it corresponds to a Wick-like rotation of Schwarzschild time in the lower half complex plane, namely $t \to  t - i \vartheta/\kappa$, which is the analyticity region of positive energy solutions of field equations. In fact:
\begin{equation}
e^{-2\kappa \tilde t} =: \frac{\tilde U}{\tilde V} \stackrel{(\ref{analytic})}{=} e^{2 i \vartheta}\, \frac{U}{V} =  e^{2 i \vartheta}\cdot e^{- 2\kappa t} =  \exp \left[- 2\kappa \left(t - \frac{i \vartheta}{\kappa}\right)\right]\;. \label{wick}
\end{equation}
One might legitimately ask for the meaning of an anti-Wick rotation, $ t \to t + i \vartheta/\kappa$. In this case, one rotates $U$ ($V$) clockwise (counter-clockwise) with consequent change of signs in Equation (\ref{analytic}). In turn, Equation (\ref{vie2}) changes in favour of
\begin{equation}
 -2\, \mbox{Im}\,  \left(\int\!\!\!\!\!\!\!\searrow d I - \int\!\!\!\!\!\!\!\nearrow d I\right) = \frac{2\pi \omega}{\kappa}\;.
\end{equation}
The plus sign at the right hand side of this equation, tells us that the object we are talking about is not a black hole, but rather its time reverse (\textit{white hole}), for which absorption is exponentially suppressed with respect to emission, $\Gamma_{{abs}} = \Gamma_{{em}} \,e^{-\frac{2\pi \omega}{\kappa}}$. 

\paragraph{The Reissner--Nordstr\"om solution} It is easy to generalise the present discussion to include charged, spherically symmetric black holes. In this case, Einstein--Maxwell equations can be solved in terms of the Reissner--Nordstr\"{o}m metric,
\begin{equation}
\label{reissner}
  d s^2 = - \left(1- \frac{2M}{r} + \frac{Q}{r^2}\right) d t^2 + \frac{d r^2}{\left(1- \frac{2M}{r} + \frac{Q}{r^2}\right)} + r^2 \left(d \theta^2 + \sin^2\theta \,d \phi^2\right)
\end{equation}
and the Maxwell one-form,
\begin{equation}
 A = \frac{Q}{r} \, d t \label{maxwell}
\end{equation}
where $M$ and $Q$ represent the black hole mass and electric charge, respectively. Without entering the details of the solution, for which we refer the reader to standard texts (\textit{e.g.} \cite{hawkingellis, wald}), it is enough to say that Reissner--Nordstr\"{o}m black hole possesses an outer (inner) event horizon $r_+$ ($r_-$) in correspondence of which we can still define a tortoise coordinate $r_*$; null retarded and advanced coordinates $u,v$; Kruskal like coordinates $U,V$; and an outer (inner) surface gravity $\kappa_+$ ($\kappa_-$).  This time, the action to be integrated on the tunnelling path is $d I=d I_0 + q A$, where $I_0$ is the free action and $q$ is the electric charge of the tunnelling particle. The form $A$ is ill-defined at the horizon, for this reason one usually makes a gauge transformation to a form $ ^{\Lambda}A=A+ d \Lambda$ which is regular there. In our case the analytic continuation takes $A$ away from the horizon so this will actually 
 be unnecessary. Using as above complex $(\tilde U,\tilde V)$ coordinates, we easily obtain 
\begin{equation}\label{main3}
\mbox{ Im}\int\!\!\!\!\!\!\!\searrow d I=\mbox{ Im}\int\!\!\!\!\!\!\!\nearrow d I+\frac{\pi}{\kappa_+} (\omega-q \Phi)
\end{equation}
where $\omega=-\partial_tI_0$ is the mechanical energy and $\Phi=Q /r_+$. The quantity $\omega-q\Phi$ is gauge invariant and conserved along the outgoing path from the past horizon to infinity. This is because for any Killing vector field $\xi$ such that the Lie derivative $({\mathcal L}_\xi \, A)_\mu \equiv \xi^\nu \partial_\nu A_\mu + \partial_\mu \xi^\nu\, A_\nu =0$, the action $I$ is invariant, to first order in $\alpha$, under transformation $x \rightarrow x + \alpha\, \xi$. The N\"{o}ther charge associated to this symmetry transformation is represented by $-\xi^\mu (p_\mu - q A_\mu)$, given that $p_\mu$ is the (mechanical) momentum of the particle. All these considerations are valid, in particular, for the special choice $\xi =\xi_{(t)} =\partial_t$. Finally
\begin{equation}
 2\, \mbox{ Im}\,  \left[\int\!\!\!\!\!\!\!\searrow (d I_0 + q A) - \int\!\!\!\!\!\!\!\nearrow (d I_0 +qA) \right] = \frac{2\pi}{\kappa_+} (\omega -q\Phi)\;,\label{vie3}
\end{equation}
which reproduces the standard result 
\begin{equation}\label{h1h1}
\Gamma_{{em}} = \Gamma_{{abs}} \,e^{-\frac{2\pi}{\kappa_+}\, (\omega - q \Phi)}\;.
\end{equation}
Note that this derivation applies equally well to particles with non vanishing angular momentum, since adding a term $\partial_{\phi}I\,d \phi$ to the differential $d I$ would not change the imaginary part. Hence equations such as eq.(\ref{hh}) and eq.(\ref{h1h1}) are actually true for any angular momentum, though of course the implicit grey body factors  do depend on it.

\paragraph{The Kerr solution} The observation that $\int\partial_{\phi}I\,d \phi$ is real will not be true for rotating black holes. We can extend the calculation to the Kerr solution by noticing that throughout the complex manifold the azimuthal angle must also be rotated in order to keep the metric regular. More precisely, one has to shift $\phi\to \phi-i\Omega^{ H}\vartheta/\kappa$, where $\Omega^{ H}$ indicates the angular velocity of the horizon. Then, adding the term $\partial_{\phi}Id \phi$ to the differential $d I$ would produce an imaginary term after a $\pi$-rotation, equal to $-i\pi\Omega^{ H}\partial_{\phi}I/\kappa$. The outgoing trajectory from the past horizon is a classical solution so $\partial_{\phi}I=J$, the conserved angular momentum. We thus obtain the result
\beq\label{main2}
\mbox{ Im}\int\!\!\!\!\!\!\!\searrow d I=\mbox{ Im}\int\!\!\!\!\!\!\!\nearrow d I+\frac{\pi}{\kappa} (\omega-\Omega^{ H} J)
\eeq
or
\beq\label{db}
\Gamma_{{ em}}=\Gamma_{{ abs}}\,e^{- \frac{2\pi}{\kappa}( \omega-\Omega^{ H} J)}\;.
\eeq
Of course in quantum theory the angular momentum is quantised. As is well known the emission and absorption probabilities for particles with energy $\omega$ and angular momentum $j$ are related to the Bogoliubov $\beta$-coefficients, whose computation is a classical problem involving only the relevant field equations. Unitarity in the space of classical solutions relates them to the transmission coefficient $T_{\omega l \mathfrak m}$ through the potential barrier around the horizon
\beq\label{unit}
\Gamma_{{ abs}}\pm \Gamma_{{ em}}=\vert T_{\omega l \mathfrak m}\vert^2
\eeq 
where the plus (minus) sign is for fermions (bosons). Together with eq.(\ref{db}) it gives the Fermi--Dirac (Bose--Einstein) spectrum. \medskip

Equation (\ref{hh}), or its cousins for the charged and  Kerr black holes, turns out to be
 perfectly consistent with equations (\ref{action})  and (\ref{ima1}) previously obtained,
 but it has been derived under the unique assumption of the analyticity of the particle
 action in the complexified Schwarzschild/Reissner--Nordstr\"{o}m/Kerr geometry, and nothing more. The simplicity of
 this result seems to us particularly important and probably deserves further discussion. In
 first place, it can be considered as the analogue of the well known \textquotedblleft path
 integral derivation of black hole radiance'' due to Hartle and Hawking \cite{hartle},
 derived uniquely from the classical action $I$. In second place, it helps us to understand
 better how the Hamilton--Jacobi method works. At this regard we immediately see that
 reference to the Hamilton--Jacobi equation and insistence on considering only null
 trajectories (on which performing near-horizon expansions) are additional ingredients of
 the method which have nothing to do with the physical phenomenon of tunnelling. Even
 if they are not absolutely necessary from a conceptual point of view, their role turns out
 to be fundamental for computational purposes. As already said, most of the times the
 analytic continuation does not even exist, since only for special manifolds there is a
 complexified extension; while considering arbitrary forbidden trajectories as $\overrightarrow{a b}$ in fig.\ref{ks1} would rule out any exact computation. \\ 
As mentioned above, the regularisation scheme typically implemented in the tunnelling
 method is the Feynman's prescription, \textit{e.g.} (\ref{action}) and (\ref{ima1}). At the
 beginning, we  introduced it with no justification, either physical or mathematical, but
 now it is clear that Feynman's prescription   
\begin{equation*}
  \int_{\mathcal Q} d x \frac{f(x)}{x} \to \int_{\mathcal Q} d x \frac{f(x)}{x-i \epsilon} = i \pi f(0) + \dots
\end{equation*}   
where $\mathcal Q$ is any interval such that $0\in \mathcal Q$, has to be the preferred since it is the only
 one consistent with analytic continuation results, when available. From now on,
 Feynman's prescription will be adopted systematically and its justification given for
 granted even in those cases where there is no analytic proof of its validity.

As a last remark, in the previous sections we repeatedly insisted on the importance of
 using well-behaved coordinates across the horizon. However, even this restriction seems
 to be less compelling in the light of analytic continuation. Equation (\ref{vie2}), which is
 the best achievement of this section, expresses the remarkable fact that it is the difference
 of Im $\int\!\!\!\!\!\!\!\searrow d I$ and Im $\int\!\!\!\!\!\!\!\nearrow d I$ which
 provides the correct ratio between the emission and absorption amplitudes. As far as we
 are concerned with regular coordinates, only the integration over $\overrightarrow{a b
 c}$ trajectories carries a non-vanishing imaginary part, being the absorption amplitude
 completely real. Insisting on singular gauges as Schwarzschild coordinates, however,
 nothing prevents the entering trajectory from picking up an imaginary contribution. It is
 exemplary, in this sense, the (Sg for) Schwarzschild gauge where both the escaping
\begin{equation}
 I_{{out}} = \int_{{Sg}}\!\!\!\!\!\!\!\!\!\searrow d I = - \omega t + \int_{{Sg}}\!\!\!\!\!\!\!\!\!\searrow \;\frac{\omega \,r}{r - r_{ g} - i \epsilon} \,d r = - \omega t + i \pi r_{ g} \omega
\label{sg out}
\end{equation}
and the classically allowed ingoing trajectory 
\begin{equation}
I_{{in}} =\int_{{Sg}}\!\!\!\!\!\!\!\!\!\nearrow d I = - \omega t - \int_{{Sg}}\!\!\!\!\!\!\!\!\!\nearrow\; \frac{\omega \,r}{r - r_{ g} - i \epsilon} \,d r = -\omega t - i \pi r_{ g} \omega\;,
\label{sg in}
\end{equation}
pick up an imaginary contribution providing in this way the correct result. This statement,
 if true for the Schwarzschild gauge, cannot be taken as a general rule valid for any
 singular gauge, as the reader may easily convince himself by taking into consideration
 isotropic coordinates. 

\subsection{Equivalence of null geodesic and Hamilton--Jacobi methods}
\label{3.2}

At this point of the discussion, it seems natural to ask what relation exists between the
 aforementioned null geodesic and Hamilton--Jacobi methods. In literature, (see for instance \cite{kernermann}) we find
 many specific examples suggesting the essential equivalence between the two methods,
 but nothing resembling a mathematical proof. To this aim, notice that in the null geodesic
 approach one starts with
\begin{equation}
\mbox{ Im}\;\int\!\!\!\!\!\!\!\searrow p_r d r
\end{equation}
which looks non covariant. However we can use the full Liouville one-form
 $\varpi=p_{\mu}dx^{\mu}$ and write the more general expression,
\begin{equation}
\mbox{ Im}\int\!\!\!\!\!\!\!\searrow \varpi
\end{equation}
which, without the ``\mbox{ Im}'',  is nothing but the reduced action. We can perform the
 analytic continuation of this integral just as we did for the complete action, first by
 writing $\varpi=p_{U}d U+p_{V}d V$, then rotating $(U,V)$ from zero to $\pi$ and
 finally integrating along the rotated curve. In this way, the imaginary part will be $i\pi(Up_{V}
-Vp_{U})$: but this is $-i\pi\omega/\kappa$, where $\omega=-p_{t}$ is the Killing
 energy as measured at infinity. In all we get
\beq\label{main4}
\mbox{ Im}\int\!\!\!\!\!\!\!\searrow\varpi=\mbox{ Im}\int\!\!\!\!\!\!\!\nearrow\varpi+\frac{\pi\omega}{\kappa}\;.
\eeq
As another example we consider the line element of the Kerr--Newman solution,
\begin{eqnarray}
 d s^2_{ KN}  = \frac{G^2 a^2 \sin^2\theta}{4 \kappa_+^2 \Sigma}& &\cdot\frac{(r-r_-)(r+r_+)}{(r^2 +a ^2) (r_+^2 + a^2)} \left[\frac{\Sigma}{r^2 + a^2} + \frac{\Sigma_+}{r_+^2 + a^2}\right] (U^2 \,d V^2 + V^2\, d U^2) + \nonumber \\
& &  +\frac{G (r-r_-)}{2 \kappa_+^2 \Sigma}\left[\frac{\Sigma^2}{(r^2 + a^2)^2} + \frac{\Sigma_+^2}{(r_+^2 + a^2)^2}\right] d U \,d V + \nonumber \\
& & + \frac{G^2 a^2 \sin^2\theta}{4 \kappa_+^2 \Sigma}\cdot\frac{(r+r_+)^2}{(r_+^2 + a^2)^2} \, (U \, d V - V \,d U)^2 + \nonumber \\
& & +\frac{G a \sin^2\theta}{\kappa_+ \Sigma (r_+^2+a^2)} \left[\Sigma_+ (r-r_-)\! +\! (r^2 +a^2)(r+r_+)\right]\!(Ud V\! -\! Vd U) d \phi_+ \nonumber \\
& & + \Sigma \,d \theta^2 + \frac{[(r^2+a^2)^2 \!- \!a^2 \sin^2\theta \, \Delta] \sin^2\theta}{\Sigma} \,d \phi_+\;, \label{max analytic ext}
\end{eqnarray}
where $U, V$ are Kruskal-like null coordinates, defined as,
\begin{equation}
 \vert U \vert = \exp \kappa (r_* - t)\;,\qquad \vert V \vert = \exp \kappa (r_* +t) \;,
\end{equation}
$d r_* = (r^2 +a^2) d r/\Delta$ and the surface gravity $\kappa$ can
 refer to $\kappa_+$ or $\kappa_-$ according to the space-time region of interest. The co-rotating angular coordinate $\phi_+$ is defined as 
\begin{equation}
 \phi_+ := \phi - \Omega^{ H} t
\end{equation}
with $\Omega^{ H}$ the horizon angular velocity and $G := (r - r_+)/ U V$. For all
 other symbols, the reader is referred to Section \ref{2.6} where a complete treatment of
 axis-symmetric space-times has been given. This metric is particularly important since in the
 appropriate limits it comprises the maximal analytic extension of Kerr--Newman, Kerr,
 Reissner--Nordstr\"{o}m and Schwarzschild black holes \cite{frolov}. \\
The analytic continuation of the coordinates in the Euclidean section requires
\begin{equation}
 t \rightarrow t - i \,\vartheta / \kappa  \qquad { and}\qquad \phi \rightarrow \phi - i\, \Omega^{ H} \vartheta /\kappa \;, \qquad \vartheta \in [0,\pi]
\end{equation}
where from now on, we consider $\kappa \equiv \kappa_+$. We have $\varpi=p_{U}d U+p_{V}d V+p_{\phi}d \phi$ and therefore 
\begin{equation}
\mbox{ Im}\int\!\!\!\!\!\!\!\searrow \varpi = \mbox{ Im}\int\!\!\!\!\!\!\!\nearrow \varpi -\frac{\pi\Omega^{ H}}{\kappa}\,p_{\phi}+\pi(Up_{U}-Vp_{V})\;.
\end{equation}
But $Up_{U}-Vp_{V}=-\kappa^{-1}p_{t}=\kappa^{-1}\omega$ is the energy and
 $p_{\phi}=J$ the conserved angular momentum. We end with our main formula,
 Equation (\ref{vie2}),
\begin{equation}\label{via}
2 \,\mbox{ Im}\left(\int\!\!\!\!\!\!\!\searrow\varpi - \int\!\!\!\!\!\!\!\nearrow \varpi \right)= \frac{2 \pi}{\kappa}\, \left(\omega - \Omega^{ H}\,J\right)
\end{equation}
which proves the equivalence of a {\it generalised null geodesic method}  with Hamilton--Jacobi {\it ansatz} for all stationary, axis-symmetric, charged black holes. In retrospect, this
 could not be otherwise because the Hawking effect is an energy conserving process, so
 that the reduced action is all one needs in a stationary geometry.

\subsection{Alternative proposals}
\label{3.3}

The tunnelling method has been explored by many authors and counts a considerable
 literature which is all but monolithic in presenting the subject. The large number of
 viewpoints is probably the best indication of the non-trivial nature of the topic.
 Conversely, it has been also the origin for a certain confusion \cite{pizzi} which made difficult to
 export its ideas outside the community of experts in the field.  

In this sense, it is paradigmatic to consider the so called  \textquotedblleft factor-two
 problem'' \cite{anm}, actually a fake problem consisting in the fact that a naive application of the tunnelling method may end with a black hole temperature which is twice the expected result.  Roughly speaking, the troubles typically arise when one tries to adopt singular coordinates (like
 Schwarzschild) and uncritically assumes that the absorption amplitude is real. Then, as it
 is clear from Equation (\ref{sg out}), one ends with a temperature equal to $1/4\pi M$.
 As we have seen in Equation (\ref{sg in}), this is simply due to the fact that we are
 missing an equal and opposite contribution coming from the absorbed particle which,
 entered the master Equation (\ref{vie2}), provides indeed the correct temperature. \\
Mitra, for example, circumvents this bug in the following way \cite{mitra}. The Hamilton--Jacobi equation is a differential equation for the action of both the ingoing and outgoing particle. Because of the staticity and spherical symmetry, we can solve the action in
 terms of the standard {\it ansatz} for the separation of variables to get, $I_{ out,in} = -
 \omega t \pm i \pi \omega r_{ g} + \dots$  It is worth to stress that in general,
 from reconstruction of the action can arise some constant of integration $\mathcal C$.
 On the other hand, it was argued, the classical theory of black holes tells us that an incoming particle is
 absorbed with probability equal to one. In order to match these apparently contradictory
 facts, we use the freedom of choosing the integration constant $\mathcal C$ to impose
 the classical constraint on the absorption probability, that is 
$ \mathcal C = \pi \omega r_{ g} $, so that
$ \mbox{ Im}\, I_{ out} = 2 \pi r_{ g} \omega$ and $\Gamma_{ em} \propto
 \exp(-8\pi M \omega)$, as expected.\\
However, if we try to repeat the same argument for the Schwarzschild geometry in
 isotropic coordinates,
\begin{equation}
 ds^2 = - \left(\frac{1-M/2\rho}{1+M/2\rho}\right)^2 d t^2 + \left(1+\frac{M}{2\rho}\right)^4[d \rho^2 + \rho^2 (d \theta^2 + \sin^2\theta\, d \phi^2)] \;,\label{isotropic}
\end{equation}
where this time the horizon is located at the minimal two-sphere $\rho_0 =M/2$, the in-out particle action derived from the
 Hamilton--Jacobi equation reads
\begin{eqnarray}
 I_{ in} &=& (-\omega t + \mathcal C) - i 4\pi M \omega \nonumber \\
 I_{ out} &=& (-\omega t + \mathcal C) + i 4\pi M \omega \;,
\end{eqnarray}
and the classical constraint  Im$\, I_{ in} =0$ fixes Im$\, \mathcal C = 4\pi M
 \omega$ implying a Hawking temperature  of $1/16\pi M$ which, this time, is a half of
 the correct, well-known, result. \smallskip

A relevant achievement in the discussion of the ``factor-two problem'' has been pursued
 by Angheben \textit{et al.} \cite{angheben} and Stotyn \textit{et al.} \cite{stotyn}, who showed that correctly using the theory of
 distributions in curved static space-time it is possible to recover the correct result for the
 Hawking temperature even working in singular gauges. In fact, the non locally integrable
 function $1/r$ does not lead to a covariant distribution $1/(r \pm i \epsilon)$; where
 instead, Feynman's regularisation of an invariantly defined distance makes unnecessary to
 work with the complex action of ingoing particles. The natural choice in substitution of
 $1/r$ is represented by the proper spatial distance, as defined by the spatial metric,
 $d\sigma^2 \equiv g_{m n} d x^md x^n$ ($m, n =1,2,3$), a quantity that is invariant under re-definitions of the spatial coordinate system and, more generally, with
 respect to the subgroup of the gauge group consisting of time re-calibrations and spatial
 diffeomorphisms, defined by
\begin{equation}
 \left\{ \begin{array}{cc}
          &t' = t'(t,x^i) \\
         &{x'}^j =  {x'}^j(x^i)\;.  
        \end{array} \right.
\end{equation}
We consider the previously discussed spherically symmetric metric in diagonal gauge,
\begin{equation}
 d s^2 = - V(r) d t^2 + \frac{d r^2}{W(r)} + r^2 (d \theta^2 + \sin^2\theta\, d\phi^2)\;,
\label{dirty}
\end{equation} 
where we suppose $V(r)$ and $W(r)$ to have both a simple pole in correspondence of the horizon $r_{ H}$,
\begin{equation}
 V(r) = V'(r) (r -r_{ H})+\dots \;, \qquad W(r) = W'(r) (r -r_{ H}) + \dots\;.
\end{equation}
The proper distance reads,
\begin{equation}
 \sigma = \int \frac{d r}{\sqrt{W(r)}} = \frac{2}{\sqrt{W(r_{ H})}} \,\sqrt{r - r_{ H}} \;,
\end{equation}
so the classical action of the outgoing massless particle is
\begin{equation}
 I_{ out} = - \omega t + \int\!\!\!\!\!\!\!\searrow d r \,\frac{\omega}{\sqrt{V(r) W(r)}} = -\omega t + \frac{2}{\sqrt{V'(r_{ H}) W'(r_{ H})}} \int\!\!\!\!\!\!\!\searrow \omega\,\frac{d\sigma}{\sigma}\;.
\label{anvz1}
\end{equation}
The integral is still divergent as soon as $\sigma \rightarrow 0$, namely the horizon is reached, but now the prescription corresponding to the Feynman's propagator $1/\sigma \rightarrow 1/(\sigma - i \epsilon)$ selects the correct imaginary part for the classical action, 
\begin{equation}
 I_{ out} = \frac{2 \pi \omega\, i}{\sqrt{V'(r_{ H}) W'(r_{ H})}} + ({ real\;\, contribution})\;.
\label{angheb}
\end{equation}
From comparison of  (\ref{angheb}) with (\ref{ima}), we soon read off the horizon surface gravity $\kappa$ for the dirty black hole (\ref{dirty}), 
\begin{equation}
 \kappa = \frac{\sqrt{V'(r_{ H}) W'(r_{ H})}}{2}\;,
\label{kappa anvz}
\end{equation}
which is amazingly in agreement with the result obtained through the conical singularity
 method \cite{entropy}. From a qualitative point of view, this coincidence can be
 ascribed to the fact that both the conical singularity and the proper distance methods
 refer the energy $\omega$ to the same Killing observer. However, we notice that, without
 any further assumption, a dirty black hole space-time is not asymptotically flat, so it is not
 completely clear how to normalise in general the Killing vector $\xi^\mu_{\; (t)}$. \par
\begin{figure}[h!!!!]
\begin{center}
\includegraphics[width=0.45\textwidth]{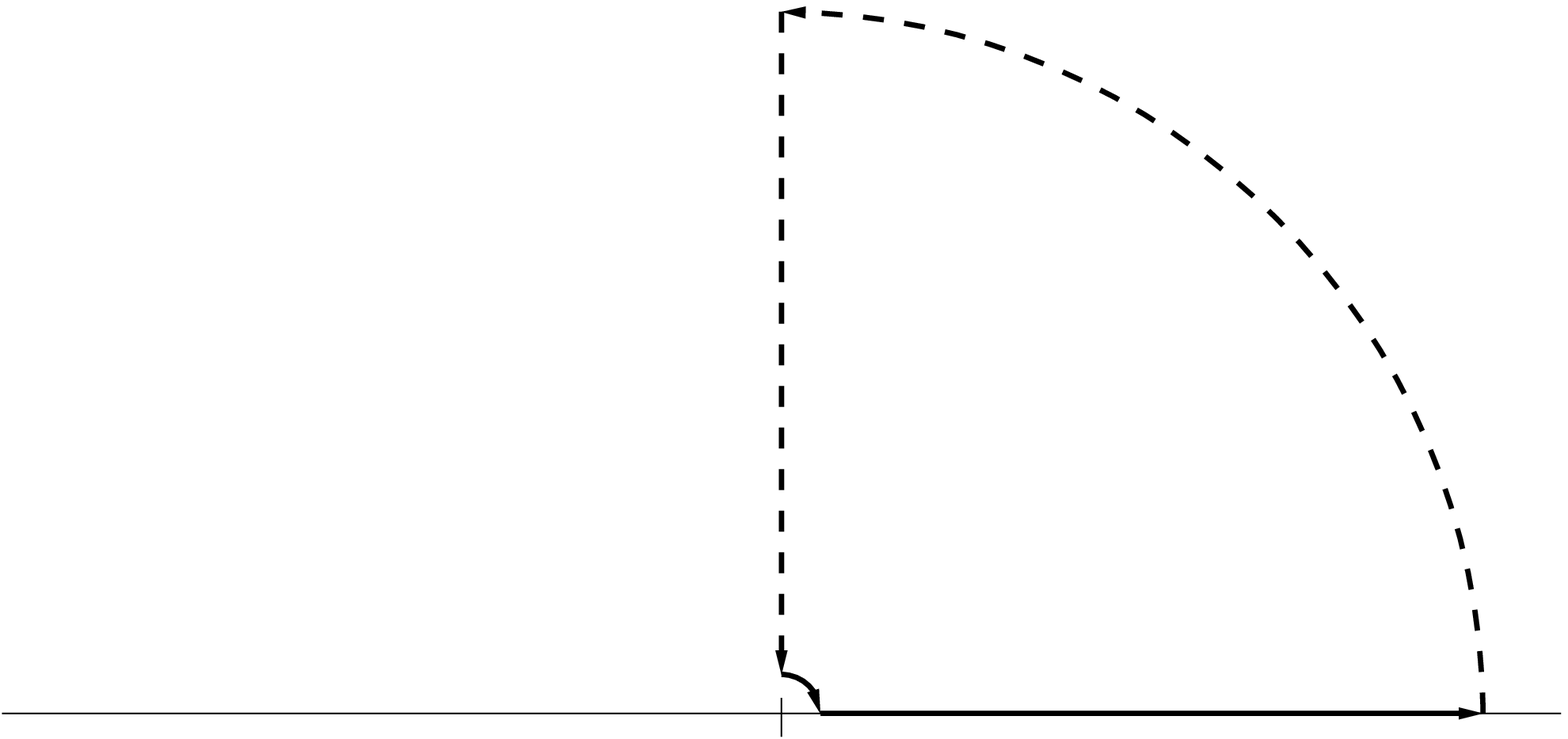} \hspace{1cm}\includegraphics[width=0.45\textwidth]{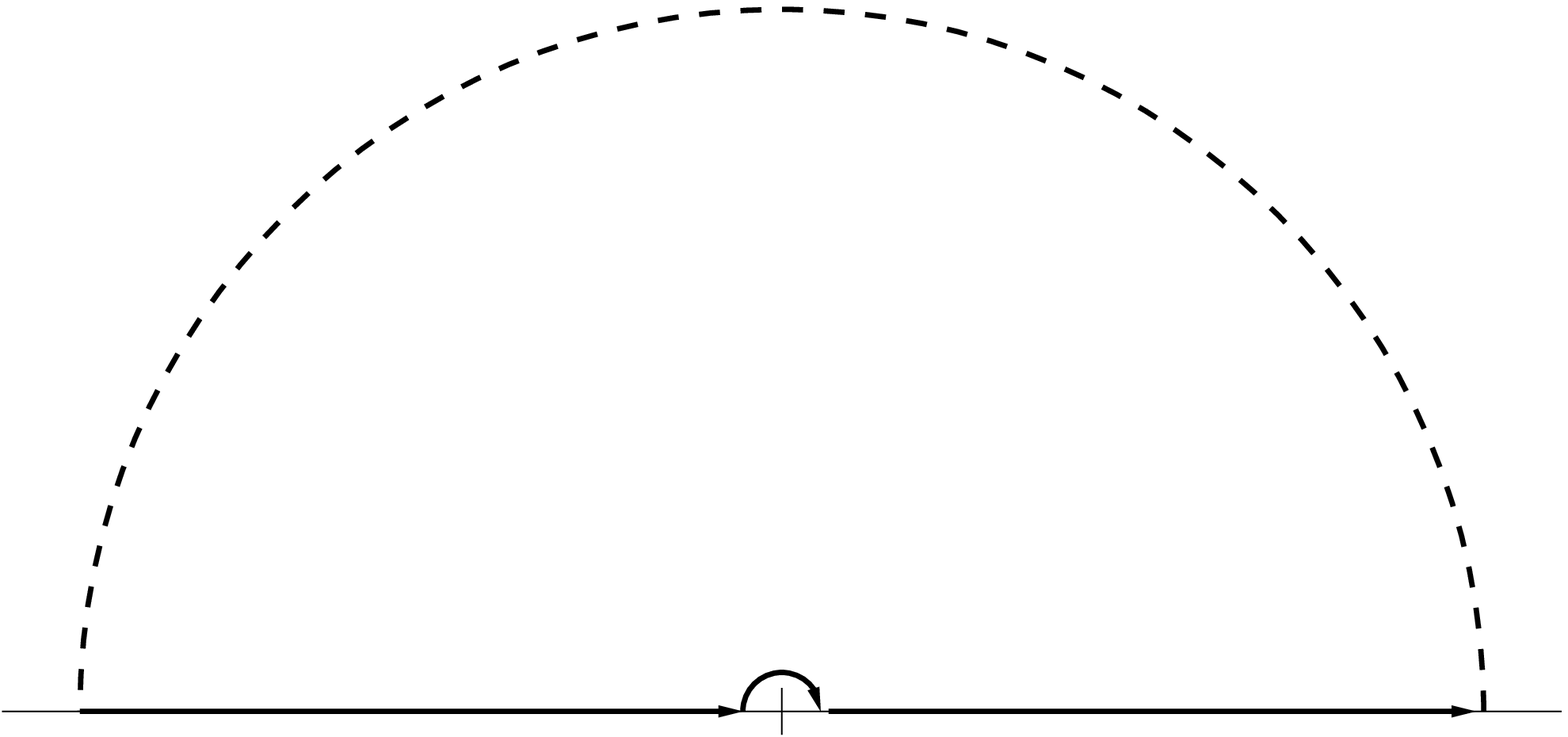}
\caption{\label{contours} Standard semi-circular contour of integration in the complex plane of $\int d\sigma/\sigma$ and the quarter-contour proposed in \cite{stotyn}.}
\end{center}
\end{figure}

 Can we say that the factor-two problem has been solved?
We have seen that, at least confining to the evaluation of the Hawking temperature of a
 Schwarzschild black hole in Schwarzschild coordinates, there are two satisfactory ways to
 circumvent the factor-two problem arisen because of the use of singular charts: the
 former, essentially based on the classical requirement that the absorption probability be
 unity; the latter, related to an appropriate way of dealing with distributions in curved
 space-times. However, the combination of these techniques as made in \cite{stotyn} has
 the deleterious drawback of making the temperature again a half of the correct result. According to these authors, the point is that to apply the Feynman prescription, the proper distance should be extended to negative values, which seems geometrically  impossible. The  
 authors then note that to recover the standard result $T_{ H} = 1/8\pi M$, one ought to use an alternative contour of integration when dealing with proper distance formulation, $\int d\sigma/\sigma$, namely the quarter-contour of fig.\ref{contours} (left) rather than the standard semi-circular contour of fig.\ref{contours} (right).
 
\smallskip

Another point of view is the one advocated in \cite{singleton} where the authors are
 among the first to point out the possible role of a ``temporal contribution'' to solve the
 factor-two problem, at least in the context of singular gauges. In fact, temporal and
 spatial contributions are on equal footing in the Hamilton--Jacobi method as formulated
 in Section \ref{2.2}, since it is covariant. For example, the master equation (\ref{vie2}) does not distinguish at all between ``time'' or ``space'' coordinates, something which makes perfectly sense according to the principles of general relativity. On the other hand, the null geodesic method as originally formulated does not involve a time integration, but appears non covariant and also non canonically invariant.  Along the line of thought
 initially formulated by \cite{canonical}, Singleton \textit{et al.} \cite{singleton} make use of closed integration paths, necessary to ensure invariance of the emission rate under canonical
 transformations. To all the extents, the integration over the closed circuit $\oint p_r d
 r$ corresponds to taking into consideration the imaginary contribution related to the
 action of both the incoming and outgoing particle, exactly as in (\ref{sg in}): 
$\oint \rightarrow \int\!\!\!\!\!\!\!\searrow - \int\!\!\!\!\!\!\!\nearrow$. The null geodesic method as formulated in Section \ref{3.2}, evidently, is covariant as well as invariant under canonical transformations in phase space. Hence, even acknowledging a certain loss of importance of canonical transformations in the passage from classical to quantum theories, this is certainly an interesting point. On the other hand, since not all the
 canonical transformations correspond to unitary transformations in the quantum theory,
 perhaps a more urgent issue is not the canonical invariance of the decay rate, but the covariance of the result -- as demanded by general relativity -- and the unitarity of the process -- as demanded by quantum mechanics -- something that nobody knows how to solve. \par
As stressed above, the Hamilton--Jacobi method was born also, surely not only, by the
 desire of disentangling Hawking radiation from back-reaction, in the sense that it is the
 existence of Hawking radiation that founds the problem of back-reaction, not vice versa.
 Given that the Hamilton--Jacobi method can prove black hole radiance without taking into
 account the back-reaction, it is natural to ask how the back-reaction can be incorporated
 in this method. As shown by Medved and Vagenas \cite{medved,others}, a tunnelled particle of
 instantaneous energy $\omega$ effectively \textquotedblleft sees'' a space-time metric of
 the form (\ref{dirty}) in which the horizon radius $r_{H}=r_{H}(M-\omega)$, $M$ being
 the original Misner--Sharp black hole mass. Because of quantum uncertainty principle, it seems too
 crude the approximation of a discontinuous jump from $M$ to $M-\omega$ for the black
 hole mass. Rather, the particle action corrected by energy conservation, turns out to be
 given by
\begin{equation}
 I'_{ out} = \int_0^\omega d\omega' I_{ out} \Big\vert_{M \rightarrow M - \omega'} \stackrel{(\ref{angheb})}{=} \int_0^\omega d\omega'\frac{2 \pi \, i}{\sqrt{V'_{ H}(M-\omega') W'_{ H}(M-\omega')}} 
\label{vag}
\end{equation}
where the subscript H is a reminder that we are evaluating this quantity at the black hole
 horizon. The quantum corrected tunnelling probability deduced from (\ref{vag}) turns
 out to be that expected from statistical mechanics and in perfect agreement with the null
 geodesic method of Section \ref{2.1}, namely $\Gamma_{ em} \sim \exp \Delta
 S_{ BH}$. A somewhat different treatment of this problem with equivalent results, within the Hamilton-Jacobi method, can also be found in \cite{Yang:2007zzm}.\par
In this section we have taken into consideration alternative proposals to the null geodesic
 and Hamilton--Jacobi methods outlined in Section \ref{2.1} and \ref{2.2}. Without
 claim for completeness, we have tried to point out their salient features, referring to
 original papers for details. We cannot deny that all of them contain small pieces of the
 complete picture, but at the same time the persistent lack of manifest general covariance is to be considered a defect, demanding a suitable reformulation. Exactly this  was our desire in writing this section.

\section{Quantum Tunnelling: Dynamical Black Holes}
\label{4:dyna}
\setcounter{equation}{0}

In the previous sections we saw that further insight into the Hawking's effect can be
 obtained by the use of the tunnelling methods. We showed that in some circumstances it is even possible to
 recover exact results by analytically continuing the integral of the classical action from a
 tunnelling path to a classical path, and how this possibility provided a foundation to the tunnelling method. But of course everything is strictly true only for stationary black holes. \par
 In this section we shall extend this work to what we shall aptly name {\it spherically
 symmetric dynamical black holes}, epitomised by the Vaidya solution \cite{vaidya:1951}
 and its generalisations by Bardeen \cite{bardeen:1981} and York \cite{york:1983}. 
As a matter of fact, it has long been felt that the usual semi-classical treatment of
 stationary black holes had to be extended to cover at least slowly changing black holes.
 By this expression, we mean black holes that can be still described in terms of few
 multipole moments such as mass, angular momentum and the charges associated to
 local gauge symmetries, except that the parameters  and the causal structure are allowed to change with time. Although a technical definition of a ``slowly varying black hole'' can be given in some cases, an example being the Booth--Fairhurst slowly evolving horizon \cite{bf04}, in general this depends on the actual physical processes involved. For example, in the case of Hawking's evaporation, conditions for slowness in the presence of a near-horizon viscous fluid have
 been given by Brevik \cite{Brevik:1999bj} in an interesting attempt to generalise 't
 Hooft's model of the self-screening Hawking atmosphere (quantum corrections to this
 model can be found in \cite{Nojiri:1999pm}). In general, it is understood that the black
 hole temperature has to be much smaller than the Planck mass, while in order to study
 the effects of the expansion, the Hubble rate $H^{-1}$ should be comparable with the
 black hole emission/absorption rates.   

Now, even for the case of a slow evolution, it was pointed out by Fredenhagen and Haag
 long ago \cite{Fredenhagen:1989kr} that by letting the mass of the black hole to change
 with time, the radiation would originate from the surface of the black hole at all
 times after its formation. This poses the question: \emph{what and where is the surface
 of a dynamical black hole}? The issue baffled scientists since the beginning of black hole physics and produced several reactions during the Nineties, which eventually culminated with the notion of {\it outer trapping horizons} by
 Hayward\cite{hayw94-,hayw94,Hayward:1997jp} and the isolated and dynamical
 horizons of Ashtekar and co-workers \cite{Ashtekar:1998sp,isohor,AK,Ashtekar:2003hk}
 (an extensive review is in \cite{Ashtekar:2004cn}). \\
One is concerned to study at first the dynamical version of the event
 horizon and to provide a mathematical definition which is able to capture a useful local
 notion of it, encompassing the time-lasting textbook definition given in Hawking--Ellis renowned book \cite{hawkingellis}. Armed with a precise notion of horizon, we shall proceed to study the instabilities occurring near the horizon of the changing black hole.
 This question looks non trivial since a changing horizon is typically embedded in a
 dynamical space-time and it is not even expected to be a null hyper-surface, though it is
 still one of infinite red-shift. Thus we shall start by reviewing what has been done on this question for spherically symmetric metrics. We can anticipate that the arrival point will be the very important local notion of {\it  future trapping horizons} and their associated {\it surface gravity}, as defined by Hayward.

\subsection{Horizons and surface gravity}
\label{4.1}

Not long after the classical definition of the event (EH) and the apparent horizons (AH) (boundaries of trapped \(3\)-dimensional space-like regions within partial Cauchy surfaces), several quasi-local notions of dynamical horizons were proposed in the 
 literature (another nice review is in \cite{Gourgoulhon:2008pu}), perhaps starting with the
 notion of perfect horizon due to H\'aji\v{c}ek \cite{hajicek}.
But this only applied to equilibrium black holes while the apparent horizon, being tied to a partial
 Cauchy surface, only represents a localisation in time. Moreover, it has proven not
 possible to formulate thermodynamic laws for AH similar to those holding good for the
 event horizons. 

The first successful attempt to go beyond the limitations imposed either by the
 instantaneous character of apparent horizons or by the global, teleological nature of
 event horizons is due to Hayward. His concept of a future outer trapping horizon
 (to be abbreviated as FOTH) then evolved either into less constrained definitions, like
 the Ashtekar--Krishnan dynamical horizons (DH), or more specialised ones, like the
 Booth-Fairhurst slowly evolving FOTH. Although the horizon as defined by Hayward will
 be central in this review, for sake of completeness we shall give an updated list of
 locally or quasi-locally defined horizons which appeared over the years, each playing
 some role in the problem of understanding dynamical black hole (more precise definitions will be given soon). At least four types of horizon have been defined over the years:
\begin{enumerate}

\item non expanding and perfect horizons (H\'aji\v{c}ek\cite{hajicek});
\item trapping horizons (Hayward \cite{hayw94-,hayw94,Hayward:1997jp});
\item dynamical horizons (Ashtekar and Krishnan \cite{AK,Ashtekar:2003hk});
\item isolated and weakly isolated horizons (Ashtekar \textit{et al.} \cite{Ashtekar:1998sp,isohor});
\item slowly evolving horizons (Booth and Fairhurst \cite{bf04}).
\end{enumerate}
Most of these newly defined
horizons have very desirable properties: they do not require a space-like hypersurface, no
 notion of interior and exterior and no conditions referring to infinity, like asymptotic
 flatness for example (all are non local conditions). Moreover, they are not endowed
 with teleological features (they do not anticipate the future, so to speak) and, given a
 solution of Einstein equations, one can find whether they exist or do not exist by purely local
 computations. Finally, unlike EH they are related to regions endowed by strong
 gravitational fields and are typically absent in weak field regions.  

We recall that the expansion $\theta$ of a bundle of null rays is the rate of change of area transverse to the bundle
\[
\frac{d A}{d v}=\int\theta\, d^2S
\]
where $v$ is a parameter along the rays. All quasi-local horizons rely on the local concept of {\it trapped or marginally trapped surface}: this is a space-like closed two-manifold $\bm{S}$ such that $\theta_{+}\theta_{-}\geq0$; if $\ell_{\pm}$ are the future-directed null normals to $\bm{S}$, normalised to
 $g_{\mu\nu}\ell_{+}^{\mu}\ell_{-}^{\nu}\equiv\ell_{+}\cdot\ell_{-}=-1$, then
 $\theta_{+}$, $\theta_{-}$ are the respective expansions or optical scalars, that is the
 expansions of the two bundles of null rays orthogonal to $\bm{S}$.  It is further
 assumed that $\ell_{+}$ is associated to an outgoing null geodesic beam, so that in a
 region of not too strong gravity $\theta_{+}>0$ and the beam is expanding, as for
 example within the exterior of a black hole at a safe distance. It follows then that $\ell_{-}$ is associated to an ingoing null geodesic beam, with $\theta_{-}<0$ and the beam
 contracting along the way. It is always possible to choose double null coordinates $x^{\pm}$ such that
 \beq\label{ex+-}
 \theta_{\pm}=\frac{2}{r}\,\partial_{\pm}r
 \eeq
 where $r$ is the areal radius, defined so that a metric sphere has area $A=4\pi r^{2}$. 
 To cover black holes rather than white holes it is further
 assumed that both expansions are negative (or non positive) on a trapped (marginally
 trapped) surface. If $\theta_{+}\theta_{-}<0$ the surface is untrapped and marginal if
 $\theta_{+}\theta_{-}=0$. In spherical symmetry with radial coordinate $r$ this means
 the co-vector $d r$ is temporal, spatial or null respectively. A further subdivision may
 be made: a trapped surface is {\it future} if $\theta_{\pm}<0$ and {\it past} if
 $\theta_{\pm}>0$. A marginal $\bm{S}$ with $\theta_{+}=0$ is {\it future} if
 $\theta_{-}<0$, {\it past} if $\theta_{-}>0$, {\it bifurcating} if $\theta_{-}=0$, {\it
 outer} if $\partial_{-}\theta_{+}<0$, {\it inner} if $\partial_{-}\theta_{+}>0$ and {\it
 degenerate} if $\partial_{-}\theta_{+}=0$
 \cite{hayw94,Hayward:1994kz,Hayward:1997jp,Hayward:1994bu}. Equivalently, the vector field $g^{r\mu}\partial_{\mu}$ is future causal (it means non-space-like) or past causal, respectively. For a better understanding of these terms we may perhaps note that 
 \[
 \partial_{r}=\partial_{+}r\partial_{+}+\partial_{-}r\partial_{-}=\frac{r}{2}(\theta_{+}\partial_{+}+\theta_{-}\partial_{-})\;.
 \]
 So, for example, on a future trapped surface with $\theta_{\pm}<0$ the radial increasing vector is past directed, confirming the intuition that one cannot escape the trapped surface without moving in the past.   
 
So much far for black holes, using the mental picture that we, the observers, are outside it at some radius larger than the radius of the trapped region. For white holes one encounters also marginally trapped surfaces with $\theta_{-}=0$ and $\theta_{+}>0$. In cosmology one may encounter again both possibilities, marginally trapped surfaces with either $\theta_{-}=0$ or $\theta_{+}=0$, but we defer their description to Section \ref{5:coho}.\par 
The abstract definitions will become much more simple and intuitive when referred to spherically symmetric space-times in Section \ref{4.2}, where only few of them will be actually used. Here we take the opportunity for few more definitions. The optical scalars can also be defined as follows: the induced metric on each $\bm{S}$ is
\begin{equation}\label{qab}
q_{\mu\nu}=g_{\mu\nu}+\ell_{+\mu}\ell_{-\nu}+\ell_{-\mu}\ell_{+\nu}
\end{equation}
for in fact $q_{\mu\nu}\ell_{\pm}^{\nu}=0$. Let
 $q^{\mu\nu}=g^{\mu\nu}+\ell_{+}^{\mu}\ell_{-}^{\nu}+\ell_{+}^{\nu}\ell_{-}^{\mu}$,
 not the inverse of $q_{\mu\nu}$. Then $q^{\mu}_{\;\,\nu}$ is the projection tensor to
 $T_*(\bm{S})$, the tangent  space to $\bm{S}$. Associated to the null vector fields
 $\ell_{\pm}$ are the  projected tensor fields
 $(\theta_{\pm})_{\mu\nu}=q_{\mu}^{\alpha}q_{\nu}^{\beta}\nabla_{\alpha}l_{\pm
 \beta}$ and  their decomposition into symmetric, anti-symmetric and trace part. They are
 tensors on $\bm{S}$ because $(\theta_{\pm})_{\mu\nu}\ell_{\pm}^{\nu}=
 (\theta_{\pm})_{\mu\nu}\ell_{\mp}^{\nu}=0$. The twists (anti-symmetric parts)
 vanish  since the geodesic beams  are normal to $\bm{S}$ (we have to assume that
 $\bm{S}$ is part of a continuous family of surfaces $\bm{S}_{t}$). The expansions  are then given by the traces
\begin{equation}\label{expan}
\theta_{+}=q_{\mu\nu}\nabla^{\mu}\ell_{+}^{\nu}, \qquad \theta_{-}=q_{\mu\nu}\nabla^{\mu}\ell_{-}^{\nu}\;.
\end{equation}
Finally, the shear is the trace-free symmetric part (note that
 $g^{\mu\nu}q_{\mu\nu}=q^{\mu\nu}q_{\mu\nu}=2$)
\begin{equation}
\sigma^{(\pm)}_{\mu\nu}=\theta_{\pm \mu\nu}-\frac{1}{2}q_{\mu\nu}\theta_{\pm}\, .
\end{equation}
As before, we indicate the Lie-derivative along a vector field $X$ by the calligraphic symbol ${\cal L}_{X}$ and ${\cal L}_{\pm}={\cal L}_{\ell_{\pm}}$. Thus, for example, 
${\cal
 L}_{-}\theta_{\mu\nu}=\ell_{-}^{\alpha}\partial_{\alpha}\theta_{\mu\nu}+\partial_{\mu}\ell_{-}^{\alpha}\theta_{\alpha\nu}+\partial_{\nu}\ell_{-}^{\alpha}\theta_{\mu\alpha}$,
 while on scalars it acts as an ordinary partial derivative. Let us describe the listed
 horizons in turn, adding comments where appropriate. A black  triangle down
 $\blacktriangledown$ will close the definitions. 

{\em Perfect and non-expanding horizons ---} A perfect horizon is a smooth 
three-dimensional \emph{null} sub-manifold $H$ of space-time with null normal
 $\ell^{\mu}$ such that its expansion $\theta_{\ell}=0$ on $H$ and which intersect
 space-like hypersurfaces in compact sets. $\blacktriangledown$ 

If in the last clause $H$ is topologically $\mathbb R\times \mathbb S^2$ and moreover the stress tensor
 $T_{\mu\nu}$ is such that $-T^{\mu}_{\nu}\ell^{\nu}$ is future causal for any future
 directed null normal $\ell^{\mu}$, then $H$ is called a non-expanding horizon
.$\blacktriangledown$ 

All stationary horizons are perfect, but the converse is not true. These horizons are not very relevant in the present context since they apply only to equilibrium black holes.
 
{\em Future outer trapping horizons ---} A future outer
trapping horizon (FOTH) is a smooth three-dimensional sub-manifold $H$
of space-time which is foliated by closed space-like two-manifolds ${\bf S}_t$, $t\in R$,
 with future-directed null normals $\ell_{\pm}$ such that: (i) the expansion $\theta_{+}$
 of the null normal $\ell_{+}$ vanishes; (ii) the expansion $\theta_{-}$ of $\ell_{-}$ is
 negative; (iii)  ${\cal L}_{-}\theta_{+} < 0$. A future inner trapping horizon (FITH) is as above except that on it ${\cal L}_{+}\theta_{-}>0$.  $\blacktriangledown$

This proved to be the most important definition since forms the basis for almost all other
 definitions.  Condition (i) requires strong fields since certainly $\theta_{+}>0$ in weak
 fields. The condition (ii) is related to the idea that $H$ is of the future type (e.~g. a black hole
 rather than a white hole); (iii) says that $H$ is of
the outer type\footnote{For example the Cauchy horizon in the Reissner--Nordstr\"{o}m
 solution is of inner type.}, since  a motion of ${\bf S}_t$ along $\ell_{-}$ makes it
 trapped.  It also distinguishes black hole horizons from cosmological ones, which for an expanding universe are FITH.  Note that no
 reference is made to any space-like hypersurface, nor to infinity. Trapping horizons are
 locally defined and have physical properties such as mass, angular momentum and
 surface gravity, satisfying conservation laws \cite{AK,bhd}. They are a geometrically
 natural generalisation of Killing horizons, which are stationary trapping horizons. A non-stationary trapping horizon is not null, but still has infinite red-shift. Unlike event and
 apparent horizon they do not require asymptotic flatness.

One can always find a scalar field $C$ on $H$ so that the vector fields
\begin{equation}\label{v}
   V^{\mu} = \ell_{+}^{\mu} - C \ell_{-}^{\mu} \quad \mbox{and} \quad
   N^{\mu} = \ell_{+}^{\mu} + C \ell_{-}^{\mu} \, , 
\end{equation}
are respectively tangent and normal to the trapping horizon. Note that $V 
\cdot V = - N\cdot N = 2C$. Hayward \cite{hayw94,Hayward:1997jp} showed that if the
 {\it null energy condition} (abbr.~NEC) holds then $C \ge 0$ on a FOTH.  Thus, the
 horizon must be either space-like or null, and it is null if and only if the shear
 $\sigma^{(+)}_{\mu\nu}$ and $T_{\mu\nu}\ell_{+}^{\mu}\ell_{+}^{\nu}$ both
 vanish across $H$. Intuitively, $H$ is space-like in the dynamical regime where
 gravitational radiation and matter are pouring into it and is null when it reaches
 equilibrium. Conversely, on a FITH $V^{\mu}$ is either null or time-like. 

It is worth mentioning that {\it the second law of the mechanics of trapping horizons}
 follows quite easily from this apparatus. Taking the Lie-derivative of $\sqrt{q}$, which is
 the area density corresponding to the metric $q_{\mu\nu}$ on the cross-sections of $H$,
 we get 
\begin{equation}\label{metdetder}
{\cal L}_{V} \sqrt{q} = - C \sqrt{q}\, \theta_{-}\;.
\end{equation}

By definition $\theta_{-}$ is negative on $H$ and we have just seen that, barring
 violations of the null energy condition, $C$ also is non-negative. Since $V$ is future
 directed we obtain the local form of the second law: {\it If the null energy condition holds,
 then the area element $\sqrt{q}$ on a FOTH is non-decreasing along future directions}.\\ 
Integrating over $\bm{S}_t$ the same law applies to the total area of the trapped
 sections. As long as the null energy condition is maintained it will be non-decreasing,
 reaching a constant value if and only if the horizon becomes a null hypersurface. 

The main difference between an apparent horizon as defined in Hawking--Ellis and the trapping
 horizon of Hayward is that the AH represents the instantaneous surface of a black hole,
 {\it i.e.} it needs a (partial) Cauchy surface $\Sigma$ and it is very sensitive to the choice of
 $\Sigma$. To compute the AH one needs only its metric and the second fundamental
 form of $\Sigma$, namely the initial data for Einstein's equations. Hayward's horizon
 instead is a null hypersurface $H$ which is insensitive to a choice of $\Sigma$ and
 does not refer to spatial infinity. The trapping horizon is a foliation of $H$.

The causal character as well as the area law required the validity of the NEC; an
 evaporating black hole violates NEC, therefore the area law will also be violated and the
 horizon will be time-like. How it could be that tunnelling along a classically forbidden
 path is still possible will be seen soon and represents the real possibility of the radiation
 process even in the temporary absence of a global event horizon.

Next come Ashtekhar and co-workers: they observe that key results, such as the area
 increase, do not depend on the sign of ${\cal L}_{-}\theta_{+}$. Hence the following
 weaker notion was introduced:

{\em Dynamical horizons ---} A smooth three-dimensional, {\it space-like} sub-manifold
 $H$ of space-time is a dynamical horizon (DH) if it can be foliated by closed space-like
 two-manifolds $\bm{S}_t$, with future-directed null normals $\ell_{\pm}$ such that: (i)
 on each leaf the expansion $\theta_{+}$ of one null normal $\ell_{+}$ vanishes; (ii) the
 expansion $\theta_{-}$ of the other
null normal $\ell_{-}$ is negative.  A submanifold $H$ such that $\theta_{+}=0$ and
 $\theta_{-}<0$ is also called a {\it marginally trapped tube} (abbr. MTT)
 $\blacktriangledown$

Like FOTHs, a DH is a space-time notion defined quasi-locally, it is
not related to a space-like hypersurface, it does not refer to infinity, it is not teleological.
 A space-like FOTH is a DH on which ${\cal L}_-\theta_{+}< 0$; a DH, which is also a
 FOTH, will be called a {\it space-like future outer horizon} (SFOTH). The precise
 properties of such horizons are fully discussed in \cite{AK,Ashtekar:2003hk}. Suitable
 analogues of the laws of black hole mechanics hold for both FOTHs and DHs. We only
 note that DHs cannot describe equilibrium black holes since they are space-like by
 definition. Likewise, DHs cannot describe evaporating black holes since for them the
 trapping horizon is time-like, but are better suited to describe how a black hole grows in general
 relativity. In general, if a MTT is space-like it is a DH. If a MTT is time-like it is a time-like
 membrane. It is of interest that in the Oppenheimer--Volkoff dust collapse the unique MTT
 on which each {\it marginally trapped surface} (MTS) is spherical is time-like
 \cite{BenDov:2004gh}! This shows that even in spherical symmetry the existence of DH
 with spherical sections is far from obvious. However, for perfect fluid collapse and
 spherical scalar field collapse the MTTs are space-like.

If $X$, $Y$ are tangent to a non-expanding horizon we can decompose the covariant
 derivative 
\[
\nabla_XY=D_XY+N(X,Y)\ell_{+}+L(X,Y)\ell_{-}
\]
where $D_X$ is the projection of the vector $\nabla_XY$ onto the
spheres $\bm{S}_t$ in $H$. If $X$ is tangent to the spheres then $D_X$
is the covariant derivative of the induced metric $q_{\mu\nu}$, and if $X$
is tangent to $H$ one may regard the operator 
$\widehat{\nabla}_X=D_X+N(X,\cdot)\ell_{+}$, acting on vector fields, as  a connection
 on $H$. If this connection is ``time independent'' then the geometry of $H$ is time
 independent too and we have Ashtekar \textit{et al.} notion of a horizon in isolation.

{\em Isolated horizons ---} A non-expanding horizon with null normal $\ell^\mu$ such that
 $[{\cal L}_{\ell},\widehat{\nabla}_X]=0$ along $H$.$\blacktriangledown$ 

These horizons were intended to model black holes that are themselves in equilibrium but
 possibly in a dynamical space-time. For a detailed description of their mathematical
 properties we refer the readers to Ashtekar--Krishnan's review \cite{Ashtekar:2004cn}. 

{\em Slowly evolving horizons ---} No matter or radiation can cross an isolated horizon, so
 the meaning of the first law for them cannot be treated in full generality. Booth and
 Fairhurst \cite{bf04} established this law for slowly evolving FOTHs as defined by
 Hayward by introducing dynamical notions of surface gravity and angular momentum.
 For this purpose, the concept of a slowly evolving FOTH was then defined.

All the horizons just introduced have their own dynamics governed by Einstein equations.
 There are for them existence and uniqueness theorems \cite{Andersson:2005gq},
 formulation of the first and the second laws
 \cite{Ashtekar:2004cn,hayw94,Hayward:2004fz} and even a ``membrane paradigm''
 analogy. In particular, they carry a momentum density which obey a Navier--Stokes-like
 equation generalising  the classical Damour's equations of EHs, except that the bulk
 viscosity $\zeta_{FOTH}=1/16\pi > 0$
 \cite{Gourgoulhon:2005ch,Gourgoulhon:2006uc}.  The newly introduced  horizons are
 also mostly  space-like or null, therefore the role they may play in the problem of black
 hole quantum evaporation is unclear. In this connection the following notion can be
 useful.

{\em Time-like dynamical horizon ---} A smooth three-dimensional, \emph{time-like} 
submanifold $H$ of space-time is a time-like dynamical horizon (TDH) if it can be
 foliated by closed space-like two-manifolds $\bm{S}_t$, with future-directed null
 normals $\ell_{\pm}$ such that: (i) on each leaf the expansion $\theta_{+}$ of one null
 normal $\ell_{+}$ vanishes, (ii) the expansion $\theta_{-}$ of the other null normal is
 strictly negative.$\blacktriangledown$\par
It must be recalled that a FOTH for which the NEC does not  hold can be time-like as well.
 What about non spherical MTS? The Vaidya metric does not admit other, non spherical
 horizons which also asymptote to the non expanding one. This is general, so a DH is
 unique, that is a space-like $3$-manifold cannot be foliated by two distinct families of
 MTS each endowed with a structure of a DH. This should also hold for a TH but we do
 not know whether there is a proof. According to Ashtekhar \textit{et al.} a better control of
 uniqueness  is the most important open issue of the classical theory of dynamical
 horizons, be trapped or in isolation.
\paragraph{Surface gravity} The surface gravity associated to an event horizon is a well known concept in black hole physics whose importance can be hardly overestimated.
 Surprisingly, a number of inequivalent definitions beyond the standard one appeared
 over the last 15 years or so in the field, with various underlying motivations. We
 have collected the following (we rely on the review of Nielsen and Yoon
 \cite{Nielsen:2007ac}):
\begin{enumerate}
\item the standard Killing surface gravity (Bardeen \textit{et al.} \cite{Bardeen: 1973gs},
 textbooks);
\item a first definition given by Hayward in \cite{hayw94};
\item the effective surface gravity appearing in Ashtekar--Krishnan
 \cite{Ashtekar:2004cn};
\item the Fodor \textit{et al.} definition for dynamical spherically symmetric space-times \cite{Fodor:1996rf};
\item the Visser \cite{Visser:2001kq} and Nielsen--Visser \cite{NV} surface  gravity;
\item still one more definition by Hayward \cite{Hayward:1997jp}, using Kodama theory
 of spherically symmetric space-times \cite{Kodama:1979vn}.

\end{enumerate}
In addition there are some more technical definitions due to Mukohyama and Hayward
 \cite{Mukohyama:1999sp} and to Booth and Fairhurst, the latter related to their notion of
 evolving horizons \cite{bf04}. Except for the last item, which is what the tunnelling
 method leads to, the remaining definitions will be less relevant in this review, so we reserve to them only some brief considerations.  
   
The Killing surface gravity is related to the fact that the integral curves of a Killing vector
 are not affinely parametrized geodesics on the Killing horizon $H$, where the norm
 $\xi^2=0$. Hence
\[
\xi^\mu\,\nabla_\mu \,\xi_\nu \cong \,\kappa\, \xi_\nu 
\]
defines the Killing surface gravity $\kappa$ on $H$, where $\cong$
means evaluation on the horizon. The Killing field is supposed to be
normalised at infinity by $\xi^2=-1$. The definition can be extended to
EHs that are not Killing horizons, by replacing $\xi$ with the null
generator of the horizon. However there is no preferred
normalisation in this case, and this is one reason of the debating
question regarding the value of the surface gravity in dynamical
situations.

Hayward's first definition was motivated by the desire to get a proof 
of the first law for THs. It is defined for a future trapping horizon without appeal to un-affinity of null geodesics,  as 
\begin{equation}\label{hayk1}
\kappa\cong\frac{1}{2}\sqrt{-n^\mu\nabla_\mu\theta_{\ell}}\, .
\end{equation}
This quantity is independent on the parametrisation of the integral curves of the vector field $\ell^\mu$, since the evaluation is on a marginal outer surface where $n\cdot\ell=-1$ and $\theta_{\ell}=0$.

Given a weakly isolated horizon $H$, Ashtekar and Krishnan showed that for any vector field $t^a$ along $H$ with respect to which energy fluxes across $H$ are defined, there is an area balance law that takes the form
\[
\delta E^t=\frac{\bar{\kappa}}{8\pi G}\delta A_{ S}+\mathrm{work\;terms}
\]
with an effective surface gravity given by
\[
\bar{\kappa}=\frac{1}{2R}\frac{d r}{d R}\, .
\]
$R$ is the areal radius of the marginally trapped surfaces, {\it i.e.} $A_{ S}=4\pi R^2$,
 the function $r$ is related to a choice of a lapse function and finally $E^t$ is the energy
 associated with the evolution vector field $t^a$. For a spherically symmetric DH a natural
 choice would be $r=R$ so $\bar{\kappa}=1/2R$, just the result for a Schwarzschild black hole.
 To illustrate the naturalness of this definition,  consider a slowly changing spherically
 symmetric black hole with mass $M(v)$, where $v$ is a time coordinate. Defining the horizon
 radius at each time by $R=2M(v)$ and  $A_{ S}=4\pi R^2$, we can differentiate $M$ 
\[
\dot{M}=\frac{\dot{R}}{2}=\frac{1}{2R}\frac{\dot{A}_{ S}}{8\pi}\, .
\]
so as to obtain $\delta M=\bar{\kappa}\delta A_{ S}/8\pi$. One recognises the usual area law in differential form with surface gravity
$\bar{\kappa}=1/2R=1/4M$. Consider, however, the more general
possibility where the horizon is the solution of the implicit equation $R=2M(v,R)$, as it
 happens for example in the Bardeen--Vaidya metric. The same computation leads to $\dot{M}=\dot A_{ S}\,(1-2M' )/(16\pi  R)$, leading to
\begin{equation}\label{comp}
\;\;
\kappa\cong \frac{1}{4M}\left(1-2M'\right)
\end{equation}
a prime denoting the radial derivative. The surface gravity here deduced does not 
conform to Ashtekar \textit{et al.} definitions, suggesting that its value depends on the definition of the black hole mass one is adopting.

The definition of Fodor \textit{et al.} looks like the Killing form of the surface gravity in that
 $\kappa\ell^\nu=\ell^\mu\nabla_\mu\ell^\nu$, where now $\ell^\mu$ is an outgoing null vector
 orthogonal to a trapped or marginally trapped surface. This is because, as a rule, such
 null vectors are not affinely parametrized, although they can always be parametrized so
 that $\kappa=0$. Fodor \textit{et al.} choose to fix the parametrization so that
\[
\kappa=-n^\mu\ell^\nu\nabla_\nu\ell_\mu
\]
with $n^\mu$ affinely parametrized and normalised to $n\cdot t=-1$ at space-like infinity, where $t^a$ is the asymptotic Killing field. Note that this definition is non local but looks like a natural generalisation of the Killing surface gravity. 

As far as we know, the Visser and Visser--Nielsen surface gravity is only defined in
 dynamical Painlev\'e--Gullstrand coordinates, so there is no guarantee that it is a
 geometrical invariant. Indeed, we shall see that for this class of metrics it is different from
 the invariant surface gravity computed within the general Hayward formalism. 

The last item is a local geometrical definition of the surface gravity for the trapping
 horizon of a spherically symmetric black hole \cite{Hayward:1997jp}, so we postpone a full discussion to Section \ref{4.2.4}. Basically, one introduces local null coordinates $x^{\pm}$ in a tubular neighbourhood of a FOTH and for any sphere of radius $r$ one defines the quantity
\begin{equation}\label{kappa}
\kappa = \frac{r}{2}\left(g^{+ -}\partial_- \theta_+ +\frac{1}{2}g^{+ -}\theta_{+}\theta_{-}\right)\, .
\end{equation} 
Evaluated on a trapping horizon, $\theta_{+}=0$, it will be positive precisely when the
 horizon is of outer type ($\partial_{-}\theta_{+}<0$, recall that with our conventions $g^{+-}<0$) and $\kappa=0$ if degenerate.
The definition may look somewhat artificial, but in fact it can be put in a form that strongly resembles the Killing surface gravity of stationary black holes. To see this,
 we anticipate a result of the next section according to which, following Kodama
 \cite{Kodama:1979vn}, any spherically symmetric metric admits a unique (up to
 normalisation) vector field $K^\mu$ such that $\nabla^\nu(K^\mu \,G_{\mu\nu})=0$, where
 $G_{\mu\nu}$ is the Einstein tensor; for instance, using the double-null form, one finds
\begin{equation}\label{kod}
K=-g^{+-}(\partial_+r\partial_--\partial_-r\partial_+)\, .
\end{equation} 
The defining property of $K$ shows that it represents a natural generalisation of the time translation Killing field of a static black hole. Now consider the expression $K_{\mu}\nabla_{[\nu}K_{\mu]}$: it is not hard to see that on H it is proportional to $K_{\nu}$. So one defines  the dynamical surface gravity as $K^\mu \nabla_{[\nu}K_{\mu]}\cong -\kappa K_\nu$. For a Killing vector field $\nabla_{\nu}K_{\mu}$ is anti-symmetric so the definition reduces to the usual one.

\subsection{Spherically symmetric fields}
\label{4.2}

In this section we discuss the general time-dependent spherically symmetric metric
 that will be the arena of our dynamical tunnelling computations. We have a twofold intent
 in doing this. If Hawking radiation proceeds by emission of discrete energy quanta, a
 continuous description of the changing metric would only be possible when their number
 is so large so as to simulate a continuous streaming of energy. In such a case,  a
 dynamical excreting black hole may be modelled by a continuously differentiable solution
 of Einstein's equations for most of its history. Given that, it would be quite ironic if there
 were not evidence of a continuously operating quantum emission process. In other
 words, if the tunnelling method only worked for the event horizons of stationary black holes
 then there would be a problem with the use of such dynamical  metrics as models of
black hole evaporation. 

The second aspect has to do with the region where the radiation originates. The
 tunnelling calculation suggests the the outgoing radiation is emitted from the trapping
 horizon, not the global event horizon. And it further suggests that the semi-classical
 probability is related to the horizon surface gravity as defined in eq.(\ref{kappa}
\cite{Hayward:1997jp}. Thus we have first of all to learn how to compute these things in a spherically symmetric, time dependent metric.

\subsubsection{Forms of the metric}
\label{tu:fm}

The paradigm of a spherically symmetric metric is of course the Schwarzschild vacuum solution of Equation (\ref{schwarzschild}).
To introduce dynamics one could think to make the replacement $M\to M(t)$; however
 the resulting metric has a curvature singularity at $r=2M(t)$; for instance, the scalar
 curvature is
\[
\mathcal R=\frac{2r(r-2M)\ddot{M}+4\dot{M}^2}{(r-2M)^3}
\]   
so only for special values of the mass function is $\mathcal R=0$; other invariants will diverge though. A more general ``Schwarzschild gauge'' can be written down
\beq\label{scga}
 d s^2=-e^{2\Phi(r,t)}\left(1-\frac{2M(r,t)}{r}\right)d t^2+\left(1-\frac{2M(r,t)}{r}\right)^{-1}d r^2+r^2(d\theta^2 + \sin^2\theta d\phi)
\eeq
which was studied in depth by Visser in his classic ``Dirty Black Holes'' \cite{visser}. We shall not make use of his metric in the following.

It was discovered by Vaidya \cite{Vaidya:1951zz} that  by taking the Schwarzschild metric in advanced  Eddington--Finkelstein coordinates $(v,r,\theta,\phi)$ (or retarded, $(u,r,\theta,\phi)$ for other purposes)
\beq\label{vaid}
d s^2=- \left(1-\frac{2M}{r}\right)d v^2+2d vd r +r^2(d\theta^2 + \sin^2\theta d\phi)
\eeq
and by making the replacement $M\to M(v)$, one obtains a regular solution of Einstein's
 equations except at the origin $r=0$, the shining star solution of Vaidya. The metric is
 still not the most general one allowed by the symmetry: in fact a spherically symmetric
 metric can depend at most on two arbitrary functions of the coordinates $(v,r)$.
 One form is due to Bardeen  \cite{bardeen:1981}
\beq\label{bard}
d s^2=- e^{2\Phi}\left(1-\frac{2M}{r}\right)d v^2+2e^{\Phi}d vd r +r^2(d\theta^2 + \sin^2\theta d\phi)
\eeq
where $\Phi$ and $M$ are functions of $v$, $r$. Another important form is obtained by passing from Schwarzschild time to the proper time of a radially infalling observer
\beq
t_{ p}=t+2\sqrt{2Mr}+4M\log\left(\frac{\sqrt{r}-\sqrt{2M}}{\sqrt{r}+\sqrt{2M}}\right)\, .
\eeq
The metric is the Painlev\'{e}--Gullstrand metric we encountered in Section \ref{2:stat}
\beq\label{pain}
d s^2=-\left(1-\frac{2M}{r}\right)d t_{ p}^2+2\sqrt{\frac{2M}{r}}\,d t_{ p}d r +d r^2+r^2\,(d\theta^2 + \sin^2\theta d\phi)\, .
\eeq
Taking now $M=M(t_{ p})$ we get a dynamical black hole long studied by Lindesay and Brown \cite{Brown:2009by,Lindesay:2008ne,Lindesay:2010uv} and others \cite{Clifton:2008sb}. There is a more general form whereby $M$ can also depend on $r$ and a second function $\Phi(\tilde t,r)$ is introduced, say
\beq
\label{vissnil}
 d s^2=-e^{2\Phi(\tilde t,r)}\left(1-\frac{2M(\tilde t,r)}{r}\right)d\tilde t^2+ 2 e^{\Phi(\tilde t,r)}\sqrt{\frac{2M(\tilde t,r)}{r}}\,d\tilde t\,d r +d r^2+r^2d\Omega^2\,,
\eeq
where $d \Omega^2 :=d\theta^2 + \sin^2\theta d\phi$. This metric is locally diffeomorphic to the metric eq.(\ref{scga}) under a change of time $t\to\tilde t(t,r)$, with the mass treated as a scalar field, {\it i.e.} $M(\tilde t,r)=M(t(\tilde t,r),r)$. It was thoroughly studied by Visser and Nielsen in \cite{NV}. In this rather impressive work the dynamics and decay of evolving horizons were investigated mainly using a dynamical version of Painlev\'e--Gullstrand coordinates.  The metric eq.(\ref{vissnil}) is also locally diffeomorphic to the metric eq.(\ref{bard}), under a change $v\to \tilde t(v,r)$, where now $M(\tilde t,r)=M(v(\tilde t,r),r)$.

All these spherically symmetric metrics (though not only these) are special cases of metrics that can locally  be expressed in the warped  form 
\beq\label{metric} 
d s^2 =\upgamma_{ij}(x^i)d x^id x^j+ R^2(x^i) d\Omega^2\,,\qquad i,j \in \{0,1\}\;, 
\eeq 
where the two-dimensional metric 
\beq 
d\upgamma^2=\upgamma_{ij}(x^i)d x^id x^j 
\label{nor} 
\eeq 
is referred to as the normal metric (in the space normal to the sphere of symmetry),
 $x^i$ are associated coordinates and $R(x^i)$ is the areal radius, considered as a scalar
 field in the normal two-dimensional space. The two-dimensional metric $\upgamma_{ij}$
 has only one degree of freedom and the function $R(x^i)$ is the second one; however the
 two functions on which the metric depends can be moved both to $d\upgamma^2$
 using the coordinates freedom. That is we can rename $x^1=r$ and put $R(x^i)=r$.
 The coordinate $r$ is known as the areal radius, and is a geometrical invariant being the
 quotient  of the area of space-like spheres by $4\pi$. This will be referred to as the 
{\it $r$-gauge}. Sometimes the normal metric can be further written, somewhat redundantly, in the expanded form
\beq\label{efg}
d\upgamma^2=-E(r,t)d t^2+2F(r,t)d td r+G(r,t)d r^2, \quad \upgamma=-(EG+F^2)
\eeq
especially if one wants to discuss certain limits, but we will always try to get metrics
 regular across the trapping horizons. If we wish, we can also locally write the line element
 in a double-null form which is conformally flat
\beq\label{dn}
d s^2=-2e^{f}d x^+d x^-+r^2(x^+,x^-)d\Omega^2
\eeq
with $f=f(x^+,x^-)$. For instance, starting with eq.(\ref{bard} we can introduce null coordinates via
\[
e^fd x^-= \frac{1}{2}\left(1-\frac{2m(r,v)}{r}\right)e^{2\Phi}d v-e^{\Phi}d r, \quad d x^+=dv
\]
where $f$ is an integrating factor.  The normal section of eq.(\ref{bard}) then takes the form eq.(\ref{dn}). One may use one spatial and one temporal direction on putting $d x^+d x^-=d t^2-d r^2$, but there is no unique choice of such directions. The remaining coordinate freedom consists of conformal diffeomorphisms (in the language of two-dimensional metrics)
\[
x^{\pm}\to\tilde{x}^{\pm}(x^{\pm})\, .
\]
The double-null form will be also referred to as the {\it conformal gauge}. 

Another form we would like to describe is the metric in the so called {\it synchronous gauge}. 
Let us consider the  Schwarzschild space-time in coordinates $(T, r, \theta, \phi)$ such that the line element can be expressed as 
\begin{equation}
 d s^2 =-d T^2 + \frac{d r^2}{B} + (r_{ g} B)^2 d\Omega^2\;, \label{lemaitre}
\end{equation}
where $r_{ g} = 2M$ is the usual gravitational radius, and
\begin{equation}
 B(T,r) := \left[\frac{3}{2r_{ g}}(r-T)\right]^{\frac{2}{3}}\, .
\end{equation}
We shall refer to these coordinates as the Lema\^{\i}tre--Rylov gauge. This is indeed an
 interesting (time-dependent) gauge since, contrary for example to isotropic coordinates,
 $(T,r)$ extend beyond the gravitational radius, $r < r_{ g}$.  Now considering
 $B(T,r)$ as an arbitrary function and replacing $r_{ g} B$ with a general function
 $R(r,T)$ we obtain the metric in synchronous gauge
\beq\label{sin}
d s^2 =-d T^2 + \frac{d r^2}{B(T,r)} + R^2(T,r) d\Omega^2
\end{equation}
in which the metric is diagonal but the areal radius is a function of $r$ and $T$. 

The last form we would like to mention is the {\it isotropic gauge} in which the spatial part of the metric is conformally flat
\beq\label{iso}
d s^2=-A(t,\rho)d t^2+B(\rho,t)(d\rho^2+\rho^2d\Omega^2)\, .
\eeq
An important example of this latter form is the McVittie solution describing in author's
 mind a point mass in a FRW flat cosmology \cite{mcv}. It reads as in eq.(\ref{iso} with
\bea\label{mcv}
A(\rho,t) &= &\left(1+\frac{M}{2a(t)\rho}\right)^{-2}\left(1-\frac{M}{2a(t)\rho}\right)^2 \\
B(\rho,t) &=& a(t)^2\left(1+\frac{M}{2a(t)\rho}\right)^4\, .
\eea
When $M=0$ it reduces to a spatially flat FRW metric with scale factor
 $a(t)$; when $a(t)=1$ it reduces to the Schwarzschild metric with mass $M$ in isotropic
 coordinates.  This solution had a strong impact on the general problem of matching
 the Schwarzschild solution with cosmology, a problem faced also by Einstein and Dirac. 
 Besides McVittie, it has been extensively studied by Nolan in a series of papers
 \cite{nolan}. To put the metric in the general $r$-gauge form, the coordinate transformation
\[
r=a(t)\rho\left(1+\frac{M}{2a(t)\rho}\right)^{2}
\]
transforms the metric in the so called Nolan gauge, in which it reads 
\beq\label{nolan}
d s^2=-\left( A_{ s}-H^2(t)r^2 \right) d t^2 - \frac{H(t)r}{2 \sqrt{A_{ s}}}\,d r\,d t+ \frac{d r^2}{A_{ s}} +r^2d\Omega^2
\eeq
where $H(t)=\dot{a(t)}/a(t)$ is the Hubble parameter and, for example, in the
charged four-dimensional case, $A_{ s}=1-2M/r+e^2/r^2$. Unlike the Schwarzschild case, $A_{ s}=0$, or $r=2M$ in the neutral case or $\rho=M/2a(t)$, is a curvature singularity rather than a global event horizon. In fact, it represents a Big Bang singularity very similar to $r=0$ in FRW models. When $H=0$ one recovers the Schwarzschild solution.  For constant $H$, it reduces to the Schwarzschild--de Sitter solution in Painlev\'e coordinates. As we shall see, the McVittie solution possesses in general black hole and cosmological trapping horizons, and the space-time is dynamical.
 
Other coordinate systems which can be used to include dynamics in the Schwarzschild metric are discussed in 
\cite{Francis:2003rj}. We may note that eq.(\ref{sin} contains as a special case the
 FRW family of metrics. We stress that all  forms discussed
 above are completely equivalent ways to describe the space-time structure of a
 spherically symmetric field, the use of any particular form being dictated only by
 computational convenience. We also note that in the warped form eq.(\ref{metric})
 the sphere of symmetry can be replaced by anyone among the two-dimensional manifolds with
 constant curvature which appear in the static topological black hole solutions
 embedded in Anti--de Sitter space. Only, the rotational isometry group must be replaced
 by the appropriate isometry group of the surface, which in the static case are torii or compact Riemann surfaces with higher genus.  

\subsubsection{Trapping horizon}
\label{4.2.2th}

To compute the trapping horizons we shall start from the Bardeen--Vaidya form eq.(\ref{bard}). We can take the two null, future directed congruences normal to spheres of constant radius as
\bea\label{null}
\ell_{+}=\left[1,\frac{1}{2}e^{\psi}\left(1-\frac{2M}{r}\right),0,0\right], \quad \ell_{-}=[0,-e^{-\psi},0,0]
\eea
where as always $\ell_{+}\cdot\ell_{-}=-1$.  A simple computation gives the optical scalars
\bea\label{exp1}
\theta_{+}=\frac{1}{r}\left(1-\frac{2M}{r}\right)e^{\psi},\quad 
\theta_{-}=-\frac{2e^{-\psi}}{r}\, .
\eea
Trapped or marginally trapped spheres have $\theta_{+}\theta_{-}\geq 0$, therefore
 there are no such surfaces in the region $r>2M(r,v)$.  We also see that $\theta_{-}<0$
 along the surface defined by $\theta_{+}=0$,  which is therefore a trapping horizon of
 the future type. Its defining equation is $r=2M(r,v)$, which defines a line $r=r(v)$ in
 normal space and therefore a hyper-surface with topology $\mathbb R^1\times \mathbb S^2$ in 
space-time. 

Consider now $\partial_{-}\theta_{+}$, where the symbol $\partial_{-}$ denotes the directional derivative along $\ell_{-}$. One obtains, using $M'=\partial_{r}M$,
\[
\partial_{-}\theta_{+}=-\frac{1}{r^2}\left(\frac{2M}{r}-2M'\right)\, .
\]
On the trapping horizon $\partial_{-}\theta_{+}\cong-(1-2M')/4M^2$ will be negative if and only if $M'<1/2$. As anticipated in Section \ref{4.2}, this is the condition which ensures the positivity of the surface gravity. Thus in this case $r=2M$ is a FOTH. The areal radius of the horizon will be denoted by $r_{ H}$ from now on.
The signs of $\theta_{\pm}$ are geometrical invariants, but their actual values are not because the null directions are defined up to an overall scale. An invariant combination is $\chi=2g^{\pm}\theta_{+}\theta_{-}$ or, using eq.(\ref{metric}),
\beq\label{comb}
\chi=\upgamma^{ij}\,\partial_{i}R\,\partial_{j}R\equiv g^{\mu\nu}\,\partial_{\mu}R\,\partial_{\nu}R\, .
\eeq
$\chi=0$ is then the condition for a trapping horizon in Hayward's sense. For example in any coordinate system where one coordinate is the areal radius $R=r$, the condition is $g^{rr}=0$. The trapping horizon of McVittie solution can also be computed using eq.(\ref{comb}): using Nolan form it is a solution of the equation $\sqrt{A_{ s}}=H r_{ H} \,,$ which in turn implies $1-2M/r_{ H}=H(t)^2r^2_{ H}$. As $M$ is a constant this a cubic algebraic equation with a priori more than one real root and in any case at most two positive roots. The situation is similar to de Sitter space, except that the horizon radius is here a function of time, $r_{ H}(t)$. For positive $H$ both horizons are of the future type, but as a rule one is outer (meaning that there is a black hole) while the larger root corresponds to an inner horizon, hence to a FITH, if and only if
\beq
\frac{M}{r_{ H}^2}-H^2r_{ H}-\frac{\dot{H}}{2H} > 0\, .
\eeq
For $M=0$, {\it i.e.} for homogeneous cosmology, only the inner cosmological horizon survives. This is the case, for example, in de Sitter space-time. We will make use of these results in Section \ref{5:coho}.

\subsubsection{Misner--Sharp--Kodama energy}
\label{4.2.3ms}

One special feature of spherically symmetric space-times is the absence of gravitational radiation. This feature makes it possible the existence of a special, privileged  notion of energy, the Misner--Sharp mass, which for spheres with areal radius $r$ is the same as the Hawking mass \cite{hawk} (a general reference for energy in GR is the review \cite{Szabados:2004}). The energy may be defined by \cite{Hayward:1994bu} (we recall that $r$ is the areal radius)
\beq\label{msmass}
E=\frac{r}{2}-rg^{+-}\partial_{+}r\partial_{-}r=\frac{r}{2}-\frac{r^3}{4}\,g^{+-}\theta_{+}\theta_{-}
\eeq
and interpreted as the energy inside a sphere of radius $r$; so, by definition, a metric sphere is trapped if and only if $E>r/2$, marginal if and only if $E=r/2$ and untrapped if and only if $E<r/2$. Note that eq.(\ref{msmass}) is a special case of the Hawking mass
\beq
E(S)=\sqrt{\frac{{ Area}(S)}{16\pi}}\left(1-\frac{1}{8\pi}\oint g^{-+}\theta_{+}\theta_{-}d^2S\right)
\eeq
and is a geometrical invariant. Two very important properties of $E$ were proved by Hayward \cite{Hayward:1994bu}: {\it in an asymptotically flat space-time, $E$ coincides with the Bondi-Sachs scalar energy at null infinity, and with the Arnowitt--Deser--Misner (ADM) mass at spatial infinity.} Using eq.(\ref{bard}) the energy takes the implicit form
\beq
g^{\mu\nu}\partial_{\mu}r\partial_{\nu}r=g^{rr}=1-\frac{2E}{r}
\eeq
which shows the relation with the Schwarzschild mass. Then $E=M$ and the  FOTH is at $r=2E$. Many important properties of $E$ can be displayed using also the Einstein gravitational field equations. We report two of them which seem remarkable, leaving the interested reader to the literature. Using double-null coordinates the variation of $E$ as determined by Einstein's equations is
\beq\label{ee}
\partial_{\pm}E=2\pi e^{-f}r^3(T_{+-}\theta_{\pm}-T_{\pm\pm}\theta_{\mp})
\eeq
where we remember that $e^{-f}=-g^{+-}$. These field equations can also be written in the Bardeen--Vaidya form, using eq.(\ref{bard}, where it is seen that $E=M$ and
\beq\label{baeq}
\partial_{v}M=4\pi r^2T^r_{\;v}\;, \quad \partial_{r}M=-4\pi r^2T^v_{\;v}\; .
\eeq
Thus in vacuo $E$ is a constant. It can then be shown that the solution is locally isometric to a Schwarzschild solution with energy $E$. This is an improvement of Birkhoff's theorem.

The second result is essentially the area law: {\it if the NEC holds on a FOTH then $E=r_{ H}/2$ is non-decreasing along the horizon.}

The question arises naturally whether $E$ is the charge associated to a conserved current. It was discovered by Kodama \cite{Kodama:1979vn} that in spherical symmetry there is a vector field $K^{\mu}$ such that $\nabla^{\mu}(G_{\mu\nu}K^{\nu})=0$; by Einstein equations it follows also the conservation equation
\beq
\nabla_{\mu}\left(T^{\mu\nu}K_{\nu}\right)=0
\eeq
and a corresponding charge. If we define the two-dimensional Levi--Civita skew tensor
\[
\epsilon_{\mu\nu}=\epsilon_{\mu\nu\alpha\beta}\,\tau_{1}^{\alpha}\,\tau_{2}^{\beta}
\]
where $\tau_{1}$, $\tau_{2}$ are tangent vectors to constant radius spheres, then the Kodama vector may be defined by
\beq\label{koda}
K^{\mu}=\epsilon^{\mu\nu}\partial_{\nu}r\, .
\eeq
It can easily be seen that
\beq
\nabla_{\mu}K^{\mu}=0\, .
\eeq
Let us give few examples: 
\begin{enumerate}
\item[(a)] for static, non-dirty, black holes it is the Killing field;
\item[(b)] for dirty black holes, eq.(\ref{ab}) or (\ref{dirty}), $K = \sqrt{W/V}\; \partial_t$;
\item[(c)] in Bardeen--Vaidya gauge $K=e^{-\Phi}\partial_{v}$;
\item[(d)] for the metrics eq.(\ref{pain}) and eq.(\ref{vissnil}), $K=\partial_{t_{ p}}$ or $K=e^{-\Phi}\partial_{\tilde t}\,$, respectively;
\item[(e)] for the metric eq.(\ref{efg}) $K=(EF+G^2)^{-1/2}\partial_{t}$;
\item[(f)] in conformal gauge $K=-g^{+-}(\partial_{+}r\partial_{-}-\partial_{-}r\partial_{+})$;
\item[(g)] in synchronous gauge $K=\sqrt{B}(R'\partial_{\tau}-\dot{R}\partial_{r})$;
\item[(h)] for FRW flat cosmology  $K=\partial_{t}-Hr\partial_{r}$.
\end{enumerate}
In every case it follows that 
\[
K^2=\frac{2E}{r}-1
\]
so that $K$ is time-like, space-like or null if and only if $r>2E$, $r<2E$ or $r=2E$, respectively. Let us also define the current
\beq\label{joda}
j^{\mu}=-T^{\mu\nu}K_{\nu}\, .
\eeq
We have just seen that $\nabla_{\mu}j^{\mu}=0$, $\nabla_{\mu}K^{\mu}=0$, and therefore there exist two conserved charges
\beq\label{qk}
Q_{j}=-\int_{\Sigma}J^{\mu}n_{\mu}d^3V
\eeq
\beq\label{qj}
Q_{K}=-\int_{\Sigma}K^{\mu}n_{\mu}d^3V
\eeq
where $\Sigma$ is a space-like three-dimensional surface with fixed boundary at some constant $r$ and future pointing time-like normal $n_{\mu}$. Here \textquotedblleft conserved'' means independent on the choice of $\Sigma$. The charges as defined will be positive in regions where $j^{\mu}$, $K^{\mu}$ are both time-like. Using eq.(\ref{ee}) one can easily see that
\beq
j=\frac{1}{4\pi r^2}(\partial_{+}E\partial_{-}-\partial_{-}E\partial_{+})\, .
\eeq
To compute the charges we pass to a synchronous gauge by choosing  coordinates $(\tau,\zeta)$ adapted to $\Sigma$, with $\partial_{\tau}$ normal and $\partial_{\zeta}$ tangent to $\Sigma$. In these coordinates we can always write the metric in the form
\beq\label{sime}
d s^2=-d\tau^2+e^{\lambda}d \zeta^2+r^2(\tau,\zeta)d\Omega^2\, .
\eeq
From eq.(\ref{koda}) and eq.(\ref{joda}) we obtain
\beq\label{kojo}
K^{\mu}=e^{-\lambda/2}(r',-\dot{r},0,0), \quad j^{\mu}=\frac{e^{-\lambda/2}}{4\pi r^2}(E',-\dot{E},0,0)
\eeq
where\ $\dot{r}=d r/d\tau$, $r'=d r/d\zeta$. Following Hayward, we shall say that the point $r=0$ is a {\it regular centre} if it is a boundary point of the normal space and $E/r\to 0$ as the centre is approached. Otherwise it is a {\it central singularity}.  We now assume that $\Sigma$ extends from a regular centre to some $r>0$:  from eq.(\ref{kojo}) and the metric we have the normal $n^{\mu}=\delta^{\mu}_{0}$ and
\[
-K^{\mu}n_{\mu}=e^{-\lambda/2}r', \quad -j^{\mu}n_{\mu}=\frac{e^{-\lambda/2}}{4\pi r^2}\,E'\, .
\]
Therefore integrating over $\Sigma$ with the invariant measure $d V=e^{\lambda/2}r^2\sin\theta d\zeta d\theta d\phi$, we finally obtain
\beq
Q_{K}=\frac{4\pi}{3}\,r^3, \qquad Q_{j}=E\, .
\eeq
We conclude that $E$ is indeed the charge associated to a conserved current. The charge $Q_{j}$ is the definition of energy of Kodama. For static solutions the Kodama vector coincides with the Killing field, the generator of the time translation symmetry.

\subsubsection{The surface gravity}
\label{4.2.4}

The  properties of the Kodama vector field discussed so far prompt for a natural definition of the surface gravity of a trapping horizon. We have seen that $K$ becomes null precisely on a trapping horizon and space-like within. The time-like integral curves of $K$ are in general not contained within the horizon. Consider the quantity  $K^{\mu}\nabla_{[\nu}K_{\mu ]}$: it can be seen that on a trapping horizon it is proportional to $K_{\nu}$
\beq\label{kK}
K^{\mu}\nabla_{[\nu}K_{\mu ]} \cong -\kappa K_{\nu}\, .
\eeq
The function $\kappa$ is, by definition, the horizon surface gravity of Hayward. For static black holes $K$ is the Killing field so $\nabla_{\mu}K_{\nu}$ is anti-symmetric and the definition reduces to the usual one. A formula to compute $\kappa$ efficiently was found by Hayward \cite{Hayward:1997jp}. Working in double-null coordinates we have $K_{+}=-\partial_{+}r$, $K_{-}=\partial_{-}r$ (see point (f) of the examples list) therefore
\beq
\partial_{+}K_{-}=-\partial_{-}K_{+}=\partial_{-}\partial_{+}r=\frac{g_{+-}}{2}\,\Box_{\upgamma} r
\eeq
where $\Box_{\upgamma}=2g^{+-}\nabla_{+}\nabla_{-}$ is the two-dimensional Klein--Gordon operator acting on scalars. Computing the left hand side of eq.(\ref{kK}) then gives the wanted formula
\beq\label{hayk}
\kappa=\frac{1}{2}\Box_{\upgamma}r=\frac{1}{2} \,r\left(g^{+ -}\partial_- \theta_+ +\frac{1}{2}g^{+ -}\theta_{+}\theta_{-}\right)\;,
\eeq
where the last form is obtained by using $\theta_{\pm}=2r^{-1}\partial_{\pm}r$. That is, $\kappa$ is the ``Box'' of $r$; when evaluated on the trapping horizon, where $\theta_{+}=0$, it is the surface gravity and is positive if and only if $\partial_{-}\theta_{+}<0$, that is if the horizon is of outer type. Its invariant character is manifest. Let us give few examples with comparison to other definitions. For the Bardeen--Vaidya metric one obtains
\beq\label{bvk}
\kappa=\frac{1}{4M}(1-2M')
\eeq
which is also the Visser dynamical  surface gravity as defined in \cite{Visser:2001kq} in a Painlev\'e--Gullstrand frame. The first Hayward's definition (see the list in Section \ref{4.1}), invoked without appeal to un-affinity, would give instead
\[
\tilde\kappa=\frac{1}{4M}\sqrt{1-2M'}
\]
which is not even correct for the static Reissner--Nordstr\"{o}m solution\footnote{For which $M'=q^2/2(2M^2-q^2+2M\sqrt{M^2-q^2})$.}. The definition of Fodor \textit{et al.}  gives
\[
\hat\kappa=\frac{e^{\Phi}}{4M}(1-2M')+\dot{\Phi}\, .
\]
For the dynamic Painlev\'e--Gullstrand metric eq.(\ref{pain}) with $M=M(t_{ p},r)$, we obtain
\beq
\kappa=\frac{1}{4M}(1-2M'+2\dot{M})
\eeq
while the Visser and Visser--Nielsen surface gravity for this kind of metrics would give the same formula without the time derivative term. Incidentally, this shows that the dynamical Bardeen--Vaidya metric is not diffeomorphic to a dynamical Painlev\'e--Gullstrand metric eq.(\ref{pain}) with the same mass function, as it would happen in the stationary case, and therefore represent physically different gravitational fields.  For the dynamic Painlev\'e--Gullstrand metric eq.(\ref{vissnil}), with mass function $m(\tilde t,r)$,  we obtain instead
\beq
\kappa=\frac{1}{4m}(1-2m'+2\dot{m}e^{-\Phi})
\eeq
again different from the Visser--Nielsen surface gravity for the same mass function, which has no time derivative terms. One may also mention the ``effective surface gravity'', $\kappa_{{ eff}}=1/2r_{ H}$ discussed, for example, in \cite{Ashtekar:2004cn} and \cite{Faraoni:2007gq}.

\subsection{Tunnelling from trapping horizons}
\label{4.3}

We now come to review what the tunnelling method has to say about dynamical, spherically symmetric black holes. We shall start by identifying the dynamical version of the tunnelling path which was displayed and discussed is Section \ref{2:stat}, see fig.\ref{ks}.  This will be accomplished by using a specific and convenient form of the metric, which we start to review a little more than already done. In absence of analytical techniques we shall make explicit use of the Hamilton--Jacobi equation, the null geodesic method being quite inconvenient to treat truly dynamical metrics (but see Clifton \cite{Clifton:2008sb}). The covariance of the method will be stressed throughout.

\subsubsection{Metrics to be used}

We shall use for the time being the Bardeen--Vaidya (BV) metric eq.(\ref{bard}), which we recall here
\beq
d s^2=-e^{2\Phi(r,v)}\left(1-\frac{2M(r,v)}{r}\right)d v^2+2e^{\Phi(r,v)}d v d r+r^2d\Omega^2\, .
\label{bv}
\eeq
The sphere of symmetry will not play any role here, though.  To illustrate the covariance of the results we shall occasionally  make use of the metric in Painlev\'e--Gullstrand (PG) form, either eq.(\ref{pain}) or eq.(\ref{vissnil}). Few things about metric (\ref{bv}) will help with the understanding.  The field equations read   
\beq\label{vbe}
\frac{\partial M}{\partial v}=4\pi r^2T^r_{\;v}, \quad \frac{\partial
  M}{\partial r}=-4\pi r^2T^v_{\;v}, \quad 
 \frac{\partial\Phi}{\partial r}=4\pi re^{\Phi}T^v_{\;r}\, ,
\eeq
and the stress tensor can be written as
\beq\label{st}
T_{\mu\nu}=\frac{\dot{M}}{4\pi r^2}\nabla_\mu v\nabla_\nu v-\frac{M'}{2\pi  r^2}\nabla_{(\mu}r\nabla_{\nu)}v \, .
\eeq
If $M$ only depends on $v$, it describes a null fluid obeying the dominant energy condition for $\dot{M}>0$. For the excreting black hole $\dot{M}<0$ so the null energy condition will also be violated. We already know that $r=2M(r,v)$ is a trapping horizon, that is a FOTH, if and only if $2M'<1$, which we shall assume from now on. Putting $r=2M$ into the metric gives
\[
d s^2=e^{2\Phi}\left(\frac{4e^{-\Phi}\dot{M}}{1-2M'}-1\right)d v^2
\]
so we conclude that the horizon is certainly time-like if $\dot{M}<0$; we know from general results that it will be space-like or null if $\dot{M}>0$, even if it is not evident from this expression. 
The Hayward surface gravity is
\beq\label{bvsg}
\kappa=\frac{1}{4M}(1-2M')
\eeq
and is positive under our assumptions. As we extensively explained, the Misner--Sharp mass, or energy for short, is the value $E(v)$ taken by $M(v,r)$ on the trapping horizon, {\it i.e.} 
\beq\label{energy}
E(v)=M(v,r_{ H}(v))=\frac{r_{ H}(v)}{2}\, .
\eeq
Using eq.(\ref{vbe}) one can show that an observer at rest at $r\gg r_{ H}$ sees a quasi-static geometry with a luminosity $L=-d E/d v$. We shall not assume, initially, that $L$ has the Hawking form $L=\hbar{\cal N}E^{-2}$, with ${\cal N}$ a constant proportional to the number of massless species radiated from the black hole. 

For sake of completeness, we note that a cosmological  constant can be introduced via
\beq\label{bardL}
d s^2=-e^{2\Phi(r,v)}\left(1-\frac{2M(r,v)}{r}-\frac{\Lambda r^2}{3}\right)d v^2+2e^{\Phi(r,v)}d v d r+r^2d\Omega^2\, .
\eeq
We shall admit that $\Lambda>0$; if $M$ is constant and $\Phi=0$ one recognises the Schwarzschild--de Sitter metric in advanced coordinates. Let us call $C$ the expression in round brackets above; the null normals and their expansions are  
\bea\label{null2}
\ell_{+}&=&\left[1,\frac{1}{2}e^{\Phi}C,0,0\right] , \qquad \theta_{+}=\frac{1}{r}e^{\Phi}C \\ 
\ell_{-}&=&[0,-e^{-\Phi},0,0], \qquad \theta_{-}=-\frac{2e^{-\Phi}}{r}\, .
\eea
Hence the horizons are located in correspondence of the roots of the equation $C=0$; the tractable case is $M=M(v)$. Then this becomes a cubic equation 
which for $0<9\Lambda M^2<1$, as is well known, admits precisely two real positive roots $r_{ c}$, $r_{ b}$, with $r_{ c}>r_{ b}$ by definition. We see that $\theta_{+}$ vanishes at both roots and $\theta_{-}<0$,  therefore the horizon spheres $r=r_{ b}$, $r=r_{ c}$, are marginally trapped  surfaces of the future type which foliate a black hole and a cosmological trapping horizon, respectively. One can easily show that 
\[
\partial_{-}\theta_{+}=  -\frac{1}{r}\,\partial_{r}C\, .
\]
Computing the radial derivatives at both horizons we see that this is negative at $r_{ b}$ and positive at $r_{ c}$. The cosmological horizon is therefore an example of a trapping horizon of inner type, the black hole horizon at $r=r_{ b}$ remaining of the outer type.

\subsubsection{Rays tracing} 
\begin{center}
\begin{figure}[h]
\vspace{-14pt}
\hspace{-20pt}\includegraphics[width=1.2\textwidth]{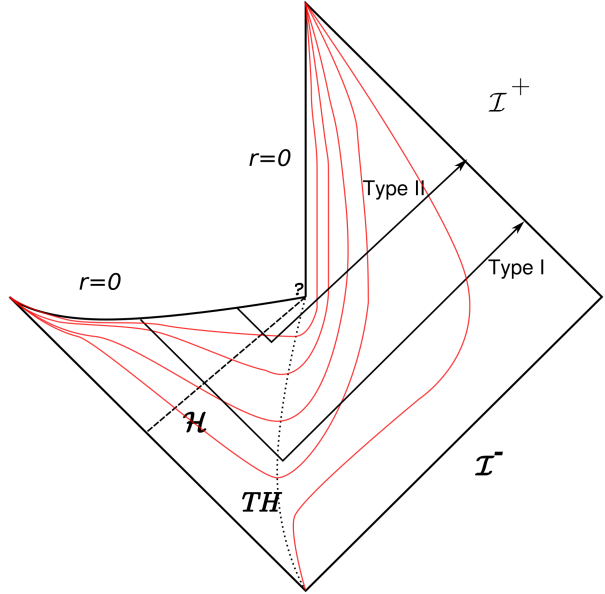}
\vspace{-80pt}
\caption{Evaporating black hole: the trapping horizon is at points where the radial outgoing rays have zero speed (dotted line). The event horizon $\mathcal H$ is represented by a dashed line. Red lines represent curves of constant $r$.}
\label{dina}
\end{figure}
\end{center}

The most important features of a dynamical black hole of the kind discussed here are: (i) the existence of the irremovable space-like singularity at the origin $r=0$ of the coordinate system; (ii) the possible existence of a global event horizon (${\cal H}$ in fig.\ref{dina}); and (iii) the time-like future trapping horizon ($TH$ in fig.\ref{dina}). 
We have seen that the BV form and the PG form are locally diffeomorphic to each other with the mass function transforming as a scalar field, so the causal structure of both solutions must be the same. A visual picture of the trapping horizon for an evaporating black hole is displayed in fig.\ref{dina}. The horizontal line represents the space-like singularity $r=0$.  Consider the radial outgoing null rays: in BV form they obey the differential equation
\beq\label{nraybv}
\dot{r}\equiv\frac{d r}{d v}=\frac{1}{2}e^{\Phi}\left(1-\frac{2M}{r}\right)
\eeq
while ingoing rays simply are $v=v_{ 0}$, for some constant $v_{ 0}$. For comparison, in the PG form eq.(\ref{pain}) the same equation takes the form
\beq\label{nraypg}
\dot{r}=\left(\pm 1-\sqrt{\frac{2m}{r}}\right)e^{\Phi}
\eeq
where $m(\tilde t,r)=M(v,r)$ and the plus (minus) sign referring to outgoing (ingoing) rays. Returning to the BV form, the event horizon is an outgoing null surface, 
\beq\label{EH}
\dot{r}_{ EH}=\frac{1}{2}\left(1-\frac{2M}{r_{ EH}}\right)e^{\Phi}<0
\eeq
so that $r_{ EH}<r_{ H}=2M$. The acceleration close to the trapping horizon is
\beq\label{acc}
\ddot{r}=-e^{\Phi}\frac{\dot{M}}{2M} + \frac{e^{\Phi}}{4M}(1-2M')\dot{r}\;\simeq \;\frac{e^{\Phi_{ H}}}{r_{ H}}\,L+\frac{e^{\Phi}}{4M}(1-2M')\dot{r}>0\, .
\eeq
We see that in both cases at the TH, which we recall is the line $r=2M(v,r)$ for BV or $r=2{\tilde M}(\tilde t,r)$ for PG, photons are only momentarily at rest, subsequently escaping on a dynamical time scale $\kappa^{-1}$, where $\kappa$ is the surface gravity eq.(\ref{bvsg}). Therefore outgoing photons (massless particles) have $\dot{r}<0$ before reaching the trapping horizon and $r$ must decrease on going to the horizon. This can be seen from the lines of constant  $r$ in fig.\ref{dina}. Implicit in the Figure is also the idea that the Misner--Sharp mass has a non zero limit as $v\to-\infty$. More complicated dynamics are possible in general depending on the mass function, including the possibility of horizon formation at finite times (some of these are discussed in Kodama's original paper \cite{Kodama:1979vn}).

{\it The fact that $\dot{r}=0$ at the trapping horizon and only there is one of the most important facts at the root of the quantum tunnelling phenomenon.}
Now we would like to consider the case where $M(r,v)$ goes to a constant at very large negative advanced time and the trapping horizon extends to past infinity, since this is the region when the black hole does not radiate yet.  In this case all radial null geodesics emerging from the trapping horizon seem to do so at $v=-\infty$, because the speed vanishes at $r=2M$ (or $2{\tilde M}$). For particles with non zero angular momentum the trapping horizon is not a surface of momentarily zero speed. For instance, in the equatorial plane the velocity is given by
\beq\label{jnull}
\dot{r}=\frac{1}{2}\left(1-\frac{2M}{r}\right)e^{\Phi}-\frac{J^2}{2r^2}
\eeq 
where $J$ is the conserved angular momentum per unit mass. The essential point is that $\dot{r}<0$ at $r=2M$, so that for a rotating particle it will even be harder to escape quantum mechanically than to a non rotating one. We will see in the next section that the presence of angular momentum will not affect the main conclusion that the radiation originates close to the trapping horizon, not the global event horizon. Actually, the global event horizon cannot even be computed without knowledge of the dynamics near the  evaporation end point. A practical definition was given by Bardeen \cite{bardeen:1981}: one looks for photons that can only reach large $r$ in a time comparable to the evaporation time $\sim E/L\gg E$. Thus, the event horizon is located by the unaccelerated photons with $\ddot{r}=0$. From eq.(\ref{nraybv}) and eq.(\ref{acc}) it follows then that
\beq
\dot{r}_{ EH}\simeq -2L/r_H\kappa, \qquad r_{ EH}\simeq 2E\left(1-\frac{2L}{r_{ H}\kappa}\right)\simeq 2E(1-4L)
\eeq
should be a solution to eq.(\ref{EH} to first order in the black hole luminosity $L$ (which is very small for large black holes) and ignoring contributions due to $\Phi$ and $M'$ which are known to be negligible to first order in $L$ \cite{york:1983}. This is the result of Bardeen--York.

\subsubsection{Tunnelling paths}
\label{tu:tupa}

We propose now to compute the probability amplitude for a quantum particle to cross a trapping horizon. We do this by making use of the Hamilton--Jacobi method. 
Recall that, in the spherically symmetric dynamical case, it is possible to introduce
the Kodama vector field $K$ such that $(K^\alpha G_{\alpha\beta})^{;\beta} =0$, actually its defining property. Given the metric (\ref{metric}), the Kodama vector components are
\beq K^i(x)=\frac{1}{ \sqrt{-\upgamma}}e^{ij}\partial_j R\,,
\qquad K^\theta=0=K^\phi \label{ko} \;. \eeq 
where $e^{ij}$ is the numerical skew tensor with $e^{01}=1$. The Kodama vector gives a preferred flow of time and in this sense it generalises the flow of time given by the Killing vector in the static case (see \cite{Abreu:2010ru} for a detailed study of the space-time foliation it determines). The conserved charges associated to $K$, eq.(\ref{qk} and eq.(\ref{qj}), are respectively the volume and the Misner--Sharp mass of space-time. We may also use the Kodama flow to define the invariant energy associated with a particle  by means of the scalar quantity on the normal space 
\beq 
\label{ener} 
\omega =- K^{i} \partial_i I\,, 
\eeq 
where $I$ denotes the classical action of the massless particle, which we assume to satisfy the reduced Hamilton--Jacobi equation 
\beq \label{hje}
\upgamma^{i j}\partial_i I \partial_j I=0\, . 
\eeq 
Thus, for example, $\omega=-e^{-\Phi}\partial_{v}I$ will be the particle energy on an extremal in a BV gauge; we may note that this gauge is only fixed up to diffeomorphisms $v\to v^{'}=v^{'}(v)$, under which the field $\Phi(v,r)$ transforms as a conformal field, $\Phi(v,r)\to\Phi^{'}(v^{'},r)=\Phi(v,r)+\log|d v^{'}/d v|$. Therefore $\omega$ will not depend on the choice of advanced time which respect to the BV form of the metric. Similarly, in Painlev\'e--Gullstrand  gauge the invariant energy is $\omega=e^{-\Phi}\partial_{\tilde t}I$, and is gauge invariant under time re-parametrisation $\tilde t\to\bar{t}=\bar{t}(\tilde t)$.
\
\\

To illustrate the method we shall work in s-wave and omit the angular dependence of $I$. We stress the importance to have at disposal an invariant definition of energy. eq.(\ref{ener} certainly satisfies this requirement if the action is a scalar. In the following our aim will be to show that there is a precise invariant prescription to deal with the imaginary part of the action, in case there is one, which is valid for all solutions in all coordinate systems which are regular across the horizon. The task is to compute
\beq\label{immI}
\mbox{Im}\int\!\!\!\!\!\!\!\searrow d I
\eeq
along a tunnelling path, say $\gamma$. A priori there are two types of such tunnelling paths, which we shall call type-I and type-II, because a pair may form on either sides of the trapping horizon. A type-I path corresponds to a pair forming outside and therefore will be a backward null ray coming out from the future singularity at $r=0$, crossing the trapping horizon and subsequently escaping to infinity, as illustrated in fig.\ref{dina}. The figure is the dynamical analogue of the corresponding figure in Schwarzschild space-time, except that it is smooth at $\dot{r}=0$, namely at the trapping horizon. A type-II path corresponds to a pair forming in the trapped region, so we may take a backward null ray crossing the event horizon from the singularity till some interior point, which subsequently escapes to infinity along a null ray crossing the trapping horizon at some other event. In some cases such a path (namely an outgoing light ray emerging from the trapping horizon) seems to emerge from the trapping horizon at $v=-\infty$ in advanced time, therefore in this case, type-II paths are those with exactly one point, the starting point, lying on the horizon. Type-II paths are absent in static geometries, since there is no region in between the two horizons. The segment of type-II paths crossing the horizon outward seems nevertheless an allowed classical path so one may wonder whether they can contribute to an emission amplitude: does the pole disappear? We will see indeed that on such paths the radial momentum vanishes, basically for the reason that the speed is zero on the trapping horizon --- and only there --- so in fact they do not contribute semi-classically. 
\
\\

We shall illustrate these results in some of the gauges we discussed above for spherically symmetric space-times, having two main purposes: to illustrate the intrinsic covariance of the result and to display how contributions to the tunnelling amplitude may or may not come from all terms in the differential form $d I=\partial_{i}Id x^i$. We also have the opportunity to test the formalism and to see how different can be the prosaic description of the same physical effect from the perspective of other coordinate systems, a fact well known in special and general relativity.
\
\\
\linebreak

\begin{center}{\it The BV gauge} \end{center}

\noindent Here the metric is
\beq
d s^2=-e^{2\Phi}Cd v^2+2e^{\Phi}d vd r+r^2d\Omega^2
\eeq 
where $M=(1-C)r/2$ is the Misner--Sharp mass discussed in Section \ref{4.2.3ms}. The Kodama vector and the invariant particle energy assume the simple expressions given by $K=(e^{-\Phi}, \mathbf{0})$ and $\omega=-e^{-\Phi}\partial_v I$, while the invariant surface gravity is
just given by $\kappa_{ H}=\partial_rC_{ H}/2$. As noted above $\Phi$ transforms as an ordinary Liouville field, {\it i.e.} $\Phi\to\Phi+\log|\partial\tilde{v}/\partial v|$, under $v\to\tilde{v}(v)$, making $\omega$ invariant under re-parametrisation of the advanced time coordinate.

Type-II tunnelling paths are easily dispensed for: they cross the horizon along a null direction which is outward with zero speed at the horizon, so $\dot{r}=0$ at and in a neighbourhood of the horizon, or $d r=0$ along the null direction; therefore on a small segment crossing the trapping horizon we have
\[
\int\partial_{r}I\,d r=0\, .
\]
On a real trajectory the energy is real and finite everywhere, which shows that the temporal $v$-contribution does not play any role, too, in what concerns us at the moment (the evaluation of $\mbox{Im}\,I$). Thus $\mbox{Im}\,I=0$ on type-II paths.  

For a type-I path the ray crosses the horizon along a null ingoing ray but in the backward direction; such a ray has $v=v_{ 0}$ (some constant) and $\delta v=0$ exactly, therefore again the $\partial_{v}Id v$ term will vanish at the horizon and $\delta r>0$. From the Hamilton--Jacobi equation we see that, for outgoing modes of the kind we consider, 
\beq 
C(\partial_rI)=  2\omega \;. 
\label{c1} 
\eeq 
Thus one has
\begin{eqnarray}
\mbox{Im}\,I&=&\mbox{Im}\int_{\gamma}\left(\partial_r I d r +
\partial_v Id v\right)=  \mbox{Im} \int_{\gamma} d r\,\frac{2\omega}{C}  \nonumber \\
&=& 2 \,\, \mbox{Im}\int_{\gamma} d r \,\frac{\omega }{\partial_r
  C\big\vert_{ H}(r-r_{ H}-i \epsilon)} \label{culo} = \frac{\pi \omega_{ H}}{\kappa_{ H}}\,,
\end{eqnarray}
where $\omega_{H}$ is $\omega$ evaluated on the horizon, the quantity $C$ has been expanded around the horizon pole, that is
\beq
 C(v,r)= \partial_r C\Big\vert_{ H} \Delta r  + \dots
\eeq
and Feynman's $i\epsilon$-prescription has been implemented in order to deal with the simple pole. $\kappa_{ H}= \partial_rC\big\vert_{ H}/2$ is the horizon surface gravity and coincides with our geometrical expectations. Unlike the stationary black holes, where $\omega$ is a constant of motion, here we get the local energy and temperature as measured near the horizon, which are connected to the quantities at infinity by the corresponding red-shift factors. We see that, in a BV coordinate system, the temporal integration does not give any contribution to the imaginary part of the  action of particles tunnelling through the trapping horizon. 

It will be very important to discuss the physical meaning of this result but we prefer to postpone this discussion to the end of the section, after we will have shown that the tunnelling amplitude is not an artefact due to a special coordinate system but holds as good as in any other. \medskip

\begin{center}{\it The conformal two-dimensional gauge}\end{center} 

\noindent A coordinate system where the temporal contribution to the action plays an essential role is the general diagonal form of a spherically symmetric metric, which
reads
\beq
d s^2=e^{\psi(t,r)}\left( -d t^2+ d r^2 \right)+R^2(t, r) d\Omega^2\,.
\label{k}
\eeq
In this form, the normal metric is conformally related to the two-dimensional Minkowski space-time. The $\chi$ function simply reads
\beq
\chi=e^{-\psi}\left[ -(\partial_{t} R)^2 +(\partial_r R)^2\right]\,,
\eeq
If $R(t,r)$ is a monotonically increasing function of $r$ (the normal case) then this leads to the future trapped horizon condition
\beq
(\partial_{t} R)_{ H} +(\partial_r R)_{ H}=0\,. 
\label{hk}
\eeq
If instead  $R(r,t)$ is decreasing as a function of $r$, then we have to take the opposite relative sign, because we expect that an evaporating horizon should have a negative time derivative, $(\partial_{t}R)_{ H}<0$.  A computation of the expansion scalars will confirm this fact. The Kodama vector and the associated invariant energy are
\beq
K=e^{-\psi}\left( \partial_r R , -\partial_{t} R , 0, 0 \right)\,,
\eeq
\beq
\omega=e^{-\psi}\left(- \partial_r R \partial_{t} I + \partial_{t} R
\partial_r I \right)\,.
\label{k2} \eeq 
The dynamical surface gravity reads 
\beq
\kappa_{ H}=\frac{1}{2} e^{-\psi_{ H}}\left( -\partial_{t}^2 R +
\partial_r^2 R \right)\Big\vert_{ H}\,. \label{kt} 
\eeq 
Due to conformal invariance, the Hamilton--Jacobi equation is the same as in
two-dimensional Minkowski space-time, namely using double null coordinates $\partial_{+}I\partial_{-}I=0$, and for the outgoing particle we may take 
\beq
\partial_+I=\partial_{t} I + \partial_r I =0\,,
\label{hjk}
\eeq
since the radial momentum $p_{r}=\partial_{r}I>0$ and $\partial_{t}I<0$ for a real outgoing particle. Given that for a type-II path the null expansion condition leads to $\delta x^-=\delta{t}-\delta r =0$, we get $d I=\partial_{+}Id x^++\partial_{-}Id x^-=0$ on account of the Hamilton--Jacobi equation, and there is no imaginary part. On reflection, this result seems to violate our basic tenet that there is no amplitude because the photon speed vanishes at the horizon. In fact, the coordinate speed $\dot{r}=\pm1$ everywhere in conformal gauge and it never vanishes. However, the speed of the wave front is
\[
\dot{R}=\partial_{r}R\,\dot{r}+\partial_{t}R
\]
and this vanishes on ${ H}$ for an outgoing photon with $\dot{r}=1$, on account of the horizon equation eq.(\ref{hk}). Thus, what really matters is that the speed of the areal radius should vanish at the horizon.

But there are also the type-I tunnelling paths, for which the null expansion gives $\delta x^+=\delta t+\delta r=0$. Therefore in this case
\beq
I=\int_{\gamma} \left(d r \,\partial_r I+ d t\, \partial_{t} I
\right) = 2 \int_{\gamma} d r\,\partial_r I \,.
\label{b0}
\eeq
Furthermore, due to (\ref{k2}) and (\ref{hjk}), one has
\beq\label{prr}
\partial_r I=\frac{\omega}{e^{-\psi} (\partial_r R+\partial_{t} R)}\,.
\eeq
and we have a pole at the horizon. Making use of near horizon
approximation along the null direction, from  (\ref{hk}) and (\ref{kt}), one has 
$ (\partial_r R)_{ H}+(\partial_{t} R)_{ H}=0$, $\delta t+\delta r =0$, thus with the understanding that we shall put $\delta r=r-r_H$, we get
\begin{eqnarray}\label{exxp}
e^{-\psi}[\partial_r R+\partial_{t} R] &=&e^{-\psi_{ H}}\left(\partial_{rr}^{2} R-\partial_{rt}^{2} R- \partial_{tt}^{2} R +\partial_{tr}^{2} R
\right)\Big\vert_{ H} (r-r_{ H}) + \dots \nonumber \\ 
&=& 2 \kappa_{ H} (r-r_{ H}) +\dots
\end{eqnarray}
and the imaginary part follows. If $R(t,r)$ is a decreasing function of $r$ then we shall have to take $p_{r}=\partial_{r}I<0$ but again $\partial_{t}I<0$ so the Hamilton--Jacobi equation implies now $\partial_{-}I=0$ along the tunnelling path. Moreover since $r$ is decreasing as $R$ increases ({\it i.e.} toward infinity), on an outgoing (ingoing) null direction we have $\delta t + \delta r=0$ ($\delta t-\delta r=0$) and the horizon condition becomes
\[
(\partial_{t}R-\partial_{r}R)\Big\vert_{ H}=0
\]
instead of eq.(\ref{hk}. It is now simple to see that a tunnelling type-II path gives no amplitude at all, but that a type-I does produce an imaginary part with the right magnitude and sign. In fact in this case we obtain first
\[
\partial_r I=-\frac{\omega}{e^{-\psi} (\partial_r R-\partial_{t} R)}\,.
\]
then, recalling $\partial_{-}I=0$,
\[
d I=\partial_{+}Id x^+=2\partial_{r}Id r=-2 \frac{\omega}{e^{-\psi} (\partial_r R-\partial_{t} R)}d r \cong-2\frac{\omega}{\kappa_{ H} (r-r_{ H})}d r\, .
\]
The integral of $d I$ has to be done on the segment of the path crossing the horizon along decreasing $r$, say from $r_{2}$ to $r_{1}$ with $r_{1}<r_{2}$, and this is of course minus the usual integral in increasing order; reversing the orientation gives back the usual result
\[
\mbox{Im}\int\!\!\!\!\!\!\!\searrow d I=\frac{\pi\omega}{\kappa_{ H}}\, .
\]\medskip

\begin{center}{\it The synchronous gauge}\end{center} 

\noindent Another coordinate system where the coordinate speed does not vanish at the trapping horizon is described by the line element 
\beq 
d s^2=-d t^2+\frac{1}{B(r,t)}d r^2+R^2(r,t)d\Omega^2=d\upgamma^2+R^2(r,t) 
d\Omega^2\,, 
\eeq  
in which the metric is diagonal, but $R$ is a function of  $r$ and $t$. In this case, one has 
\beq \chi=-(\partial_tR)^2+B (\partial_r R)^2\,, 
\eeq 
If $R(r,t)$ is an increasing function of $r$, the horizon  $\chi_{ H}=0$ should be defined by  
\beq\label{horsyn}
(\partial_t R)_{ H}=-\sqrt{B_{ H}} (\partial_r R)_{ H}\,, 
\eeq 
in which we are assuming again a regular coordinate system on the horizon, namely that  $B_{ H}$ and the partial derivatives are non-vanishing. For $R$ decreasing with $r$ one should change the sign to the right of eq.(\ref{horsyn}. The Kodama vector reads  
\beq 
K=(\sqrt{B} \partial_r R, -\sqrt{B} \partial_t R, 0 , 0)\,, 
\eeq 
and the invariant energy 
\beq 
\omega=\sqrt{B} (\partial_r R \partial_t I- \partial_t R \partial_r I)\,. 
\eeq 
The dynamical surface gravity is evaluated to be
\beq 
\kappa_{ H}=\frac{1}{4}\left(- 2\partial^2_t R_{ H}+2 B_{ H} \partial^2_r R_{ H}+ 
\frac{1}{B_{ H}}\partial_t R_{ H}  \partial_t B_{ H} +\partial_r R_{ H} 
\partial_r B_{ H} \right)\,. 
\label{sd}
\eeq
Making use of the horizon condition, we may rewrite it
\beq
\kappa_{ H}=\frac{1}{4}\left(- 2\partial^2_t R_{ H}+2 B_{ H} \partial^2_r R_{ H} -
\frac{1}{\sqrt{B_{ H}}}\partial_r R_{ H}  \partial_t B_{ H} +\partial_r R_{ H}
\partial_r B_{ H} \right)\, .
\eeq
In this case, the Hamilton--Jacobi equation reads simply
\beq
-(\partial_tI)^2+B (\partial_r I)^2=0
\eeq
and for an outgoing particle we have to choose the root with $\partial_rI>0$. Therefore, in a type-II path crossing the trapping horizon we have $\sqrt{B_{ H}}\delta t=\delta r$, in a neighbourhood of TH, because $\dot{r}=\sqrt{B}$. As a consequence the outgoing temporal contribution will cancel the radial one. Type-II paths will then give no tunnelling amplitude. On reflection, this happens because although the coordinate speed is non vanishing at the trapping horizon, the areal velocity indicating the speed of the wave front 
\[
\dot{R}=\partial_{r}R\,\dot{r}+\partial_{t}R
\]
does vanish on TH for an outgoing photon, which has $\dot{r}=\sqrt{B}$, on account of eq.(\ref{horsyn}. On a type-I path instead $\sqrt{B_{ H}}\Delta t=-\Delta r$, and the two contributions add so that  
\beq
I=2\int_{\gamma} d r \partial_r I\,.
\eeq
The Hamilton--Jacobi equation and the expression for the invariant energy lead to
\beq
\partial_r I =\frac{\omega} { B \partial_rR+ \sqrt{B}\partial_t R}\,.
\eeq
Making the expansion along the null curve, for which $\sqrt{B_{ H}}\Delta t+\Delta r =0$, in the near-horizon approximation, one gets
\beq
\mbox{Im }\, I=2 \cdot \mbox{Im }\, \int_{\gamma} d r \frac{\omega}{2
  \kappa_{ H} (r-r_{ H}-i \epsilon)} = \frac{\pi 
  \omega_{ H}}{\kappa_{ H}}\,, 
\eeq
leading to the desired amplitude. But notice that in this gauge, the temporal contribution is essential in order to obtain the correct result.\medskip

\begin{center}{\it The $r$-gauge} \end{center}

The fact that type-II paths never contributed to the tunnelling amplitude suggest that it is a general fact. Our last example will use a general metric in the $r$-gauge to confirm this. The normal metric here is non-diagonal, but as in BV gauge $R=r$. We have 
\beq 
d s^2=d\upgamma^2+r^2d\Omega^2\,, \label{tnd} 
\eeq  
where the reduced normal metric is now taken in the (redundant) form 
\beq 
d\upgamma^2=-E(r,t)d t^2 +2 F(r,t)d td r +
G(r,t)d r^2\,, \qquad F \neq 0\,. 
\label{nd} 
\eeq  
The horizon is located at the zeroes of
\beq
\chi(t,r)=\upgamma^{ij}\,\partial_i R \,\partial_j R = \upgamma^{rr}(t,r)=\frac{E}{EG + F^2}=0
\label{l}
\eeq
{\it i.e.} at  $E_{ H}=E(t,r_{ H})=0$, provided $F_{ H}\neq 0$. The Misner--Sharp mass inside a sphere of radius $r$ is
\beq\label{MSS}
M=\frac{r}{2}\left(1-\frac{E}{EG+F^2}\right)
\eeq
and its value on the horizon is the black hole mass by definition, $M=r_{ H}/2$. The other ingredient, the Kodama vector, reads 
\beq K=
\left(\frac{1}{\sqrt{F^2 +EG}},0,0,0\right)\,, 
\eeq 
and the invariant particle energy follows
\beq
\omega=-\frac{\partial_t I}{\sqrt{F^2 + EG}}\,.
\label{ro}
\eeq
The dynamical surface gravity is computed to be 
\beq\label{krgauge}
\kappa_{ H}= \frac{1}{2}\Box_{\upgamma}r\Big\vert_{ H}= \left[\frac{1}{2 F^3}\left( E' F  -\frac{1}{2} \dot{E}
  G\right)\right]\Big\vert_{ H} \,, 
\eeq
where an overdot and a prime denote differentiation with respect to $t$ and $r$, respectively. Now the horizon will be future and of outer type provided $F>0$ and $\kappa_{ H}>0$. We expect the mass within the sphere to be a decreasing function of $r$; also, for an excreting black hole the time derivative $\dot{M}$ should be negative as well. From eq.(\ref{MSS}) this will be true if $E'>0$ and $\dot{E}>0$, although these are only sufficient conditions. It follows that for $F<0$, $\kappa_{ H}>0$ always but the horizon is foliated by past marginally trapped surfaces ($G>0$ of course), while for $F>0$ it is future and outer if the inequality
\beq
E'\Big\vert_{ H}>\left(\frac{\dot{E}G}{2F}\right)\Big\vert_{ H}
\eeq   
is true, which we shall assume from now on. The outgoing null rays at the horizon satisfy either $\dot{r}=0$ or, if $F<0$, $\dot{r}=-2F/G$. Only the first type will concern us, since photons from the past horizon should always be allowed classically. In fact, from the Kodama energy expression (\ref{ro}) and equation (\ref{l}) as well, the Hamilton--Jacobi equation reads 
\beq\label{HJrg}
\chi (\partial_r I)^2-2\frac{\omega F}{\sqrt{EG+F^2}}\partial_r I-\omega^2 G=0\,,
\eeq 
Solving this, one sees that $\partial_{r}I$ has no pole at ${ H}$ (which is past for $F<0$) and therefore no imaginary part is found. Returning to our case with $F>0$, on a type-II path $\dot{r}=0$ at the horizon therefore, as before, the radial momentum vanishes and no imaginary term can come from $\int\partial_{r}I\,d r$. 

It remains to consider type-I paths. From the metric the null radial expansion across the segment crossing the horizon outward gives $\delta t=-(G/2 F)\vert_{ H} \delta r$, now with $F>0$, so we can expand the $\chi$ function along this ingoing null direction to first order in $\delta r=r-r_{ H}$. We obtain easily
\beq\label{congo}
\chi  \cong \dot{\chi}\delta t + \chi'\delta r = \left(\chi'-\frac{G}{2F}\dot{\chi}\right)\Big\vert_{ H}(r-r_{ H}) +\dots = 2\kappa_{ H}(r-r_{ H}) +{ O}\left((r-r_{ H})^2\right)
\eeq
where $\cong $ means evaluation on the horizon and eq.(\ref{krgauge} has been used. Also, $\partial_t I=-F_{ H} \omega$ from definition eq.(\ref{ro}) and the horizon condition $E=0$. Hence we end up with $I$ given by the sum of a real term and a possibly imaginary part coming from the horizon: 
\begin{eqnarray}
 I =\int_{\gamma} (d r \partial_r I + d t \partial_t I) = \int_{\gamma} 
d r \,\left[\partial_rI +\frac{1}{2} G_{ H} \omega_{ H} \right]\;.
\end{eqnarray}
What is remarkable is that in this gauge the temporal part is
present but, being regular, it does not contribute to the imaginary part of the action. From the Hamilton--Jacobi equation eq.(\ref{HJrg}) we get, choosing the solution with positive radial momentum,  
\beq
\partial_r I=\frac{\omega F}{\sqrt{EG+F^2}\,\,
  \chi}\left(2+{ O}(\chi)\right)\,. 
\eeq
and we see that this has a pole at $\chi=0$. Making use of this equation, Feynman's prescription and eq.(\ref{congo}),
one has once more
\bea 
 \mbox{Im }\, I & = &\mbox{Im}\,
\int_{\gamma} d r \,\partial_r I   = \mbox{Im} \int_{\gamma} d r\, \frac{\omega F}{\sqrt{F^2 +EG}} \cdot \frac{1+ \sqrt{1+{ O}(\chi)}}{2\,\kappa_{ H}\;(r-r_{ H}-i \epsilon)} =\frac{\pi \omega_{ H}}{\kappa_{ H}}\,.
\eea
We have shown in general that  type-II paths will not contribute a tunnelling amplitude to future trapping horizons ($F>0$), because $\delta r=0$ to first order in
 $\chi$ and we are assuming the energy to be real and finite. As to type-I paths the situation also is clear, they give a amplitude fully controlled by the value of the
 geometrical surface gravity $\kappa_{H}$. This example also shows the possibility of a quantum tunnelling through past trapping horizons of outer type, as for $F<0$ there is an amplitude for a photon to cross with negative radial momentum, {\it i.e.} for a photon to be absorbed (we are using the mental picture of a past horizon as one from which it is impossible to enter but very easy to escape). The fact that there is no imaginary part on
 type-II paths either because the coordinate speed of a photon vanishes or because the
 temporal contribution cancels the radial one was the source of much confusion in the past
 literature. For instance, Belinski ventured to say that there is no Hawking radiation from
 black holes just from this fact \cite{Belinski:2009bc}, since he apparently forgot about the existence of type-I paths.\par
 We end this section by noticing once more the special role of the trapping horizon. The imaginary part being produced on crossing it, it should be natural to think that this is the place where most of radiation forms.  This has been confirmed in \cite{Clifton:2008sb}, where it is shown that asymptotic observers register a radiation flux that starts increasing at a time they see the collapsing shell crosses the surface $r=2M$ of the model used (a special case of Painlev\'e-Gullstrand model with a time dependent mass, see eq.(\ref{pain}). The stress-tensor for Vaidya space-time has been considered in \cite{Hisc,Balb} and should be consistent with this result.

{\it A note ---}  In the previous computations various choices of signs have been applied in
 such a way that it may seem they  were chosen somewhat {\it ad hoc} in order to get the
 wanted result. This is not so. Once the future sheet of the trapping horizon has been
 chosen, and the sign of the Kodama vector determined so that it is future directed, no
 other sign uncertainties will occur for either outgoing or ingoing particles if one uses
 consistently the Hamilton--Jacobi equation for outgoing particles on every segment of the
 tunnelling path; furthermore, questions such as the right form of the equations
 determining the horizon, the orientation of the space-time coordinates and the sign of the
 metric coefficients are carefully established once for all.

On the other hand, if there exists a past sheet in the trapping horizon then using the
 tunnelling picture we may as well compute the action along an inward directed curve at
 the horizon (the ambiguity inherent in this and analogous terms is easily resolved if the
 manifold is asymptotically flat). Then there will be again a non-vanishing imaginary part,
 but we can interpret it as a small absorption probability, as we did for the static case in
 relation to white holes.   


\section{Cosmological Horizons, Decays  and Naked Singularities}
\label{5:coho}
\setcounter{equation}{0}

In this section we review the formalism of horizon tunnelling as applied to cosmological horizons, decay of unstable particles that in the absence of gravity would be otherwise stable and radiation from naked singularities. Especially the subject of horizon tunnelling in cosmology attracted much interest recently, leading to a related stream of papers. A sample of these articles which seemed relevant to us can be found in \cite{Cai,cinesi,volovik}, although we rely mainly on the papers \cite{bob09,DiCriscienzo:2010gh} and the references cited therein. We shall follow mainly the Hamilton--Jacobi version of the tunnelling picture, although the null geodesic method can be still applied. Results in this direction appeared recently in \cite{gaowu}, where the Parikh--Wilczek and Hamilton--Jacobi methods are compared and showed to agree. An early study of the evolution of evaporating black holes in inflationary cosmology is in \cite{mallet}. 
\
\\

\subsection{The FRW space-time}
\label{5.1}

Consider a generic FRW space-time, namely one with constant curvature spatial sections. Its line element can be written as 
\begin{equation}
d s^2=-d t^2+\frac{a^2(t)}{1-k r^2}\, d r^2 + [a(t)r]^2 d \Omega^2 
\label{frw nflat} 
\end{equation}
where $r$ is measured in units of the curvature radius and, as usual, $k=0, -1, +1$ 
labels flat, open and closed three-geometries, respectively. In this gauge, the normal reduced metric is diagonal and  the horizon is implicitly given by the equation $\chi=0$, where
\begin{equation}
\chi(t,r)=1- [a(t)r]^2\left[H^2(t) + \frac{k}{a^2(t)}\right]\,,
\end{equation}
namely 
\begin{equation} 
R_{ H}:= a(t) r_{ H} = \frac{1}{\sqrt{ H^2(t)+\frac{ k}{a^2(t)}}}\,
,\quad\mbox{with}\quad  H(t)=\frac{\dot a(t)}{a(t)}\;, 
\label{h}
\end{equation}
provided the space-time energy density $\varrho(t)$ is positive. As always it corresponds to the vanishing of the wave front velocity whose radius is $R=a(t)r$.  
The surface $R_{ H}(t)$ coincides also with the Hubble radius as defined by astronomers for vanishing curvature, but we shall call it Hubble radius in any
case. The important case $k = 0$ deserves special attention. The horizon is the surface
$r\dot{a}(t) = 1$; substituting into the metric reveals the causal character of the horizon: it will be null if and only if either $p=\varrho/3$ (radiation) or $\varrho = -p$ (vacuum energy), $\varrho$  and $p$ being the energy density and pressure, respectively. It will be time-like or space-like according to whether $|(\varrho + 3p)/2\varrho|$ is less or greater than one. In the former case $-\varrho<p<\varrho/3$, which covers almost the totality of cosmological models, while the space-like character can only be achieved if $p < -\varrho$ (dubbed phantom energy) or $p > \varrho/3$, which includes stiff matter. These results actually hold for non zero curvature too. In Section \ref{4.1} we defined future outer trapping horizons. The situation in flat cosmology is easily spelled out: the horizon is a surface with $\theta_{-}=0$, $\theta_{+}>0$ that for $-\varrho<p<\varrho/3$ satisfies also $\partial_{+}\theta_{-}>0$. In the terminology of Section \ref{4.1} it can be classified as a past inner trapping horizon (abbr. PITH), because all metric spheres at larger radii are  trapped with $\theta_{\pm}>0$. It means that given a comoving observer, a spherical light beam approaching him from cosmic distance will have increasing area due to cosmological expansion. One has to be careful here that the areal radius is $R=a(t)r$, not $r$, so the area of the wave front is $4\pi R^{2}$ and the horizon can be written as the condition $HR_{ H}=1$. Examples of a FITH are provided by the de Sitter horizon and the collapsing Vaidya solution in de Sitter space, eq.(\ref{bardL}.\par  
The dynamical surface gravity reads 
\begin{equation}
\kappa_{ H}= \frac{1}{2}\Box_{\upgamma}[a(t)r]= - \left( H^2(t) +\frac{1}{2} \dot H(t) + \frac{k}{2a^2(t)}\right)\,R_{ H}(t) <0 \,,
\label{frw_dyn_sg}  
\end{equation}
and the minus sign refers to the fact the Hubble horizon is of the inner type. Similarly, the Kodama vector is 
\begin{equation}\label{kv}
K= \sqrt{1- k r^2}(\partial_t -r H(t)\partial_r )
\end{equation} 
so that the invariant Kodama energy of a particle is equal to  
\begin{equation}
\omega = \sqrt{1- k r^2}(-\partial_t I + r H(t) \partial_r
I) \equiv \sqrt{1-k r^2}\,\tilde{\omega}\, .
 \label{b2}
\end{equation} 
Notice that $K$ is space-like for $ra>(H^2+k/a^2)^{-1/2}$,
{\it i.e.} beyond the horizon, so that  particles will eventually tunnel from outside to the inner region, $r< r_{ H}$.

The next ingredient will be the reduced Hamilton--Jacobi equation for a relativistic
particle with mass parameter $m$,  which reads
\begin{equation}
 -(\partial_t I)^2 + \frac{(1- k r^2) }{a^2(t)} \,(\partial_r I)^2 + m^2=0\,.
\label{b}
\end{equation}
Solving eq.(\ref{b2} and eq.(\ref{b}) for $\partial_r I$ and $\partial_tI$ we obtain
\beq\label{g0}
\partial_tI=-\tilde\omega+rH\partial_rI
\eeq
\begin{equation}
\partial_r I=-\frac{a H \tilde \omega (a r) \pm a\sqrt{\omega^2 -m^2 +
    m^2\,\left(H^2 + \frac{ k}{a^2}\right)\,(a r)^2}}{1-\left(H^2
  + \frac{ k}{a^2}\right) \,(a r)^2}\,,  
\label{g}
\end{equation}
with the sign chosen according to which direction we think the
particle is propagating. The effective mass here defines two important and complementary energy scales: if one is interested in the horizon tunnelling then only the pole matters (since the denominator vanishes), and we may neglect
to all the extents the mass parameter setting $m=0$ (since its
coefficient vanishes on the horizon).  On the opposite, in investigating other effects in the bulk away from the horizon, such as the decay rate of composite particles, the  role of the effective mass becomes relevant as the energy of the particle can be smaller than the energy scale settled by $m$, and the square
root can possibly acquire a branch cut singularity.  eq.(\ref{g} will be the starting point of our considerations since it embraces all semi-classical quantum effects we are interested in.
\
\\

\subsection{Horizon tunnelling}
\label{5.2}

Using eq.(\ref{g} we may derive, following \cite{bob09}, the cosmic 
horizon tunnelling rate. To this aim, as we have anticipated, the energy scale is such that near the  horizon, we may neglect the particle's mass, and note that radially moving massless particles follow a null direction. The horizon region being a region with strong gravity we expect indeed some ``particle creation''. To a pair created near the horizon in the inner region $r<r_{ H}$ corresponds a type-I path, one segment crossing the horizon backward in cosmological time. To a pair created outside the horizon corresponds a type-II path, the segment crossing the horizon inward being classically allowed. As in Section \ref{tu:tupa} only type-I paths will contribute to horizon tunnelling. 

Take then a null radial direction crossing the horizon to the inner region: we have
\begin{equation} 
\delta t = \frac{a(t)}{\sqrt{1-k r^2}} \,\delta r\, .
\label{ne}
\end{equation}
The action for particles coming out of the horizon towards the inner region  is to be integrated on a type-I path: then we must choose the solution with negative radial momentum, $\partial_{r}I<0$ and we obtain
\begin{equation}
I = \int d t\,\partial_t I + \int d r\,\partial_r I  =  2 \int d r \partial_r I
\,,\label{+action} 
\end{equation}
upon solving the Hamilton--Jacobi equation eq.(\ref{b}) with zero mass and using
(\ref{ne}). Note that the time derivative of the action contributes to the total imaginary part which, as we have seen, is a general feature of the tunnelling method. For $\partial_rI$ we use now Equation eq.(\ref{g}), which exhibits a
pole at the vanishing of the function $F(r,t):=1-(a^2H^2+\hat k)r^2 $,
defining the horizon position. Expanding $F(r,t)$ again along the  null direction eq.(\ref{ne}) we obtain
\begin{equation}
F(r,t) \approx  + 4 \kappa_{ H} a(t) (r-r_{ H}) +\dots \;,
\label{hay sg}
\end{equation}
where $\kappa_{ H}$ given in eq.(\ref{frw_dyn_sg} represents the
dynamical surface gravity associated with the horizon. 
In order to deal with the simple pole in the integrand, we implement Feynman's $i\epsilon$~{--}~prescription. In the final result, beside a real (irrelevant) contribution, we  obtain the following imaginary part
\begin{equation} 
\mbox{Im}\, I =-\frac{\pi \omega_{ H}}{\kappa_{ H}}\, .
\label{im} 
\end{equation}
This imaginary part is usually interpreted as arising because of a non-vanishing tunnelling probability rate of massless particles across the cosmological horizon, 
\begin{equation}
\Gamma \sim \exp\left(-2\mbox{Im} \,I\right) \sim e^{-\frac{2 \pi}{(-\kappa_{ H})}\cdot \omega_{ H}}\, .
\end{equation}
Notice that, since $\kappa_{ H} <0$ and $\omega_{ H} >0$ for physical
particles, (\ref{im}) is positive definite. Due to the invariant character of the quantities involved, we may interpret  the scalar $T=-\kappa_{ H}/2 \pi$ as the dynamical temperature parameter associated to FRW space-times. In particular, this gives naturally a positive temperature for de Sitter space-time, a long debated question years ago, usually resolved by changing the sign of the horizon's energy.
It should be noted that in literature, the dynamical temperature is usually given in the form  $T=H/2\pi$ (exceptions are the papers \cite{Wu:2008ir}).  
Of course this is the expected result for de Sitter space in inflationary coordinates, but it ceases to be correct in any other coordinate system. In this regard, the $\dot H$ and $k$ terms are crucial in order to get an invariant temperature. The horizon's temperature and the ensuing heating of matter was  foreseen several years ago in the interesting paper \cite{Brout:1987tq}.   
\
\\

\subsection{Decay rate of unstable particles}
\label{5.3}

We consider the decay rate of a composite particle in a regime where the energy of the decaying particle is lower than the proper mass $m$ of the decayed products. A crucial point is to identify the energy of the particle before the decay with the Kodama energy. We also denote $m$ the effective mass parameter of one of the decay products (recall it may contain a curvature terms). The relevant contribution to the action comes from the radial momentum given by  equation  (\ref{g}). If we introduce the instantaneous radius $r_{ 0}$ by
\begin{equation}
[a(t) r_{ 0}]^2 = R_{ 0}^2 := \left(1- \frac{\omega^2}{m^2}\right) R_{ H}^2\;,
\label{ir} 
\end{equation}
where $R_{ H}$ is the horizon radius given by Equation eq.(\ref{h}), then
the classically forbidden region is  $0< r < r_{ 0} $. From eq.(\ref{g}),
we see that for the unstable particle sitting at
rest at the origin of the comoving coordinates, one has an imaginary
part of the action as soon as the decay product is tunnelling into this region to escape beyond $r_{ 0}$, 
\begin{equation}
\mbox{Im}\, I= m R_{ H} \int_0^{R_{ 0}} d R \frac{\sqrt{R_{ 0}^2 - R^2}}{R_{ H}^2 - R^2}\,.
\end{equation}
The integral can be computed exactly at fixed $t$, and the result is
\begin{equation}
\mbox{Im}\, I = \frac{\pi}{2} \,R_{ H} \,(m-\omega) >0\,,
\end{equation}
leading to a rate which, assuming a two-particle decay, takes the form
\begin{equation}
\Gamma=\Gamma_{ 0} e^{-2\pi\,R_{ H}\,(m-\omega)}   \,,
\label{g1}
\end{equation} 
where $\Gamma_{ 0}$ is an unknown pre-factor depending on the coupling
constant of the interaction responsible of the decay (for instance, for a $\lambda\phi^3$ interaction one should have $\Gamma_{ 0}\sim\lambda^2$). Of course, each newly produced particle will itself decay, leading possibly to the instability mechanism first discussed by Myhrvold \cite{Myhrvold:1983hx} in de Sitter space. Since the tunnelling process locally conserves energy one
should put $\omega=m/2$, so that the tunnelled particle will emerge
in the classical region at $r=r_{ 0}$ with vanishing momentum. Furthermore, the result is again invariant against coordinate changes, since both $\omega$ and $R_{ H}$ are invariantly defined quantities. 

A particularly interesting case is represented by de Sitter space. Taking for example the line element in the static patch 
\begin{equation}
d s^2=-(1- H_{0}^2r^2) d t^2+\frac{d r^2}{(1- H_{ 0}^2 r^2 )}+ r^2 d \Omega^2\,, 
\label{ds s} 
\end{equation}
for the imaginary part (\ref{g1}) we obtain
\begin{equation}
\mbox{Im}\, I=\frac{\pi}{2H_{ 0}}(m-\omega)
\label{g11}
\end{equation}
a result actually  independent of the coordinate system in use. Putting  $\omega=m/2$, the above result has been obtained by Volovik \cite{Volo09} using the so-called \textquotedblleft fluid'' static form of de Sitter space
\begin{equation}
d s^2=-d t^2+(d R-H_{ 0} R d t)^2+ R^2 d \Omega^2\,.
\label{nolan nflat} 
\end{equation}
in agreement with the asymptotic approximation of the exact result due to \cite{ugo}.  
\
\\

\subsection{Particle creation by black holes singularities}
\label{5.4}

One may also use the tunnelling formalism to investigate whether particle creation in the bulk of space-time is possible due to the presence of space-time singularities, for example  due to static black holes.  With regard to this, we consider the exterior region of a spherically symmetric static 
black hole space-time and repeat the same argument. Quite generally, we
can write the line element as  
\begin{equation}
d s^2 = -  e^{2\Phi(r)} C(r)d t^2 + C^{-1}(r) d r^2 + r^2
d\Omega^2.\label{clifton} 
\end{equation}
From the Hamilton--Jacobi equation, the radial momentum turns out to be

\begin{equation}
\int d r\,\partial_r I = \int d r\, \frac{\sqrt{\omega^2 - m^2 C(r)
    e^{2\Phi(r)}}}{C(r)e^{\Phi(r)}}\;.  
\end{equation}
If we are interested in particle creation we should set $\omega=0$: in fact,
according to the interpretation of the Kodama energy we gave before, this approximation simulates the vacuum condition. Then 
\begin{equation}
\int d r\,\partial_r I =  m \int_{r_1}^{r_2} d r \frac{1}{\sqrt{-C(r)}}\,,
\label{inf} 
\end{equation}
where the integration is performed in every interval $(r_1,r_2)$ in
which $C(r) >0$.   eq.(\ref{inf} shows that, under very general conditions, in
static black hole space-times there could be a production rate whenever a
 region where $C(r)>0$ exists.

As a first example,  let us analyse the Schwarzschild black hole.  For the exterior (static) solution, one has  $C(r)=1-2M/r>0$ and $\Phi(r)=0$, thus the imaginary part diverges since the integral has an infinite range. We conclude that the
space-like singularity does not create particles in the semi-classical
regime. In the interior the Kodama vector is space-like, thus no energy
can be introduced. A similar conclusion has been obtained also for the
Big Bang cosmic singularity, the only scale factor leading to particle emission being $a(t)\sim t^{-1}$. This is like a big rip in the past.

The situation is different when a naked singularity is present. Consider a neutral particle in the Reissner--Nordstr\"om solution with mass $M$ and charge $Q>0$ (for definiteness) given by the spherically symmetric line element  
\begin{equation}
d s^2=-\frac{(r-r_-)(r-r_+)}{r^2}d t^2+\frac{r^2}{(r-r_-)(r-r_+)}\,d r^2+r^2d \Omega^2  \,.
\end{equation}
Here $r_{\pm}=M\pm\sqrt{M^2-Q^2}$ are the horizon radii, connected to
the black hole mass and charge by the relations  
\begin{equation}
M=\frac{r_+ + r_-}{2}, \qquad Q=\sqrt{r_+r_-}\,.
\end{equation}
The Kodama energy coincides with the usual Killing energy and
\begin{equation}
 C(r) = \frac{(r-r_-)(r-r_+)}{r^2}\;.
\end{equation}
The metric function $C(r)$ is negative in between the two horizons, where the Kodama vector is space-like, so there the action is real. On the other hand, it is positive within the outer communication domain, $r>r_+$, and also within the region contained in the inner Cauchy horizon, that is $0<r<r_-$. Thus,
because of eq.(\ref{inf}) and assuming the particles come created in pairs, we obtain
\begin{equation}
\label{ps}
 \mbox{Im}\, I=-m\int_0^{r_-}\frac{r}{\sqrt{(r_--r)(r_+-r)}}\;d r 
=  m q -\frac{mM}{2}\ln\left(\frac{M+Q}{M-Q}\right)\;.
\end{equation}
Modulo the pre-factor over which we have nothing to say, with the usual interpretation there is a probability 
\begin{equation}\label{p1}
\Gamma \sim \exp(-2\mbox{ Im}\, I) =
\left(\frac{M-Q}{M+Q}\right)^{mM}\!\!\!\!e^{-2 m Q }\, .
\end{equation}
Pleasantly, eq.(\ref{p1}) vanishes in the extremal limit $M=Q$. Being computed
for particles with zero energy, we can interpret this as an effect of particle
creation by the strong gravitational field near the singularity. Since the electric field
 is of order $Q/r^2$ near $r=0$, there should be also a strong Schwinger's effect. In that
 case one should write the Hamilton--Jacobi equation for charged particles.  

\begin{wrapfigure}{l}{0.6\textwidth}
\hspace{2cm} \includegraphics[scale=0.6]{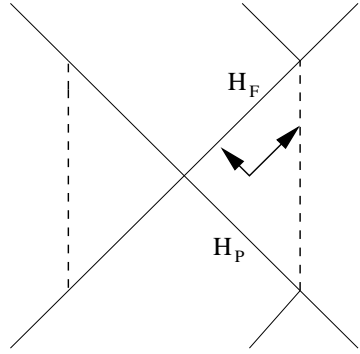}
 \begin{flushleft}
 \caption{  \label{rnn} A pair of particles is created in the compact region inside the inner horizon.}
 \end{flushleft}
\end{wrapfigure}

The processes just discussed should bear a bit on the question of the stability of the
 Cauchy horizons. Due to infinite blue-shift of
perturbations coming in from the asymptotically flat exterior regions
both sheets of the Cauchy horizon (${ H_F}$ and ${ H_P}$ in fig.\ref{rnn}) are
believed to be classically unstable. Of course, if the naked
singularity is formed from collapse 
of charged matter, one asymptotically flat region (say, the left one)
disappears.
Taking into account  particle creation, it can be easily seen that escaping
 particles will reach the future portion ${ H_F}$ of the inner horizon with
infinite blue-shift, or infinite Kodama energy, as
measured by an observer on a Kodama trajectory. 
 Hence, the future sheet
will probably become unstable, this time by quantum effects. On the
other hand, the particle reaching the singularity will do so with
infinite red-shift, that is with zero energy, giving a negligible back-reaction and substantially not
changing the nature of the singularity (its time-like character, for example). 
Thus, the only process potentially capable of modifying the singularity would be its conjectured screening by absorption of charged particles. Of course the mass inflation phenomenon is an issue here but, in the context of perfect spherical symmetry, exact solutions of the Einstein--Maxwell system exist having both event and Cauchy horizons and describing collapse of spherical shells. For these what we said should apply. The present formalism also predicts the absence of particle creation in the region in between the two horizons, despite the metric there is dynamical. The possibility that the naked singularity itself radiates away its mass will be taken up in the next section.  
\
\\

\subsection{Naked singularities}
\label{5.5}

A general reference on the physics of naked singularities is \cite{Harada:2001nj}. Particle emission from naked singularities in higher dimensions has been studied also in \cite{Miyamoto:2010vn}. Sticking to two-dimensional models for simplicity, consider the following metric \cite{vaz1993}
\beq
\label{ns}
 d s^2=\sigma^{-1}d x^+d x^-, \quad
\sigma=\lambda^2x^+x^--a(x^+-x_{ 0}^+)\Uptheta(x^+-x_{ 0}^+) 
\eeq
where $\lambda$ is related to the cosmological constant by $\Lambda=-4\lambda^2$. 
\
\\

\begin{wrapfigure}{l}{0.5\textwidth}
\begin{center}
\includegraphics[scale=0.50]{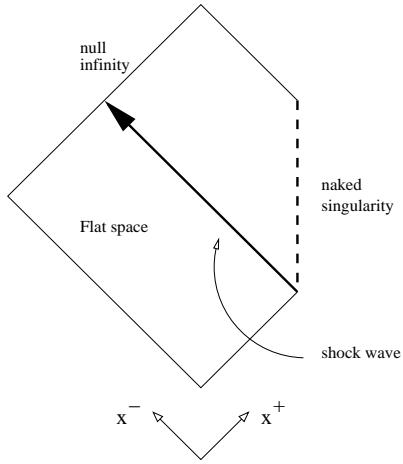}\hspace{-10pt}
\caption{\label{nsing} The naked singularity formed by the shock wave.}
\end{center}
\vspace{-10pt}
\end{wrapfigure}

This metric arises as a solution of two-dimensional dilaton gravity coupled to a bosonic field with stress tensor $T_{++}=2a\delta(x^+-x_{ 0}^+)$, describing a shock wave.  A look at fig.\ref{nsing} reveals that $\sigma=0$ is a naked singularity partly to the future of a flat space region, usually named the linear dilaton vacuum. The heavy arrow represents the history of the shock wave responsible for the existence of the time-like singularity. The Hamilton--Jacobi equation implies either $\partial_+I=0$ or $\partial_-I=0$, $I$ being the action. To find the ingoing flux we integrate along $x^+$ till we encounter the naked singularity, using $\partial_-I=0$, so that
\beq
I=\int d x^+\partial_+I = \int\omega\, \frac{d x^+}{2\sigma}
=\int\frac{\omega \,d x^+}{2(\lambda^2x^--a)(x^++ax_{ 0}^+/C-i\epsilon)}   
\eeq
where $C = C(x^-):=(\lambda^2x^- - a)$ and $\omega=2\sigma\partial_+I$ is the familiar Kodama's energy. The imaginary part immediately follows on using $(x-i\epsilon)^{-1}=P\frac{1}{x}+i\pi\delta(x)$, giving the absorption probability as a function of retarded time 
\beq
\Gamma(\omega)=\Gamma_{ 0}\, e^{-2 \mbox{ Im}\,I}=\Gamma_{ 0}\,e^{-\pi\omega/C(x^-)}\;,
\eeq
$\Gamma_{ 0}$ being some pre-factor of order one.
\
\\

The flux is computed by integrating the probability over the coordinate frequency (that is, the variable conjugated to the coordinate time)
$\hat{\omega}=\omega/\sigma$, with the density of states measure
$d\hat{\omega}/2\pi$, which gives
\begin{align}\label{iflx}
T_{++}&=\frac{\Gamma_{ 0}}{2\pi}\int\Gamma(\sigma\hat{\omega})\,\hat{\omega}\,
d\hat{\omega}\nonumber
\\
&= \Gamma_{ 0}\frac{(\lambda^2x^--a)^2}{2\pi^3\sigma^2}\;.
\end{align}
Similarly, in order to find the outgoing flux we
integrate along $x^-$ starting from the naked singularity, this time
using $\partial_+I=0$. A similar calculation gives 
\beq
\mbox{Im} I=\frac{\pi\omega}{2\lambda^2x^+}\;,
\eeq
then, integrating the probability over the coordinate frequency, we obtain
\beq\label{oflx}
T_{--}=\Gamma_{ 0}\frac{\lambda^4 (x^+)^2}{2\pi^3\sigma^2}\, .
\eeq
The outgoing flux is $2(T_{++}-T_{--})$. The conservation equations  
\begin{equation}
\sigma\partial_+T_{--}+\partial_-(\sigma T_{+-}) = 0, \qquad \sigma\partial_-T_{++}+\partial_+(\sigma T_{+-})=0
\end{equation}
determine the components only up to arbitrary functions $B(x^-)$ and $A(x^+)$, something which
 corresponds to the freedom in the choice of a vacuum. For instance, requiring the fluxes to vanish in the linear dilaton vacuum fixes them uniquely. As it is well known, $T_{+-}$ is given by the conformal anomaly: $T=4\sigma T_{+-}=R/24\pi$ (for
 one bosonic degree of freedom). Matching to the anomaly gives the pre-factor
 $\Gamma_{ 0}=\pi^2/24$, of order one indeed. These results agree with the
one-loop calculation to be found in \cite{vaz1993}. Note that the stress tensor diverges while 
 approaching the singularity, indicating that its resolution will not be possible within
 classical gravity but requires quantum gravity instead
 \cite{Harada:2000me,Iguchi:2001ya}. \par
We return now to the Reissner--Nordstr\"{o}m solution. Could it be that the naked singularity emits particles? In the four-dimensional case one easily sees that the action has no imaginary part along null trajectories either ending or beginning at the singularity. Formally this is because the Kodama energy coincides with the Killing energy in such a static manifold and there is no infinite red-shift from the singularity to infinity. Even
considering the metric as a genuinely two-dimensional solution, this
would lead to an integral for $I$ 
\beq
 I=\int\omega\,\frac{(r-r_+)(r-r_-)}{r^2}\,d x^+
\eeq
where $x^{\pm}=t\pm r_*$, with
\beq
 r_*=r+\frac{r_+^2}{r_+-r_-}\ln\left(\frac{r_+-r}{r_+}\right)-\frac{r_+^2}{r_+-r_-}
\ln\left(\frac{r_--r}{r_-}\right)=\frac{x^+-x^-}{2}\;.
\eeq
But close to the singularity
\beq
r^2=\left[\frac{3r_+r_-}{2}(x^+-x^-)\right]^{2/3}+\cdots
\eeq
not a simple pole, rather, an integrable singularity. It is fair to say that the  Reissner--Nordstr\"{o}m naked
singularity will not emit particles in this approximation. This seems to be coherent with quantum field theoretical results. 
With the customary $u=x^-$ and $v=x^+$, there is a map $u\to v=G(u)$ which gives the family of  ingoing null geodesics, characterised by constant values of $v$, which after reflection in the origin emerge as the the family of outgoing null geodesics with constant $u$. According to \cite{Ford:1978ip}, the radiated s-wave power of a minimally coupled scalar field is given in terms of the map $G(u)$ by the Schwarzian derivative 
\beq
\mathcal W=\frac{1}{24\pi}\left[\frac{3}{2}\left(\frac{G''}{G'}\right)^2-
\frac{G'''}{G'}\right]\;.
\eeq
The $(u,v)$ section of the Reissner--Nordstr\"{o}m metric is conformally flat, hence the
above map is trivial ({\it i.e.} linear) and $\mathcal W=0$. 

\section{Conclusions}
\label{6:conc}

We have completed our journey into the realm of quantum gravitational effects that can be understood as a result of tunnelling processes. In this journey, we visited black hole radiation, de Sitter and cosmological radiation, decays and naked singularities, but we have not included tunnelling between different gravitational field configurations, nor topology changing processes, nor sonic black holes or other analogue objects. Thus we may generically qualify the subject as covering semi-classical effects in a given background field. From this point of view it seemed remarkable that it could be extended to cover back-reaction effects to first order in the ratio $E/M$, where $E$ is the energy scale of the tunnelling process and $M$ the black hole mass or some generalisation thereof. We would like to conclude by summarising the achievements, the drawbacks and the prospects. 

{\it Achievements.} We list the results which seem to us more robust and have a chance to last longer. And also what is to be expected, reasonably, by pursuing this approach still further. Results  that were obtained in Section \ref{2:stat} and Section \ref{3:found} comprise:

\begin{enumerate}
\item The ability of the method to include first order back-reaction effects in order to comply with energy conservation and the predicted correlations among successive emission events is one of the key results.
\item In the stationary case, it is possible to provide a foundation of tunnelling methods using analytic continuation of the classical action alone, avoiding in this way the seemingly impossible task of crossing the horizon. The analytic continuation could be extended to cover white holes and to associate with them a well defined geometric temperature, a point not always appreciated in literature.
\item  In the same vein, one can prove the complete equivalence of the null geodesic method with the Hamilton--Jacobi method for stationary fields. We also gave a ``covariantised'' version of the null geodesic method, which allows to avoid questions such as  covariance or canonical invariance. 
\item Many special cases have been considered, including rotating black holes, G\"odel space-times, higher dimensional solutions, de Sitter and anti--de Sitter spaces, Taub and Taub--NUT solutions; also, we considered some supergravity solutions and fermion tunnelling, including gravitinos. The temperature of these solutions were computed and showed to agree with those obtained by geometric methods, confirming the universal character and kinematics of the Hawking effect. It should also be noticed that the Hawking temperature computed from the tunnelling method does not receive higher order corrections in $\hbar$, contrary to past suggestions.
\end{enumerate}
 
In Section \ref{4:dyna}, using a local notion of horizon and dynamical surface gravity for time varying spherically symmetric black holes, the main results comprise:
\begin{enumerate}
\item The application of the tunnelling method to non stationary, spherically symmetric  black holes and the fact that it naturally selects a trapping horizon of the future non degenerate type, either inner or outer. It can also be applied to past horizons under the same condition and by interchanging emission with absorption. In the limiting situation where the surface gravity vanishes it is consistent with the idea that extremal black  holes have zero temperature.
\item  One of the principal consequences of the tunnelling method is the observation that the radiation seems to originate near the local trapping horizon, not the global event horizon.
\item  Using the Kodama--Hayward theory of spherically symmetric space-times, and noting that one can possibly associate an observable temperature to a black hole only if it is a coordinate scalar, we showed that the tunnelling probability depends on the ratio, $2\pi E/\kappa$, of  the Kodama invariant charge taken as energy and  the geometrical surface gravity as defined by Hayward.
\item  One can include non interacting fermions in the formalism, as  demonstrated in several papers. Also noticeable is the fact that any mass term gets strongly suppressed by the horizon pole, which dominates the rate well over mass contributions. 
\item  The time-like nature of the trapping horizon of an excreting black hole is consistent with the physical interpretation of the formalism. It is true that it can be crossed either ways, but it remains the fact that there is a class of paths for which an imaginary part exists, and these paths precisely correspond in Feynman diagram language to particle creation.
\end{enumerate}
In Section \ref{5:coho} we addressed other themes -- cosmology, decays and naked singularities. Since FRW spaces are spherically symmetric, the general theory outlined in Section \ref{4.2} applies. We showed that, with obvious modifications, the tunnelling picture works equally well. In particular the type-II paths never contribute, while type-I paths do.  We may consider these findings for cosmological horizons as a generalisation of the well known facts discovered for de Sitter space-time, in complete analogy with the parallel extension of the theory from stationary to dynamical black holes. We also considered the decay of particles which in the absence of gravity would not occur, and find that the Hamilton--Jacobi equation can manage them through the presence of a branch point singularity in the radial momentum, that would not be present in absence of an external field. From this point of view the tunnelling method has universally valid features. Again, its value is to
  be found where it provides approximate results when exact calculations are in general impossible, at the same time being consistent with the exact calculations when available. Such is the case of  de Sitter space. We showed also that the presence of strong fields in a certain space-time region, even in the absence of horizons, can result  in particle production by means of a tunnelling process ``from nothing''. It can be considered as the gravitational analogue of the Schwinger  effect in a strong external electric field.  
Finally, it is of interest that the tunnelling method can also be applied to naked singularities, although in this case there is nothing to tunnel through, the singularity being the boundary  of space-time. We showed nevertheless that the emerging picture is quite in agreement with quantum field theory in two-dimensional models, making one confident of its general validity.

As regard the drawbacks, in the stationary case one can perhaps complain that the method gives only approximate results. But for black holes with really intricate metrics exact calculations are illusory even in stationarity, so we feel that this is not a serious drawback. Even if there is little doubt that the method is correct, still it remains of a hypothetical nature. However, we think that also in case of a failure due to some internal inconsistency, that should be nonetheless an important message.  In fact, tunnelling processes are predictions of quantum theory, so a failure of them in gravity theory would certainly signal something interesting. Despite all the desirable properties discussed above, the tunnelling method is and remains an essentially semi-classical procedure carrying along with it all the limitations inherent to its nature. In particular, it applies only to free particles while in principle quantum field theory methods certainly have a wider scope.

{\it Prospects.} The tunnelling method is not a closed subject. For instance, in striking contrast with the stationary case, an important missing point of the dynamical case is the absence of an extension of Kodama--Hayward's theory to dynamical axis-symmetric black holes, to be used as a tool for studying tunnelling. We mention here the papers \cite{jingWang,Yang:2003tc} where a tentative theory is developed. We feel that this is perhaps the most important missing point. The first back-reaction corrections are also of great interest, as they are not so well developed in dynamical situations. Processes where small black holes are emitted by large ones should also be within the range of problems where the tunnelling picture could work, and that also is an important ``to do''. 

In cosmology, the physical interpretation of Hayward's dynamical surface gravity as a temperature is not as clear as in de Sitter or black hole case, but it is strongly favoured by the tunnelling picture. In particular, an Unruh--DeWitt detector moving in de Sitter or black hole space-time always thermalises at the corresponding Gibbons--Hawking temperature even if it is not on a geodesic path, while for cosmology there is not an analogous result. One may conjecture that the cosmological temperature given by eq.(\ref{frw_dyn_sg} is the one that a point-like detector registers when moving on a Kodama trajectory, whose  four-velocity is the Kodama vector field. In general, such a trajectory is accelerated relative to geodesic observers so one expects a mixed contribution from the horizon and the acceleration together. These expectations were actually confirmed years ago in de Sitter and anti--de Sitter space and resulted in beautiful formulas \cite{Narnhofer:1996zk,Deser:1997ri}. It would be very interesting to apply tunnelling methods also to Unruh's like detectors (but see \cite{deGill:2010nb}). Finally, two more intriguing prospects are the application to field theories where fields, rather than particles, tunnel across forbidden regions; and to sonic black holes \cite{Unruh:1980cg}, or dielectric black hole analogues \cite{Schutzhold:2001fw}, since both can be studied in laboratory experiments. But these arguments would easily cover another review.

\end{document}